\documentclass[superscriptaddress,groupedaddress,nofootnoteinbib,11pt]{article}

\usepackage{graphicx}  % needed for figures
\usepackage{dcolumn}   % needed for some tables
\usepackage{bm}        % for math
\usepackage{jcappub}

\def\be{\begin{equation}}
\def\ee{\end{equation}}
\def\ba{\begin{eqnarray}}
\def\ea{\end{eqnarray}}

\def\d{\mathrm{d}}
\def\p{{\cal P}}
\def\S{\Sigma}

\def\L*{{\cal L}_*}
\def\L{\mathcal{L}}
\def\({\left(}
\def\){\right)}

\def\p{\partial}

\def\p{\partial}

\def\<{\langle}
\def\>{\rangle}

\def\cs2{c_{s}^{2}}

 \def\p{\partial}

 \def\be   {\begin{equation}}   \def\ee   {\end{equation}}
 \def\ba   {\begin{array}}      \def\ea   {\end{array}}
 \def\bea  {\begin{eqnarray}}   \def\eea  {\end{eqnarray}}
 \def\bean {\begin{eqnarray*}}  \def\eean {\end{eqnarray*}}
\renewcommand{\thesubsection}{\arabic{section}.\arabic{subsection}}
\begin{document}

\title{On the Trispectrum of Galileon Inflation}
\author{\hspace{-10pt} Frederico Arroja$^a$, Nicola Bartolo$^{a,b}$, Emanuela Dimastrogiovanni$^c$\\ \hspace{-10pt} and Matteo Fasiello$^d$}
\affiliation{$^a$ INFN, Sezione di Padova, via Marzolo 8,  I-35131 Padova, Italy}
\affiliation{$^b$ Dipartimento di Fisica e Astronomia ``G. Galilei'', Universit\`{a} degli Studi di
Padova, via Marzolo 8,  I-35131 Padova, Italy}
\affiliation{$^c$School of Physics and Astronomy, University of Minnesota, Minneapolis, 55455, USA}
\affiliation{$^d$Department of Physics, Case Western Reserve University, Cleveland, 44106, USA}
\affiliation{}
\affiliation{\hspace{-10pt } {\bf Email}: arroja@pd.infn.it, bartolo@pd.infn.it, emanuela1573@gmail.com, matte@case.edu.}
%\date{\today}
\leftline{UMN-TH-3214/13}

\abstract{We present a detailed study of the trispectrum of the curvature perturbation generated within a stable, well defined and predictive theory which comprises an inflationary phase. In this model the usual shift symmetry is enhanced up to the so-called \textit{Galileon} symmetry. The appeal of this type of theories rests on being unitary and stable under quantum corrections. Furthermore, in the specific model under consideration here, these properties have been shown to approximately hold in realistic scenarios which account for curved spacetime and the coupling with gravity.

\noindent In the literature, the analysis of the bispectrum of the curvature perturbation for this theory revealed non-Gaussian features which are shared by a number of inflationary models, including stable ones.
It is therefore both timely and useful to investigate further and turn to observables such as the trispectrum. We find that, in a number of specific momenta configurations, the trispectrum shape-functions present strikingly different features as compared to, for example, the entire class of the so-called $P(X,\phi)$ inflationary models.}

\maketitle

\section{Introduction}
The very recent release of the Planck mission data \cite{Ade:2013ktc,Ade:2013zuv,Ade:2013uln,Ade:2013nlj,Ade:2013ydc,Ade:2013xla} has been eagerly awaited because it does provide an improvement on the bounds we can place on fundamental cosmological observables such as non-Gaussianities \cite{Bartolo:2004if}. The value of the bispectrum amplitude $f_{NL}$ and, in general, higher order correlators of curvature fluctuations, provides an invaluable window on the physics of the early universe. Already the simplest standard single-field slow-roll scenario of inflation \cite{Lyth:1998xn} has been spectacularly successful in solving a host of cosmological puzzles and it is in agreement with observations \cite{Ade:2013uln,Ade:2013ydc}.

Quite naturally then, one wonders if the inflationary mechanism can be generated within UV finite theories such as string theory. This effort has culminated in \cite{Alishahiha:2004eh,Chen:2004gc,Chen:2005ad}, a model which has been studied in great details, often within the realm of the so-called $P(X,\phi)$ models,  in the literature \cite{Seery:2005wm,Chen:2006nt,Seery:2006vu,Huang:2006eha,Seery:2006js,Byrnes:2006vq,Arroja:2008ga,Langlois:2008wt,Langlois:2008qf,Arroja:2008yy,Seery:2008ax,Gao:2009gd,Mizuno:2009cv,Chen:2009bc,Arroja:2009pd,Mizuno:2009mv} (for other interesting specific realizations and reviews see \cite{Kachru:2003sx,Silverstein:2008sg,McAllister:2008hb,Flauger:2009ab,Berg:2009tg,Cline:2006hu,Kallosh:2007ig,Burgess:2011fa,Burgess:2013sla}). This theory is unitary and stable under quantum corrections and is embedded in string theory. As to the observational side, at least the single field $DBI$ model \cite{Alishahiha:2004eh}, in the range in which it can be trusted, is dangerously close to be excluded by data \cite{Ade:2013ydc}.

Recently in \cite{Burrage:2010cu}, a compelling effective field theory model which accounts for an inflationary phase has been put forward, the so-called {\it Galileon Inflation}. We will discuss the properties that characterize this theory at great length in \textit{Section 2}. The predictions extracted from this model at the level of non-Gaussianities amount so far to the calculation of the bispectrum of curvature perturbations \cite{Burrage:2010cu}. The predicted parameter $f_{NL}$ is compatible with values that fit within the bounds provided by Planck data. On the shape-function side though, the momentum dependence of the three-point function reveals no peculiar or highly distinctive features  which could be used as clear-cut signatures of this specific model. Indeed, there exist in the literature plenty of (mostly phenomenologically-oriented) models which share with the model of \cite{Burrage:2010cu} a detectably large bispectrum which peaks in the equilateral configuration, see e.g. \cite{Chen:2006nt,Langlois:2008wt,Langlois:2008qf}.

One would like to retain the interesting stability feature of the Galileon inflation model and supplement the bispectrum analysis with the study of other observables in the hope to remove some of the degeneracy  among inflationary models with similar predictions. It is therefore in this spirit that we present here a detailed analysis of the amplitude and shape-function of the different contributions to the trispectrum.
Alternatively, one might want to consider Galileon-inspired models \cite{Creminelli:2010qf,Bartolo:2013eka} which still possess to some degree the properties of the model under study here and offer more in the way of bispectrum signature as well as trispectrum ones.

There is some sort of ``conservation of desirable features'' one must deal with: the model we study here is compelling from a formal effective field theory perspective in that it qualitatively stands next to $DBI$ inflation, at least in terms of stability. It is, perhaps, also the model more in need of a trispectrum analysis considering the fact that the bispectrum one did not reveal outstanding signatures. It is certainly worth mentioning that another attempt at preserving the typical features of Galileon theories and yet  generate distinct prediction for the bispectrum was presented in \cite{Gao:2011qe,DeFelice:2011uc}. The models introduced there do indeed possess interesting features but the non-renormalization properties which we will briefly discuss and put at work in \textit{Section 2} no longer hold.

For the sake of the impatient reader, we summarize the main result of this work: performing the trispectrum analysis did pay back in terms of signatures. As compared with $DBI$ inflation and more generic $P(X,\phi)$ models \cite{Chen:2009bc,Arroja:2009pd}, the shape functions analysis shows novel features in our model for two of the four different momenta configurations analyzed. To mention just one of them, in the so-called  \textit{equilateral} configuration, the shape-function of a number of quartic interaction terms is markedly different from their $P(X,\phi)$ counterpart. The amplitudes are calculated and, considering the relative freedom in the coefficients which regulate the various interaction terms, it is no surprise that, where observational constraints are available, there is indeed room for a four-point function which fits within current experimental bounds \cite{Ade:2013ydc,Fergusson:2010gn,Sekiguchi:2013hza,Giannantonio:2013uqa}.

The paper is organized as follows. In \textit{Section 2} we give an introduction to our set up: the Galileon inflation model of \cite{Burrage:2010cu}. We go into some detail in order to explain what makes it interesting and in what aspects it differs from the prescriptions of \cite{Creminelli:2010qf,Bartolo:2013eka,Gao:2011qe}. In \textit{Section 3} we proceed to calculating non-Gaussian observables for our model: first we give a brief summary of the bispectrum study in the literature and then offer a thorough analysis of the trispectrum. We offer some comments on the results in the \textit{Conclusions}. In the \textit{Appendices} we give the analytical expressions for the various contributions to the trispectrum.

\section{The Theory}
On top of the shift symmetry, directly linked to the slow-rolling of the inflationary potential, Galileon theories are endowed, at least in flat space, with the symmetry on the scalar degree of freedom $\phi$ $: \phi \rightarrow \phi+ c + b_{\mu}x^{\mu}$ where $c, b_{\mu}$ are constant and $x$ is the spacetime coordinate. It mimics the coordinate transformation  between non-relativistic inertial frames, hence the name. Clearly, just about any term $\partial \partial..\, \phi$ with two or more derivatives is automatically Galilean invariant, at least in flat space. What we will employ here is a rather special (among other things, these terms give second order equation of motion) set of Galilean invariant terms. These are ubiquitous in the literature \cite{Dvali:2000hr,Nicolis:2008in,deRham:2010eu,Gabadadze:2012tr,Chow:2009fm,Silva:2009km,Kobayashi:2009wr,Creminelli:2010ba,DeFelice:2010pv,Ali:2010gr,DeFelice:2010nf,Kobayashi:2010cm,Mizuno:2010ag,RenauxPetel:2011uk,Renaux-Petel:2013ppa,Frusciante:2013haa} and keep attracting considerable attention. The Galileon terms under consideration are a highly restricted number; in four space-time dimension there are five of them (generically there are D terms in D dimensions plus the tadpole):
\bea
S= \int d^4 x \sqrt{-g}\Big[ c_1 \mathcal{L}_1+c_2\mathcal{L}_2+ c_3\mathcal{L}_3+c_4 \mathcal{L}_4+c_5\mathcal{L}_5 \Big].
\eea
The explicit expression for the $\mathcal{L}_n$ is:
\begin{eqnarray}
{\cal L}_1 & = & \pi  ,\label{lag1} \nonumber \\
{\cal L}_2 & = & -\frac{1}{2} \,  \partial \pi \cdot \p \pi  , \nonumber \\
{\cal L}_3 & = & - \frac{1}{2} \,  [\Pi] \, \partial \pi \cdot \partial \pi  , \nonumber \\
{\cal L}_4 & = & - \frac{1}{4} \big( [\Pi]^2 \,  \partial \pi \cdot \partial \pi - 2 \,  [\Pi] \, \partial \pi \cdot \Pi \cdot \partial \pi - [\Pi ^2 ] \, \partial \pi \cdot \partial \pi + 2 \, \partial \pi \cdot \Pi ^2 \cdot \partial\pi \big) ,
\label{galilean_flat} \nonumber \\
{\cal L}_5 & = & -\frac{1}{5} \big(
[\Pi]^3 \, \partial \pi \cdot \partial \pi- 3 [\Pi]^2 \,  \partial \pi \cdot \Pi \cdot \partial \pi
-3 [\Pi] [\Pi^2] \,  \partial \pi \cdot \partial \pi
+6 [\Pi] \, \partial \pi \cdot \Pi^2 \cdot \partial\pi
 \nonumber
 \\
&&
+2 [\Pi ^3] \, \partial \pi \cdot \partial \pi
+3 [\Pi ^2] \, \partial \pi \cdot \Pi \cdot \partial \pi
- 6 \,  \partial \pi \cdot \Pi^3 \cdot \partial\pi \big),
\end{eqnarray}
where  $(\partial_\mu \partial_\nu \pi)^n \equiv [ \Pi ^n ]$, $[...]$ stands for the trace operator and the $c_n$ are constants. Here we use the scalar field $\pi$ to adhere to the notation of \cite{Nicolis:2008in}, but could just as well have replaced $\pi$ with $\phi$ .
\noindent We discuss the flat space properties of this theory first, we then move on to clarify to what degree they hold in curved space.

\subsection{Flat Theory Properties}

- The theory has \textit{second-order equations of motion}: this guarantees it is, by construction \footnote{There are ways around this problem when one is dealing with higher derivative terms which first appear in the interactions. A crude way of putting it is that the derivative expansion is organized around a perturbative expansion parameter $\epsilon$ and a safe sector of the phase space is chosen \cite{Jaen:1986iz}. To accept this method one must think in terms of effective theories and be prepared to trust the perturbative expansion up to a certain point.  See also \cite{Woodard:2006nt,Chen:2012au}}, free of the Ostrogradski ghost \cite{Ostro,Chen:2012au}. At the quantum level, this translates into having a theory which is \textit{unitary}.\\
- The \textit{non-renormalization} properties of Galileon theories guarantee that the coefficients of each term is not largely renormalized \cite{Nicolis:2008in,Burrage:2010cu}, they are protected by the symmetry, therefore one can focus on a finite, small number of interactions and deal with a theory which is \textit{predictive}. The model in \cite{Gao:2011qe} represents an interesting generalization of the theory we study here but, whilst unitarity is preserved, the coefficients within the terms of their $\mathcal{L}_{n}$ can receive large corrections from renormalization.\\
The properties just discussed pertain to the flat space theory. In order to get to a realistic inflationary phase, several modifications need to be accounted for: inflation must end at some point and  we therefore ought to break the exact shift symmetry; de Sitter (dS) space, and eventually quasi dS, must take over Minkowski space; the coupling to gravity. Let us see if and how these modifications affect the nature of the theory.

\subsection{Towards Curved Spacetime and Gravity: the Route is Safe}

\noindent In order for inflation to end we need to break the shift symmetry. Switching on a linear potential $V(\phi)=V_0 -\lambda^3 \phi$ in dS space seems like the easiest way to do it.  Interestingly, upon going to the following decoupling limit:
\bea
M_{Pl}\rightarrow \infty, \qquad \,\, {\rm and}\,\, \qquad 3H^2 M_{Pl}^2= V_0  , \label{decoupling1}
\eea
\noindent one quickly realizes that the shift symmetry is still exact and therefore any later breaking is accompanied by $1/M_{Pl}$ coefficients. The authors of \cite{Burrage:2010cu} also show that in this limit the background $\dot \phi(t)$ is constant, a result which will be crucial in what follows \footnote{This is of paramount importance for the renormalization properties of the theory: the Lagrangian in Eq.~(\ref{full}) does receive renormalization contributions proportional to higher $\phi$ time derivative. Having $\ddot \phi_0=0$ in dS makes those term innocuous.}.

So far, at least in the region (\ref{decoupling1}), the shift symmetry is preserved. What about non-renormalization properties? Recall that Galilean invariance was the key ingredient for the coefficients of the various $\mathcal{L}_n$ not to be renormalized. One can prove \cite{Burrage:2010cu} that, in addition to those terms, also a linear and quadratic (no further) piece in the potential, despite breaking the symmetries, do not receive large renormalization. Because of what we see in the decoupling limit then, we rest assured that inflation can end and renormalization does not render the theory unpredictive. At this stage, once again, we use the fact that a breaking due to the coupling with gravity is bound to have a $\Lambda/M_{Pl}$ suppression \footnote{Here $\Lambda$ is generically a mass that represents the cutoff scale of the underlying theory.}  in front to infer that the theory remains  manageable also in more realistic regimes.

The other property that  needs discussing is classical stability; in other words we want to stay clear of Ostrogradski ghosts also in curved space. The work in \cite{Deffayet:2009wt}  (see also \cite{Deffayet:2009mn,Deffayet:2010zh}) does precisely what we need here: a covariantization procedure is  spelled out that does guarantee second order equations of motion.
The resulting action (besides the \textit{Einstein-Hilbert} term and the potential) includes:
 \bea
S =  \int \d^4 x \; \sqrt{-g} \, \Big[	- \frac{c_2}{2} (\nabla \phi)^2
				+ \frac{c_3}{\Lambda^3} \Box \phi ( \nabla \phi )^2
				- \frac{c_4}{\Lambda^6} ( \nabla \phi )^2 \Big\{
					(\Box \phi)^2 - (\nabla_\mu \nabla_\nu \phi)
					(\nabla^\mu \nabla^\nu \phi)
					- \frac{1}{4} R (\nabla \phi)^2
				\Big\}	+\nonumber \\ \frac{c_5}{\Lambda^9} ( \nabla \phi )^2 \Big\{
					(\Box \phi)^3 - 3 (\Box \phi)( \nabla_\mu \nabla_\nu \phi)
					(\nabla^\mu	 \nabla^\nu \phi)
					+ 2 ( \nabla_\mu  \nabla_\nu \phi)
					(\nabla^\nu	 \nabla^\alpha \phi)
					(\nabla_\alpha \nabla^\mu \phi)
					- 6 G_{\mu \nu} \nabla^\mu \nabla^\alpha \phi
					\nabla^\nu \phi \nabla_\alpha \phi
			\Big\}	\Big] .  \label{full} \nonumber \\
\eea
\noindent
\noindent Note that this is not the result of the most intuitive, straightforward $(\partial \rightarrow \nabla)$ covariantization procedure, the latter would have left us with a theory plagued by ghosts \cite{Deffayet:2009wt}. The constants $c_2$, $c_3$, $c_4$ and $c_5$ are model parameters and one of them can be set to one by a redefinition of the scalar field.

\subsection{The Interesting Regime}

By considering Eq.~(\ref{full}) at the background level in the decoupling limit of (\ref{decoupling1}), one finds \cite{Burrage:2010cu}:
\bea
S_0 = \int \d^4 x \; a^3 \left( \frac{c_2}{2} \dot{\phi_0}^2+ \frac{2c_3 H}{\Lambda^3} \dot{\phi_0}^3+\frac{9c_4H^2}{2\Lambda^6} \dot{\phi_0}^4+ \frac{6c_5H^3}{\Lambda^9} \dot{\phi_0}^5+ \lambda^3 \phi_0	\right),
\eea
\noindent which, upon defining $Z\equiv H \dot \phi_0 / \Lambda^3 $, can be written as:
\bea
S_0 = \int \d^4 x \; a^3 \Bigg[\dot{\phi_0}^2 \left( \frac{c_2}{2}+ 2c_3\,Z\,  +\frac{9c_4 }{2} \,Z^2\, + 6c_5\,Z^3\, \right)+ \lambda^3 \phi_0\Bigg]	.
\eea

\noindent  Interesting information can already be extracted at this level. Using the fact that $\ddot \phi_0=0$ in this limit, as mentioned before, one realizes that the dynamics of $\dot \phi$ can occur in two different regimes. According to the value of $Z$, the theory is in a  weakly coupled regime ($Z\ll 1$), where the theory approaches the canonical inflation model, or a regime ($Z\gtrsim 1$) where the $c_n$ coefficients (with $n\geq 3$, that is, the Galileon interactions) are increasingly important.

It is easy to show \cite{Burrage:2010cu} at this point that for $Z\gtrsim 1$ the main contribution to the fluctuations comes from the non-linearities represented by the Galileon terms and therefore the metric fluctuations can be disregarded. As a side note, we can add that one could have considered at the onset non-minimal coupling with gravity, such as via a Gauss-Bonnet term, without spoiling much in the model. On the other hand, as shown in \cite{Burrage:2010cu}, this type of terms are sub-leading precisely in the $Z\gtrsim1$ regime where Galileon interactions play the main role; they can therefore be discarded in this context.

In what follows we will neglect metric perturbations, this is because we will work in the decoupling limit \cite{Burrage:2010cu}. The latter is completely analogous  to what happens  within the effective theory approach to inflation of \cite{Cheung:2007st} (see also \cite{Bartolo:2010bj,Bartolo:2010di,Bartolo:2010im,Fasiello:2011fj,Renaux-Petel:2013wya}): there the large non linearities of the pseudo-goldstone boson $\pi$ win over the mixed $\pi$--metric perturbations above a certain energy regime $E_{mix}$ \footnote{$E_{mix}$ itself depends upon which is assumed to be the leading quadratic kinetic term in the Lagrangian (it can in principle be non-standard) and therefore, ultimately, on the proper canonical normalization.}.
This scale $E_{mix}$ must be paired with the scale $\Lambda$ of the underlying theory when identifying  the regime of validity for the effective field theory.

Where we part ways with the approach of \cite{Cheung:2007st} to Galileon inflation, such as the interesting work in \cite{Creminelli:2010qf}, is  on the fact that in \cite{Creminelli:2010qf,Bartolo:2013eka} a Galileon theory of fluctuations around FLRW is developed, while a well-behaved (from a renormalization perspective) background is considered as an assumption (which, in light of the results of \cite{Burrage:2010cu} and of several previous works, it certainly seems to be quite a reasonable one).

A note of caution here: it is important to keep in mind that the decoupling limit of Eq.~(\ref{decoupling1}) is completely different from the one taken in \cite{Cheung:2007st}: the former has been used here to infer properties of the theory at hand in the $M_{Pl}\rightarrow \infty$ dS limit, and parametrically extend them to the full theory.  The decoupling limit of  \cite{Cheung:2007st} is in direct correspondence to the simplifications that occur in the $Z\gtrsim1$ regime, chief among which \cite{Burrage:2010cu} the possibility to consistently neglect metric fluctuation, which are known to be subdominant in the corresponding \textit{effective field theory of inflation} regime of  \cite{Cheung:2007st}. In the following sections we will employ this last decoupling limit.

\section{Non-Gaussianities}
We set out to calculate non-Gaussian observables now. We are working in a regime where metric fluctuations can be neglected; the gauge choice is equivalent, to first order in slow roll, to expressing the observable $\zeta$, comoving curvature perturbation, as $\zeta=-\frac{H}{\dot{\phi}_0} \delta \phi$, where $\phi(t,\vec x)=\phi_0 (t)+\delta\phi(t,\vec x)$. Implementing perturbations in Eq.~(\ref{full}) in the uniform curvature gauge, one obtains the quadratic, cubic and quartic action in fluctuations.

We start from the quadratic action, determine the wavefunction and normalize it according to the Bunch-Davies (B-D) vacuum prescription. The quadratic action is \footnote{We have set $M_{Pl}=1;\,\, \hbar=1;\,\, c=1.$}:
\begin{eqnarray}
S^{(2)}&=&\int d^4x\bigg[
a^3(\dot{\delta\phi})^2\left(\frac{c_2}{2} +6\frac{H\dot\phi_0}{\Lambda^{3}}c_3+27\frac{H^2(\dot\phi_0)^2}{\Lambda^{6}}c_4+60\frac{H^3(\dot\phi_0)^3}{\Lambda^{9}}c_5\right)
\nonumber\\
&&\quad\qquad-a(\partial\delta\phi)^2\left(\frac{c_2}{2}+4 \frac{H\dot\phi_0}{\Lambda^{3}} c_3+ 13\frac{H^2(\dot\phi_0)^2}{\Lambda^{6}} c_4+24 \frac{H^3(\dot\phi_0)^3}{\Lambda^{9}}\,c_5\right)
\bigg],  \label{3.1}
\eea
where from now on we neglect all boundary terms that appear after integration by parts when simplifying the action.\\
\noindent The quadratic action serves in the IN-IN formalism \cite{Schwinger:1960qe,Calzetta:1986ey,Jordan:1986ug,Weinberg:2005vy} as the free theory, above which one switches on interactions. Upon B-D normalization, one obtains for the wave function at the leading order in slow-roll:
\bea
\zeta_{k}=-\frac{H^2}{2\dot{\phi}_0\sqrt{A(kc_{s})^3}}\left(1+i k c_{s}\tau\right)e^{-ikc_{s}\tau},
\eea
where the sound speed $c_{s}^{2}=-B / A$, $\,$ $Z\equiv H \dot{\phi}_0/ \Lambda^3$, and
\bea
A= \frac{1}{2}\left( c_2+12 c_3 Z + 54 c_4 Z^2 + 120 c_5 Z^3 \right), \quad B=- \frac{1}{2}\left( c_2+8 c_3 Z + 26 c_4 Z^2 + 48 c_5 Z^3 \right) , \nonumber \\
\eea
\noindent where slow-roll suppressed contributions, such as those $\sim \ddot \phi$ , have been omitted.
Note here that, as we will see below, $Z$ plays the role of a coupling constant for the Galilean non-linearities. An $A>0$ grants  no ghost appears and a $B<0$ no Laplace instability.
Proceeding similarly for  interactions, one has the cubic order action,
\begin{eqnarray}
S^{(3)}&=&\int d^4xa^3\bigg[
2H\left(\frac{c_3}{\Lambda^{3}}+9\frac{\dot\phi_0H}{\Lambda^{6}}c_4+30\frac{(\dot\phi_0)^2H^2}{\Lambda^{9}}c_5\right)(\dot{\delta\phi})^3
\nonumber\\
&&\quad\qquad\qquad-2a^{-2}\left(\frac{c_3}{\Lambda^{3}}+6\frac{H\dot\phi_0}{\Lambda^{6}}c_4+18\frac{H^2(\dot{\phi_0})^2}{\Lambda^{9}}c_5\right)(\dot{\delta\phi})^2\partial^2\delta\phi
\nonumber\\
&&\quad\qquad\qquad-2Ha^{-2}\left(\frac{c_3}{\Lambda^{3}}+7\frac{H\dot\phi_0}{\Lambda^{6}}c_4+18\frac{H^2(\dot{\phi_0})^2}{\Lambda^{9}}c_5\right)\dot{\delta\phi}(\partial\delta\phi)^2
\nonumber\\
&&\quad\qquad\qquad+a^{-4}\left(\frac{c_3}{\Lambda^{3}}+3 \frac{H\dot\phi_0}{\Lambda^{6}} c_4+6\frac{(\dot{\phi_0})^2 H^2}{\Lambda^{9}} c_5\right)\partial^2\delta\phi(\partial\delta\phi)^2
\bigg] .\label{esse3}
\end{eqnarray}
The above action as well as Eq.~(\ref{3.1}) have been obtained in \cite{Burrage:2010cu} ($\dot{\phi}_0\,\xi= \delta \phi $ for the conversion) up to second-order in slow-roll and used to study  the three-point function amplitude and profile. For later convenience, we define linear combinations of the coefficients multiplying each one of the four operators in Eq.~(\ref{esse3}) above as $\mathcal{O}_1,\mathcal{O}_2,\mathcal{O}_3,\mathcal{O}_4$, this according to the order in which they are written above (see Eq.~\ref{009}).

What one is after in cosmological setups is the so-called \textit{equal-time correlator}; for a generic observable $\mathcal{O}(t)$ then:
\begin{eqnarray}
\langle \mathcal{O}(t) \rangle &=&\langle 0| \left[\bar{T}\exp\left(i\int_{t_0}^t H_I(t)\,dt\right)\right]\,\mathcal{O}^I(t)\,\left[T\exp\left(-i\int_{t_0}^t H_I(t)\,dt\right)\right] |0 \rangle \;, \label{def}
\end{eqnarray}
where the index $I$ stands for \textit{interaction picture} operators and $\bar T$ is the anti time-order operator. We  will shortly introduce  the main ingredients of the machinery that is employed in calculating higher order correlators, namely the IN-IN formalism, but already from Eq.~(\ref{def}) we can see the starting point is the fluctuations Hamiltonian in the interaction picture, $H_I(t)$.

As it is well known, up to third order in perturbations, going from the Lagrangian to the Hamiltonian density usually consists of a mere sign flip: $\mathcal{H}_n=-\mathcal{L}_n\,; \,\, n\leq 3$. Things are slightly more involved at higher orders \cite{Huang:2006eha}. Here we  give directly the final expression for the quartic action and Hamiltonian without the steps in between.
\begin{eqnarray}
S^{(4)}=\int d^4xa^3\Bigg[
&H^2&(\dot{\delta\phi})^4\left(\frac{9}{2}\frac{c_4}{\Lambda^{6}}+30H\dot\phi_0\frac{c_5}{\Lambda^{9}}\right)\nonumber \\
&+&a^{-2}\left(
H(\dot{\delta\phi})^2\partial\dot{\delta\phi}\partial\delta\phi(12\frac{c_4}{\Lambda^{6}}+72H\dot\phi_0\frac{c_5}{\Lambda^{9}})
-H^2(\dot{\delta\phi})^2(\partial\delta\phi)^2(7\frac{c_4}{\Lambda^{6}}+36H\dot\phi_0\frac{c_5}{\Lambda^{9}})
\right)
\nonumber\\
&+&  a^{-4}\bigg[
(\partial\delta\phi)^4\left(2H^2\frac{c_4}{\Lambda^{6}}+6H^3\dot\phi_0\frac{c_5}{\Lambda^{9}}\right)+6H\dot{\delta\phi}(\partial\delta\phi)^2\partial^2\delta\phi(\frac{c_4}{\Lambda^{6}}+4H\dot\phi_0 \frac{c_5}{\Lambda^{9}})
\qquad \qquad \nonumber\\
&+&3(\dot{\delta\phi})^2\left((\partial^2\delta\phi)^2-\partial_i\partial_j\delta\phi\partial^i\partial^j\delta\phi\right)(\frac{c_4}{\Lambda^{6}}+4H\dot\phi_0\frac{c_5}{\Lambda^{9}})
\Bigg]
\nonumber \\&-&a^{-6}(\partial\delta\phi)^2\left((\partial^2\delta\phi)^2-\partial_i\partial_j\delta\phi\partial^i\partial^j\delta\phi\right)\frac{c_4}{\Lambda^{6}}
\Bigg],
\end{eqnarray}
where we have kept the explicit dependence on the $c_n$ coefficients. After some algebra, one finds the quartic interaction Hamiltonian density to be:
\begin{eqnarray}
\mathcal{H}_I^{(4)}&=&
V_1a^3(\dot{\delta\phi})^4
+V_2a(\dot{\delta\phi})^3\partial^2\delta\phi
+V_3a(\dot{\delta\phi})^2(\partial\delta\phi)^2
+V_4a^{-1}
(\dot{\delta\phi})^2(\partial^2\delta\phi)^2\nonumber\\
&&+V_5a^{-1}(\dot{\delta\phi})^2\partial_i\partial_j\delta\phi\partial^i\partial^j\delta\phi +V_6a^{-1}\dot{\delta\phi}(\partial\delta\phi)^2\partial^2\delta\phi
+V_7a^{-1}(\partial\delta\phi)^4
\nonumber\\
&&
+V_8a^{-3}(\partial\delta\phi)^2\left((\partial^2\delta\phi)^2-\partial_i\partial_j\delta\phi\partial^i\partial^j\delta\phi\right),
\label{FourthOrderHamiltonian}
\end{eqnarray}
where the $V_n$ coefficients are simple functions of the $c_n$ parameters as well as the background quantities $\dot{\phi}_0,H$ and the scale $\Lambda$. The precise definition is given below:
\begin{eqnarray}
V_1=\frac{9\mathcal{O}_1^2}{4A}-\alpha,\quad
V_2&=&\frac{3\mathcal{O}_1\mathcal{O}_2}{A}-\beta_1,\quad
V_3=\frac{3\mathcal{O}_1\mathcal{O}_3}{2A}-\beta_2,\quad
V_4=\frac{\mathcal{O}_2^2}{A}-\gamma_1,\quad
V_5=\gamma_1,\nonumber\\
V_6&=&\frac{\mathcal{O}_2\mathcal{O}_3}{A}-\gamma_2,\quad
V_7=\frac{\mathcal{O}_3^2}{4A}-\gamma_3,\quad
V_8=-\Delta,
\end{eqnarray}
\noindent  where:
\begin{eqnarray}
&& \mathcal{O}_1\equiv \frac{2}{\dot\phi_0}\left(Z c_3+9\,Z^{2}c_4+30\,Z^{3}c_5\right),\quad\quad\quad\quad\,\,\,\,\,
\mathcal{O}_2\equiv -\frac{2}{H\dot\phi_0}\left(Zc_3+6\,Z^{2}c_4+18\,Z^{3}c_5\right),
\nonumber\\
&& \mathcal{O}_3\equiv -\frac{2}{\dot{\phi}_{0}}\left(Z c_3+7\,Z^{2} c_4+18\,Z^{3}c_5\right),\quad\quad\quad\quad
\mathcal{O}_4\equiv \frac{1}{H\dot\phi_0}\left(Z c_3+3\,Z^{2}c_4+6\,Z^3 c_5\right),
\nonumber\\
&& \alpha \equiv\frac{1}{\dot{\phi}_{0}^{2}}\left(\frac{9}{2}Z^{2}c_4+30\,Z^{3}c_5\right),\quad\quad\quad\quad\quad\quad\quad\,\,\,
\beta_1\equiv -\frac{4}{H\dot{\phi}_{0}^{2}}\left(Z^{2} c_4+6\,Z^{3}c_5\right),\nonumber\\
&& \beta_2\equiv -\frac{1}{\dot{\phi}_{0}^2}\left(7\,Z^{2}c_4+36\,Z^{3}c_5\right),\quad\quad\quad\quad\quad\quad\quad
\gamma_1\equiv \frac{3}{H^{2}\dot{\phi}_{0}^{2}}\left(Z^{2}c_4+4\,Z^{3} c_5\right),\nonumber\\
&& \gamma_2\equiv \frac{6}{H\dot{\phi}_{0}^{2}}\left(Z^{2}c_4+4\,Z^{3}c_5\right),\quad\,
\gamma_3\equiv  \frac{2}{\dot{\phi}_{0}^2}\left(Z^{2} c_4+3\,Z^3 c_5\right),\quad\,
\Delta\equiv -\frac{1}{H^{2}\dot{\phi}_{0}^{2}}\left(Z^{2}c_4\right),
\label{009}
\end{eqnarray}
\noindent with the last expressions above to be used at leading order in the slow-roll approximation. The interested reader may find a (necessarily partial, the two theories are clearly not equivalent)  dictionary between the Galileon inflation fluctuations Lagrangian and the $P(X,\phi)$ one in \textit{Appendix E}.

\noindent The observables we intend to calculate, non-Gaussianities, are respectively  the three and the four-point function for $\zeta$ in Fourier space. It is however convenient to start by giving the definition of the power spectrum for $\zeta$. One usually isolates a momentum-conservation Dirac delta from the quantities to be handled, in the case of the two-point function:
\bea
\langle \zeta(\tau,{\bf k}_1) \zeta(\tau,{\bf k}_2) \rangle \Big|_{\tau\rightarrow 0} = (2\pi)^3 \delta^{(3)}\left({\bf k}_1+{\bf k}_2\right) P({k_1})\,\,,
\eea
it is also useful for what follows to introduce the quantity $\mathcal{P}_{\zeta}=P(k) \, k^3/(2\pi^2)=H^4/(8\pi^2 A \dot{\phi_0}^2 c_s^3)$.

\noindent The bispectrum reads:
\bea
\langle \zeta(\tau,{\bf k}_1) \zeta(\tau,{\bf k}_2) \zeta(\tau,{\bf k}_3) \rangle  \Big|_{\tau\rightarrow 0}  =(2\pi)^7 \delta^{(3)}({\bf k}_{1}+{\bf k}_{2}+{\bf k}_{3})  \mathcal{P}^{2}_{\zeta}\, \mathcal{B}(k_1,k_2,k_3)\,\,\,,
\eea
where $\mathcal{B} \times  \prod_i k_i^2 $ is the bispectrum shape-function one usually plots.

\noindent The bispectrum amplitude, $f_{NL}$ is defined in the equilateral limit as:
\bea
f_{NL}=-\frac{10}{3} \frac{\mathcal{B}(k_1,k_2,k_3)}{\big[ \frac{1}{k_1^3 k_2^3}+ \frac{1}{k_2^3 k_3^3}+ \frac{1}{k_1^3 k_3^3} \big]}\Bigg|_{k_1=k_2=k_3=k} \,\,\,.
\eea

\noindent Moving on to the trispectrum, the amplitude $t_{NL}$ \cite{Chen:2009bc} in the regular tetrahedron limit $(k\equiv k_1=k_2=k_3=k_4=k_{12}=k_{14})$ is extracted from the formula below
\bea
\langle \zeta({\tau,{\bf k}_1}) \zeta({\tau,{\bf k}_2})\zeta({\tau,{\bf k}_3})\zeta({\tau,{\bf k}_4}) \rangle_{{\rm reg.\,tetra}}  \Big|_{\tau\rightarrow 0} \equiv (2\pi)^9 \mathcal{P}_{\zeta}^3\, \delta^{(3)}\left(\sum_{i=1}^{4} {\bf k}_i \right) \frac{1}{k^9}\, t_{NL}\,\,\,.
\label{tnl}
\eea
\noindent Here we note that we will give below (Eq.~\ref{exchange} and \ref{contact}) the expressions for each contribution $t_{NL}$, originated by the various interactions terms, to the  total $t_{NL}$ . \\The trispectrum form factor $\mathcal{T}$ is defined in what follows:
\bea
\langle \zeta({\tau,{\bf k}_1}) \zeta({\tau,{\bf k}_2})\zeta({\tau,{\bf k}_3})\zeta({\tau,{\bf k}_4}) \rangle \Big|_{\tau\rightarrow 0} \equiv (2\pi)^9 \mathcal{P}_{\zeta}^3\, \delta^{(3)}\left(\sum_{i=1}^{4} {\bf k}_i \right) \prod_{i=1}^{4} \frac{1}{k_i^3}\mathcal{T}(k_1,k_2,k_3,k_4,k_{12},k_{14})\,\,\, . \nonumber \\ \label{trishapedef}
\eea
Notice that, in the following, we will plot the most interesting $\mathcal{T}_{(i)}$ among the leading contributions to the total $\mathcal{T}$. The contact interaction (CI) contributions, which come from the fourth-order Hamiltonian, are to be found in Eqs.~(\ref{6.3}-\ref{6.10}) of \textit{Appendix B} while the scalar exchange (SE) contributions, which come from the third-order action, can be found in Eqs.~(\ref{7.2}-\ref{7.11}) of \textit{Appendix C}. The final total trispectrum of \textit{Galileon inflation} is the sum of all the CI and SE contributions.

\subsection{Summary on the Bispectrum}

The bispectrum analysis for the theory under investigation in this paper has been performed in \cite{Burrage:2010cu}. It suffices here to say that the shape-functions associated to the interactions terms that contribute to the three-point function peak in the equilateral configuration \footnote{However, it is important to mention that it is possible to adjust the coupling constants so as to obtain  differently shaped contributions to the bispectrum.}. The equilateral profile characterizes also third order interactions in $P(X,\phi)$ \cite{Chen:2006nt} models as well as other inflationary mechanisms.

\noindent We plot below in Fig.~(\ref{default1})  a sample  shape-function that represents a typical \footnote{More precisely, the contribution plotted here is the one given by Eq. (\ref{5.4}) (\textit{$Shape_3$}), one of the three obtained in \cite{Burrage:2010cu}, which all peak in the equilateral configuration. Their analytical expressions are given in \textit{Appendix A}.} contribution to the bispectrum.
\begin{figure}[htbp]
\begin{center}
\includegraphics[scale=0.70]{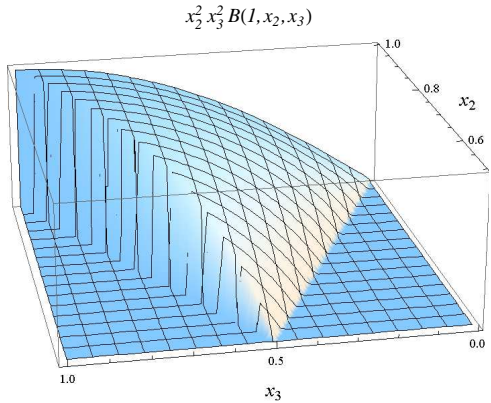}
\caption{The shape-function peaks in the equilateral ($k_{2,3}/k_1\equiv x_{2,3}=1$) limit.}
\label{default1}
\end{center}
\end{figure}

\noindent  The realization that this theory produces a somewhat common bispectrum profile, naturally calls for an analysis of the four-point function. The latter is  an observable  which now becomes paramount in (possibly) enabling one to remove the degeneracy between the model we study here and  other appealing realizations such as $DBI$ inflation. It is to this aim that we now move to the trispectrum analysis. \\

\subsection{The Trispectrum}
\subsubsection{Amplitudes}
The contributions to the trispectrum amplitude originate from cubic and quartic interactions. In the IN-IN formalism one uses usual Feynman diagrams to organize the different terms. Just as in standard field theory then, one has at tree-level for the 4-point function a contact interaction diagram (whose contributions come from $\mathcal{H}_4$) and a scalar exchange diagram (the latter is fed by cubic interactions).

All interactions contribute to the total trispectrum. On the other hand, it is also useful to single out and compare the different contributions to the four-point function from the various interaction terms. In particular, we give below the corresponding $t_{NL}$ amplitudes obtained in the \textit{regular tetrahedron} limit as detailed in Eq.~(\ref{tnl}). The ones generated by terms populating the cubic action, the scalar exchange contributions, are:
\begin{eqnarray}
&&t_{NL}^{\mathcal{O}_1\mathcal{O}_1}=0.063\frac{\mathcal{O}_1^2\dot\phi_0^2}{A^2},\quad
t_{NL}^{\mathcal{O}_3\mathcal{O}_3}=0.31\frac{\mathcal{O}_3^2\dot\phi_0^2}{A^2c_s^4},\quad
t_{NL}^{\mathcal{O}_1\mathcal{O}_3}=-0.21\frac{\mathcal{O}_1\mathcal{O}_3\dot\phi_0^2}{A^2c_s^2},\quad
t_{NL}^{\mathcal{O}_1\mathcal{O}_2}=0.25\frac{H\mathcal{O}_1\mathcal{O}_2\dot\phi_0^2}{A^2c_s^2},\quad
\nonumber\\
&&
t_{NL}^{\mathcal{O}_1\mathcal{O}_4}=-0.33\frac{H\mathcal{O}_1\mathcal{O}_4\dot\phi_0^2}{A^2c_s^4},\quad
t_{NL}^{\mathcal{O}_2\mathcal{O}_3}=-0.37\frac{H\mathcal{O}_2\mathcal{O}_3\dot\phi_0^2}{A^2c_s^4},\quad
t_{NL}^{\mathcal{O}_2\mathcal{O}_4}=-0.50\frac{H^2\mathcal{O}_2\mathcal{O}_4\dot\phi_0^2}{A^2c_s^6},\quad
\nonumber\\
&&
t_{NL}^{\mathcal{O}_3\mathcal{O}_4}=0.85\frac{H\mathcal{O}_3\mathcal{O}_4\dot\phi_0^2}{A^2c_s^6},\quad
t_{NL}^{\mathcal{O}_2\mathcal{O}_2}=0.23\frac{H^2\mathcal{O}_2^2\dot\phi_0^2}{A^2c_s^4},\quad
t_{NL}^{\mathcal{O}_4\mathcal{O}_4}=0.52\frac{H^2\mathcal{O}_4^2\dot\phi_0^2}{A^2c_s^8} , \label{exchange}
\end{eqnarray}
where generically $t_{NL}^{\mathcal{O}_{m} \mathcal{O}_{n}}$ stands for the scalar exchange contribution to the trispectrum amplitude which is generated by the interactions $\mathcal{O}_{m}$ and $\mathcal{O}_{n}$ at the two vertices of the diagram.

\noindent As for the quartic action terms generating contribution to the contact interaction diagram, one has:
\begin{eqnarray}
&&
t_{NL}^{V_1}=-0.035\frac{V_1\dot\phi_0^2}{A},\quad
t_{NL}^{V_2}=-0.079\frac{HV_2\dot\phi_0^2}{Ac_s^2},\quad
t_{NL}^{V_3}=0.051\frac{V_3\dot\phi_0^2}{Ac_s^2},\quad
t_{NL}^{V_4}=-0.19\frac{H^2V_4\dot\phi_0^2}{Ac_s^4},\quad
\nonumber\\
&&
t_{NL}^{V_5}=-0.031\frac{H^2V_5\dot\phi_0^2}{Ac_s^4},\quad
t_{NL}^{V_6}=0.10\frac{HV_6\dot\phi_0^2}{Ac_s^4},\quad
t_{NL}^{V_7}=-0.20\frac{V_7\dot\phi_0^2}{Ac_s^4},\quad
t_{NL}^{V_8}=0.16\frac{H^2V_8\dot\phi_0^2}{Ac_s^6} .
\nonumber\\ \label{contact}
\end{eqnarray}

\noindent Note the common structure $1/(A^m c_s^n)$ in (\ref{exchange}) and again in (\ref{contact}) . The formulas above are necessarily compact, explicit expressions for these quantities as functions of the $c_n$ coefficients in the initial Lagrangian can be found in \textit{Appendix D}.  It is clear then that by a judicious use of the freedom on the $c_n$ coefficients (which are also the basic blocks of $c_s$) one can span a large  spectrum of values for $t_{NL}$, subject to the constraints available at present~\cite{Ade:2013ydc,Fergusson:2010gn,Sekiguchi:2013hza,Giannantonio:2013uqa}.~\footnote{Note however that the constraints in Refs.~\cite{Ade:2013ydc,Fergusson:2010gn,Sekiguchi:2013hza,Giannantonio:2013uqa} in fact apply just to trispectra of local type or to a typical example of ``equilateral'' trispectrum generated in models with non-standard kinetic term~\cite{Fergusson:2010gn}.
As we discuss in the next section, the trispectrum shapes produced in the Galileon models can be very different from these two classes of trispectra analyzed so far.}

An interesting feature which is evident from the results of the trispectrum amplitudes~(\ref{exchange}) and (\ref{contact}) is their peculiar dependence on the sound speed: some of the amplitudes scale like $c_s^{-6}$ or as $c_s^{-8}$ which is markedly different with respect to, e.g., the case of models with non-standard kinetic terms, $P(X,\phi)$, where some of the amplitudes scale at most as $c_s^{-4}$ (see~\cite{Chen:2009bc,Arroja:2009pd}) for details).

\noindent These results parallels similar findings first obtained for the bispectrum of Galileon models in~\cite{Burrage:2010cu}.  A scaling like $c_s^{-6}$ was also found for some of the trispectrum interaction terms studied within the effective field theory approach in~\cite{Bartolo:2010di}.

\subsubsection{Shape Analysis}

\noindent From Eq.~(\ref{trishapedef}), we see that the form factor $\mathcal{T}$ depends on six variables. To get a flavor of these higher-order non-Gaussianities, one can plot the shape-function in different momenta configurations. This has already been done for several inflationary theories and, for the sake of a better comparison of the different signatures in the various models, we also follow suit and present our results in the same fashion as, for example, \cite{Chen:2009bc}.\\

\noindent The momenta configurations can be understood pictorially by looking at the tetrahedron  in Fig.~(\ref{default2}) (the momentum-conservation Dirac delta warrants a regular, closed polyhedron).\\

\begin{figure}[htbp]
\begin{center}
\includegraphics[scale=0.65]{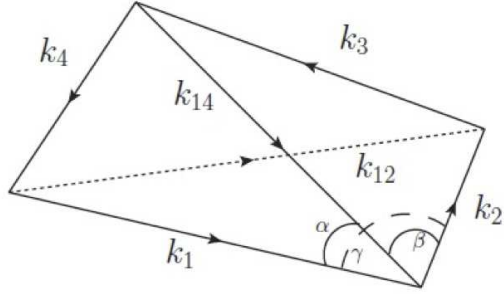}
\caption{According to the relative size of the various $k_n$, the tetrahedron above describes different configurations.}
\label{default2}
\end{center}
\end{figure}

\noindent More into details, below we give a precise description of the four configurations we consider in this paper, and for each configuration we plot a number of representative shape-functions associated to various contributions to the trispectrum. These account for some of the interaction terms generating both the \textit{scalar-exchange} (SE) and the \textit{contact-interaction} (CI) diagrams. \\
\noindent Whenever appropriate,  just for the sake of comparison, we have also included a plot of one or both of the so-called \textit{local} form factors (\textit{Local1,Local2}) \cite{Chen:2009bc}, which are not generated at all within our setup nor within $P(X,\phi)$ models. Also, due to the large number of plots, we include in the text only the more distinctive ones.\\
\noindent To mention just one interesting feature, we anticipate here that one of the operators which most strikingly differs from the predictions of inflationary models with non-standard kinetic terms $P(X,\phi)$
is the one which appears in $\mathcal{H}_4$ regulated by the $V_8$ coefficient.\\
\noindent Intuitively, one would expect the main differences between Galileon inflation trispectra and $P(X,\phi)$ ones to be more visible at the level of the contact interaction diagram contributions. This is because the key role in determining the profile is played by the different contractions of the $k$ vectors \footnote{Clearly, there is more room for such contractions in \textit{Galileon inflation} than $P(X,\phi)$ as in the former model the number of derivative per scalar degree of freedom is not limited to one.}.
Generally, the more fields the contractions involve, the more chances that the profile will have a non-trivial overall $k$-dependence. Having one more field at one's disposal, the contractions  originating from the fourth-order Hamiltonian are more likely to generate distinct form-factors than those originating from third order. Indeed, as we shall see, whenever our findings differ from the $P(X,\phi)$ ones, it is mainly because of CI-type contributions.\\

\noindent $\textit{I}$ ) \textit{Equilateral limit}: it is defined by $k_1=k_2=k_3=k_4$. We plot the trispectra $\mathcal{T}_n$ as functions of $k_{12}/k_1$ and $k_{14}/k_1$ in this configuration.

\noindent Specifically, in Fig.~(\ref{SEequilateral}) we plot the SE trispectra and in Fig.~(\ref{CIequilateral1},\ref{CIequilateral2}) the CI trispectra including for comparison, in the last two plots, the two local-model trispectra.

\noindent In Fig.~(\ref{SEequilateral}) we see that, at least qualitatively, in the \textit{equilateral} configuration the SE-type shape-functions are similar to the corresponding $P(X,\phi)$ result\footnote{At leading order in the $c_s\ll 1$ regime, the $P(X,\phi)$ model predicts the shapes we label as $O_1O_1, O_1O_3,O_3O_3.$} \cite{Chen:2009bc,Arroja:2009pd}.
\noindent One should also keep in mind that, as clear from the figure, here the overall sign of the form-factor can change.

Next, we move on (Fig.~\ref{CIequilateral1}, \ref{CIequilateral2}) to the various CI-generated  shape-functions in the same, \textit{equilateral}, configuration. In this limit the significantly different shapes in comparison with $P(X,\phi)$ are generated by terms such as e.g. the $V_8$-driven interaction in Fig.~(\ref{CIequilateral1}).
\noindent The equilateral configuration plots (Fig.~\ref{CIequilateral1}, \ref{CIequilateral2})  are indeed quite interesting: $P(X,\phi)$ models predict a simple plateau, which is what we obtain from terms such as the ones driven by the coefficients\footnote{At leading order in the $c_s\ll 1$ regime, $P(X,\phi)$ produces only $V_1$.} $V_1,V_2,V_3,V_4,V_6$ and also from one local-type of trispectrum. On the other hand,  this is clearly not the case for terms such as those proportional to $V_5 , V_7, V_8$, as well as the other local-type profile.
\noindent These latter $V_n$ coefficients are then the most intriguing as they drive interactions which evidently have a quite different signature than any $P(X,\phi)$ realization.

It is important to keep in mind that what we have here is a $k$-configuration dependent signature and that one would need to probe the trispectrum in its entirety in order to be on par with what is done for the bispectrum analysis. However, what we found is nevertheless a distinct signature, and, as we shall see, we will have some further noteworthy results  in the \textit{double-squeezed} configuration as well.\\
\\

\noindent $\textit{II}$ ) \textit{Folded limit}: $k_{12}=0$. In this limit $k_1=k_2$ and $k_3=k_4$. The trispectra are plotted as functions of $k_4/k_1$ and $k_{14}/k_1$.

\noindent  In Fig.~(\ref{SEfoldedlimit}) we plot the scalar-exchange trispectra and in Fig.~(\ref{CIfoldedlimit1},\ref{CIfoldedlimit2}) the contact-interaction trispectra including, in the last  plot, the so-called \textit{Local2}-type trispectra (\textit{Local1} is divergent in this limit).\\
\noindent Just as we saw for the SE \textit{equilateral} configuration profiles, from the plots in Fig.~(\ref{SEfoldedlimit}) one can conclude that the patterns that emerge in $k$-space for the folded configuration are very similar to those in  the analysis of  \cite{Chen:2009bc} for $P(X,\phi)$ models.
\noindent As to the CI-generated plots in the \textit{folded} configuration, we limit the graphical representations in the main text to the more illustrative cases. In this specific $k$ arrangement, it is again hard to see any qualitative difference with the work in  \cite{Chen:2009bc}. \\
\noindent It is also worth mentioning here that the $V_8$ regulated CI term does not contribute to any of the configurations  in $II, III, IV$ and is therefore not plotted in the corresponding figures. \\

\noindent $\textit{III}$ ) \textit{Specialized planar limit}: $k_1=k_3=k_{14}$ and the tetrahedron lies on a plane. One can solve for $k_{12}$ \cite{Chen:2009bc}, to find
$k_{12}=\left[k_1^2+\frac{k_2k_4}{2k_1^2}\left(k_2k_4+\sqrt{(4k_1^2-k_2^2)(4k_1^2-k_4^2)}\right)\right]^\frac{1}{2}$. The trispectra are plotted as functions of $k_2/k_1$ and $k_4/k_1$.

\noindent Following the presentation pattern of the previous two configurations, in Fig.(\ref{SEspecializedplanarlimit}) we have plotted some representative operators contribution to the scalar exchange diagram, namely the couples $\{O_1 O_1\}, \{O_3 O_3\}, \{O_1 O_3\}, \{O_2 O_4\}$, in the \textit{specialized planar} configuration. In Fig.~(\ref{CIspecializedplanarlimit1},\ref{CIspecializedplanarlimit2}) we plot the contact-interaction trispectra including, as usual, the two local-model trispectra.\\
\\The same considerations we made for the \textit{folded} configuration hold true for all the plots (SE as well as CI) in the \textit{specialized planar} configuration. Upon a more quantitative analysis, one finds that the shape-functions are indeed analytically different  from those in~\cite{Chen:2009bc} (and from each other as well), but this fact in itself is not sufficient to claim a real significance for the difference. Indeed, in the case of qualitatively similar shape-functions a more detailed treatment, such as a shape scalar product analysis, would reveal small differences. On the other hand, for both the \textit{folded} and the \textit{specialized planar} configurations findings in this model we do not anticipate that the shape profiles differ from those of \cite{Chen:2009bc} to a sufficient degree so as to motivate such a detailed analysis.\\

\noindent $\textit{IV}$ ) \textit{Near the double-squeezed limit}: $k_3=k_4=k_{12}$ and the tetrahedron is lying on a plane. In this limit $k_2$ \cite{Chen:2009bc} can be written in terms of the other variables as
\[k_2=\frac{1}{\sqrt{2}k_4}\sqrt{k_1^2(-k_{12}^2+k_3^2+k_4^2)-k_{s1}^2k_{s2}^2+k_{12}^2k_{14}^2+k_{12}^2k_{4}^2+k_{14}^2k_{4}^2-k_{14}^2k_{3}^2-k_4^4+k_{3}^2k_{4}^2},\]
\\ where the variables $k_{s1}$ and $k_{s2}$ are defined as\\
\[k_{s1}^2=2\sqrt{(k_1k_4+\mathbf{k}_1\cdot\mathbf{k}_4)(k_1k_4-\mathbf{k}_1\cdot\mathbf{k}_4)} \,\, ,\qquad k_{s2}^2=2\sqrt{(k_3k_4+\mathbf{k}_3\cdot\mathbf{k}_4)(k_3k_4-\mathbf{k}_3\cdot\mathbf{k}_4)} \,\, .\]
The scalar products can be written as $\mathbf{k}_1\cdot\mathbf{k}_4=(k_{14}^2-k_1^2-k_4^2)/2$ and $\mathbf{k}_3\cdot\mathbf{k}_4=(k_{12}^2-k_3^2-k_4^2)/2$.
The trispectra $\mathcal{T}_n$ are plotted as functions of $k_{12}/k_1$ and $k_{14}/k_1$ but are this time further divided by $k_1k_2k_3k_4$ for enhancement in shape-comparison.

\noindent In Fig.~(\ref{SEdoublesqueezedlimit}) we report the plots of the scalar exchange trispectra while Fig.~(\ref{CIdoublesqueezedlimit1},\ref{CIdoublesqueezedlimit2}) reproduce  the plots of the contact interaction trispectra and the two local-model trispectra.

\begin{figure}
\center
\includegraphics[width=0.3\textwidth]{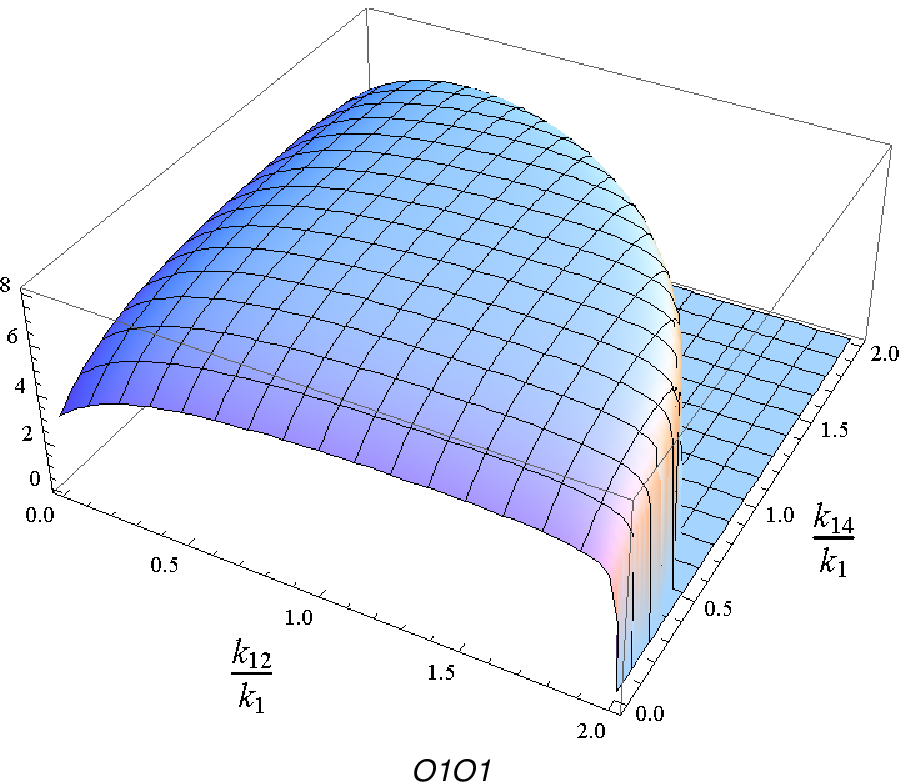}
\hspace{0.05\textwidth}
\includegraphics[width=0.3\textwidth]{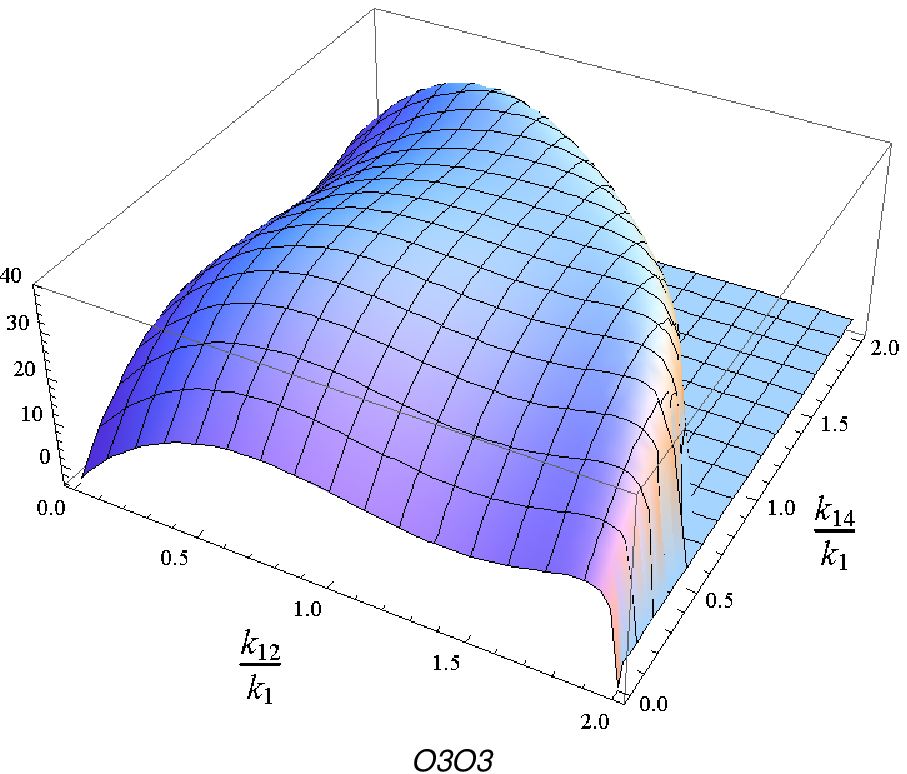}

\includegraphics[width=0.3\textwidth]{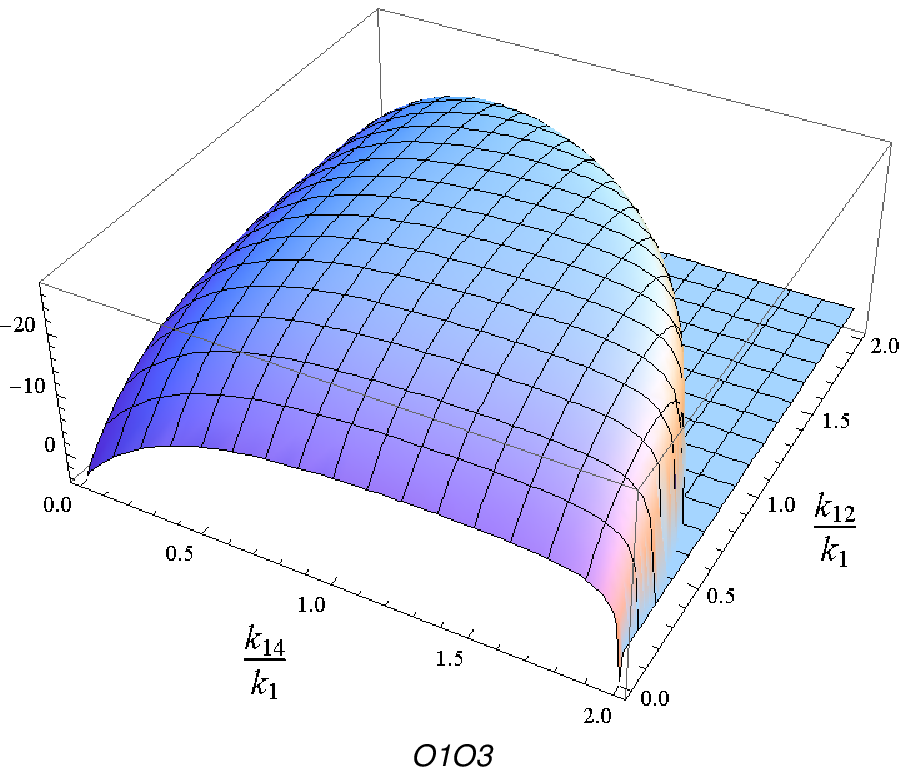}
\hspace{0.05\textwidth}
\includegraphics[width=0.3\textwidth]{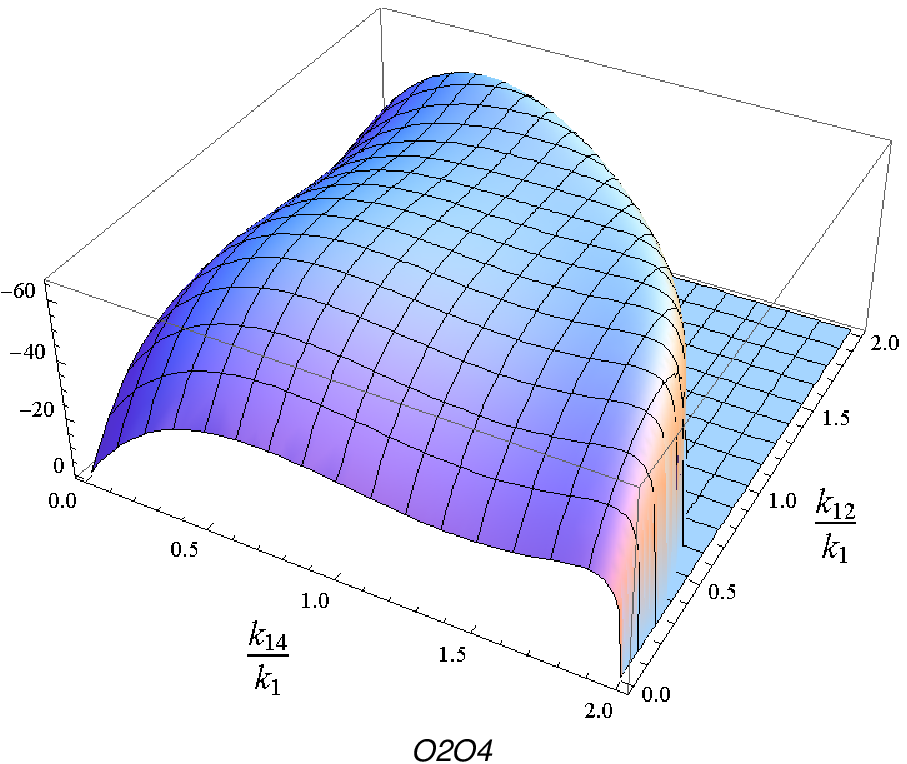}
\caption{\label{SEequilateral} Some of the different SE trispectra shapes in the \textit{equilateral} limit as functions of $k_{12}/k_1$ and $k_{14}/k_1$. The $\mathcal{O}_m \mathcal{O}_n$ at the bottom of each graph signifies that the contribution plotted is coming from the SE vertices regulated by the interactions $\mathcal{O}_m$ and $\mathcal{O}_n$. The normalization is arbitrary.  Note that the plots have different \textit{Mathematica} ``viewpoint" and ``view vertical" options. This is to emphasize their similarity. The other contributions do not qualitatively differ from the ones above.}
\end{figure}
\begin{figure}
\center
\includegraphics[width=0.3\textwidth]{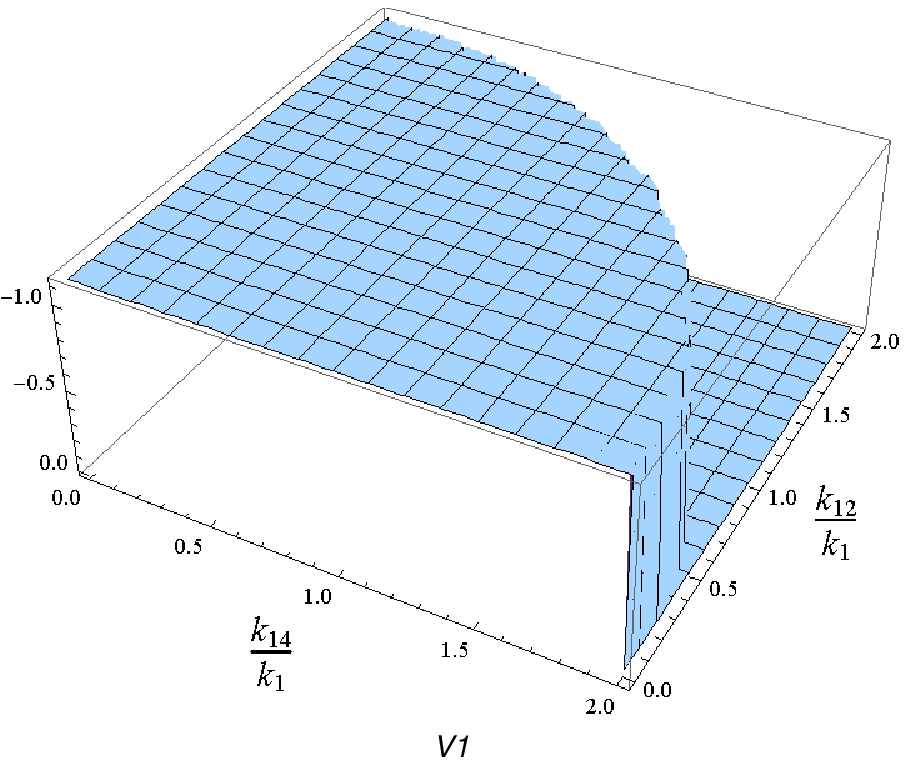}
\hspace{0.05\textwidth}
\includegraphics[width=0.3\textwidth]{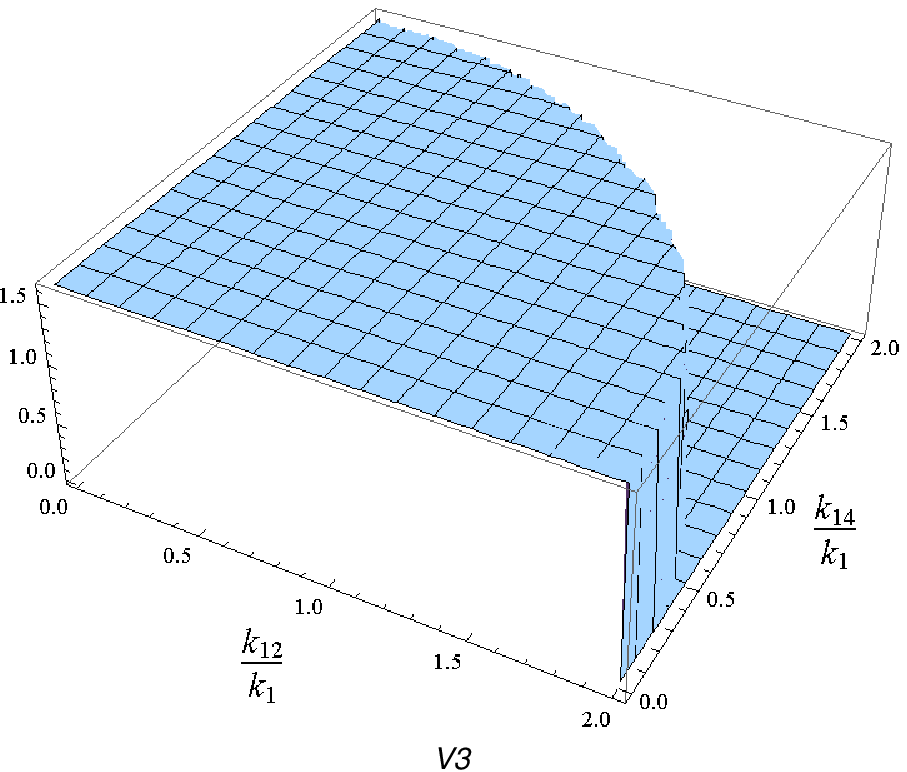}

\includegraphics[width=0.3\textwidth]{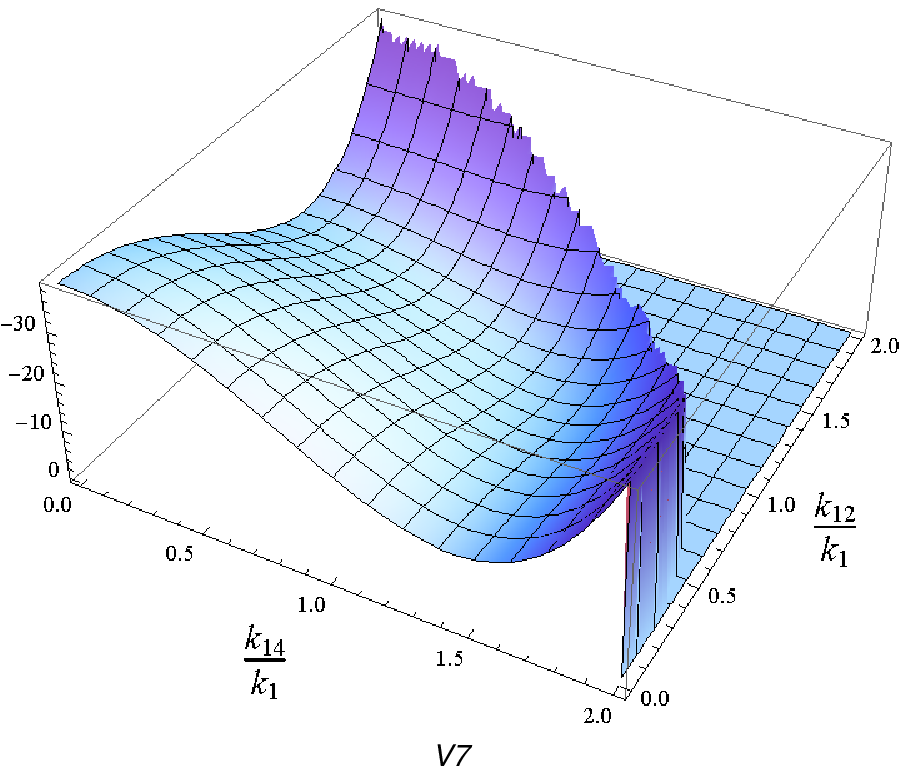}
\hspace{0.05\textwidth}
\includegraphics[width=0.3\textwidth]{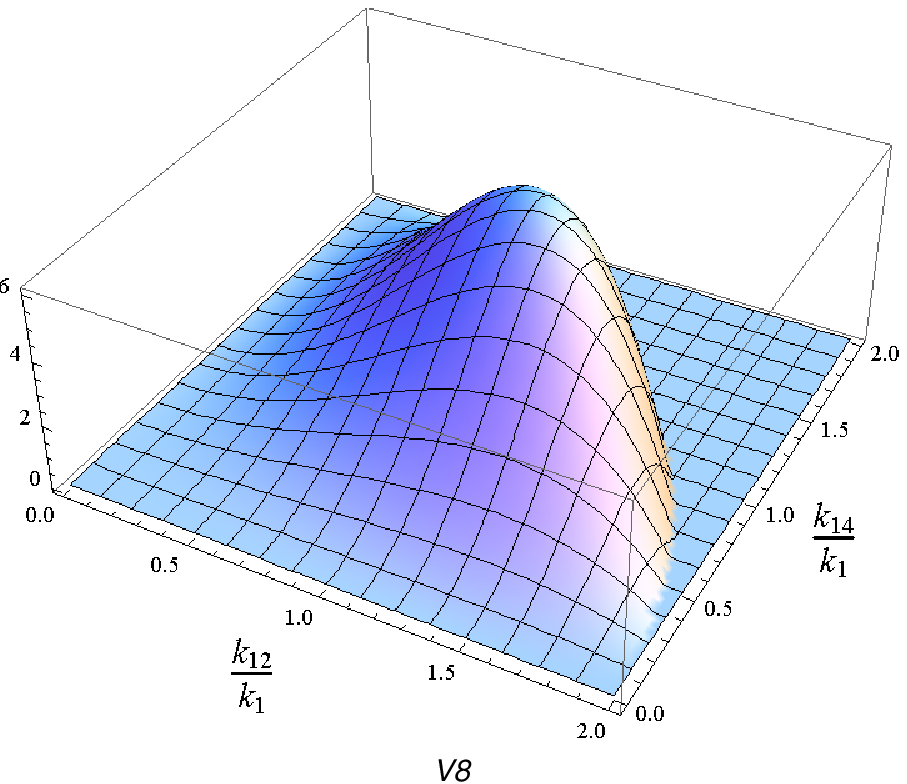}
\caption{\label{CIequilateral1} Some of the different CI trispectra shapes in the \textit{equilateral} limit as functions of $k_{12}/k_1$ and $k_{14}/k_1$. The label $V_i$ at the bottom of each graph signifies that the contribution plotted is originating from the CI terms proportional to the coefficients $V_i$.  More in Fig.(\ref{CIequilateral2}) .}
\end{figure}
\begin{figure}
\center
\hspace{0.05\textwidth}
\includegraphics[width=0.3\textwidth]{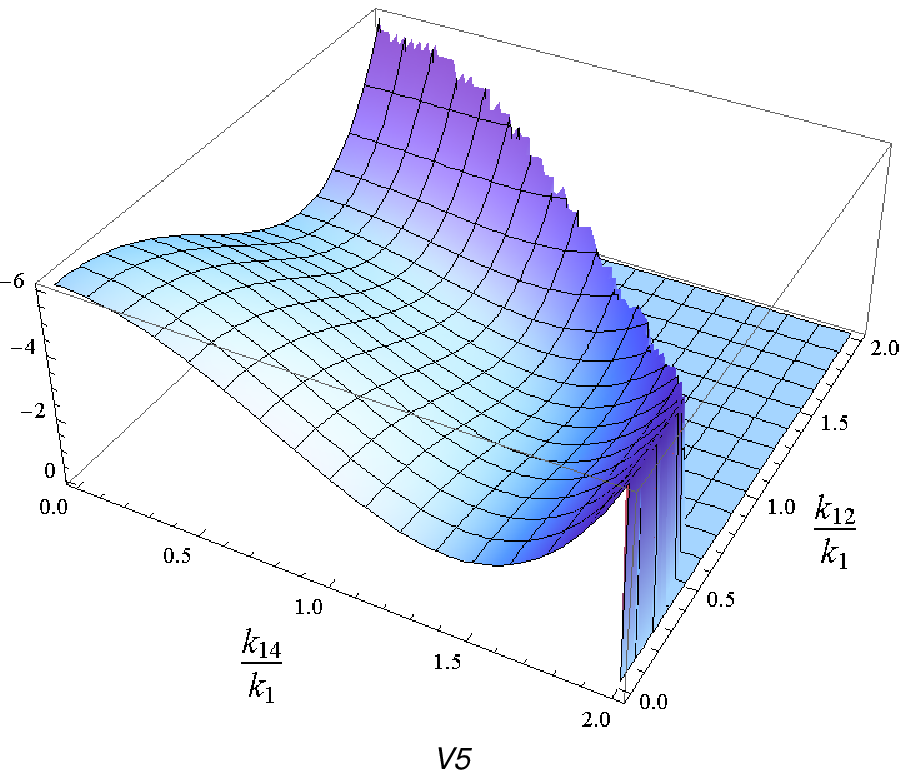}
\includegraphics[width=0.3\textwidth]{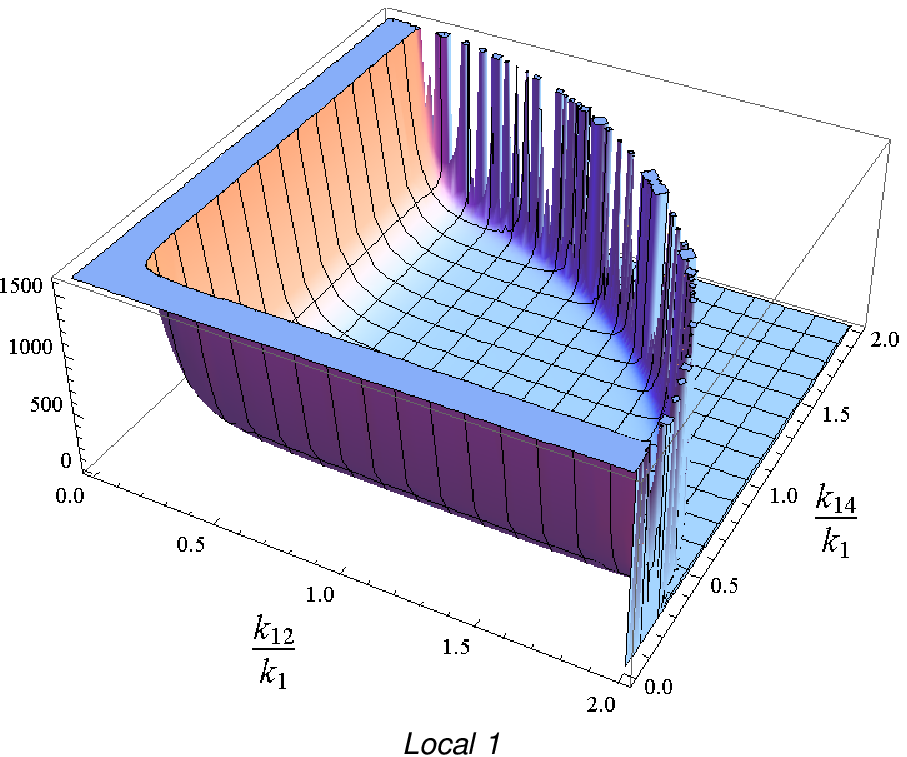}
\hspace{0.05\textwidth}
\includegraphics[width=0.3\textwidth]{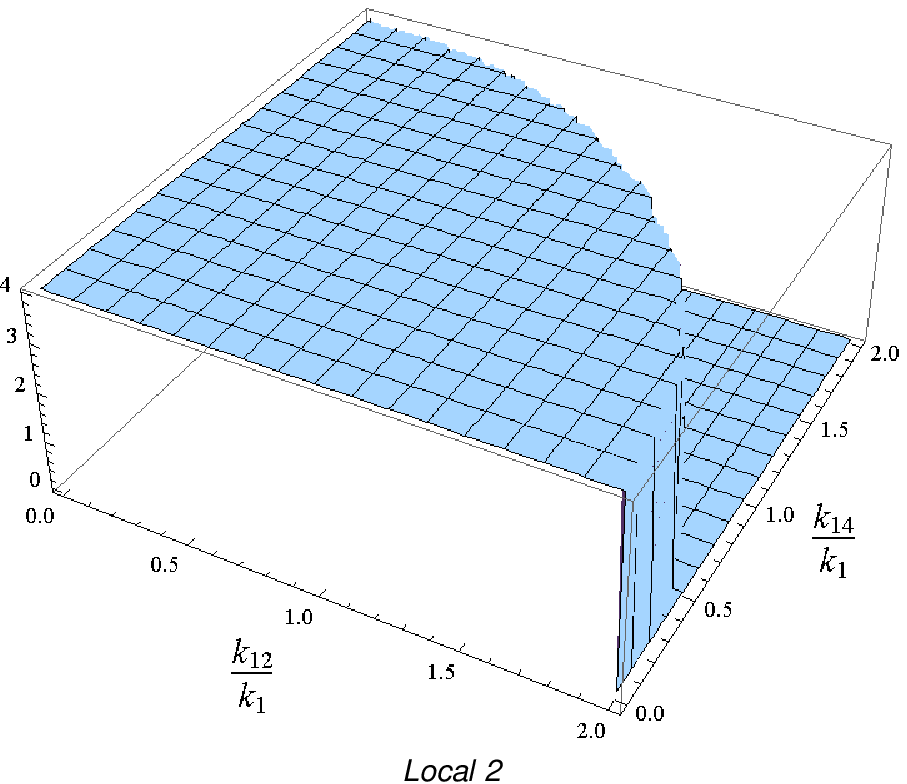}
\caption{\label{CIequilateral2} Other CI trispectra shapes in the \textit{equilateral} limit as functions of $k_{12}/k_1$ and $k_{14}/k_1$. The last two plots are the local trispectra shapes.}
\end{figure}
\begin{figure}
\center
\includegraphics[width=0.3\textwidth]{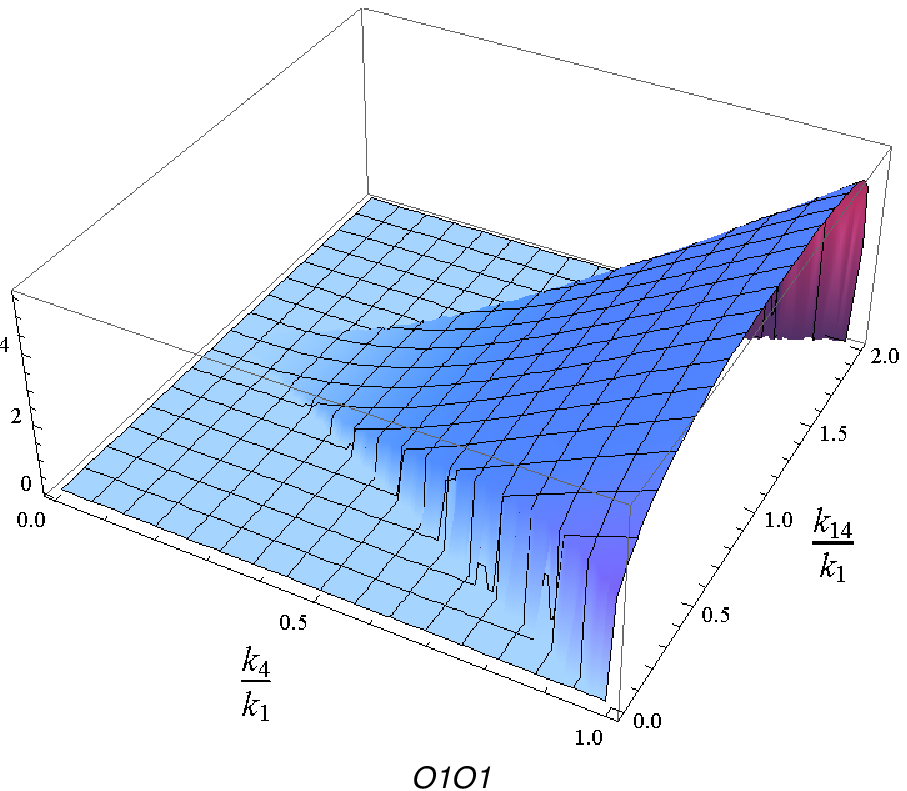}
\hspace{0.05\textwidth}
\includegraphics[width=0.3\textwidth]{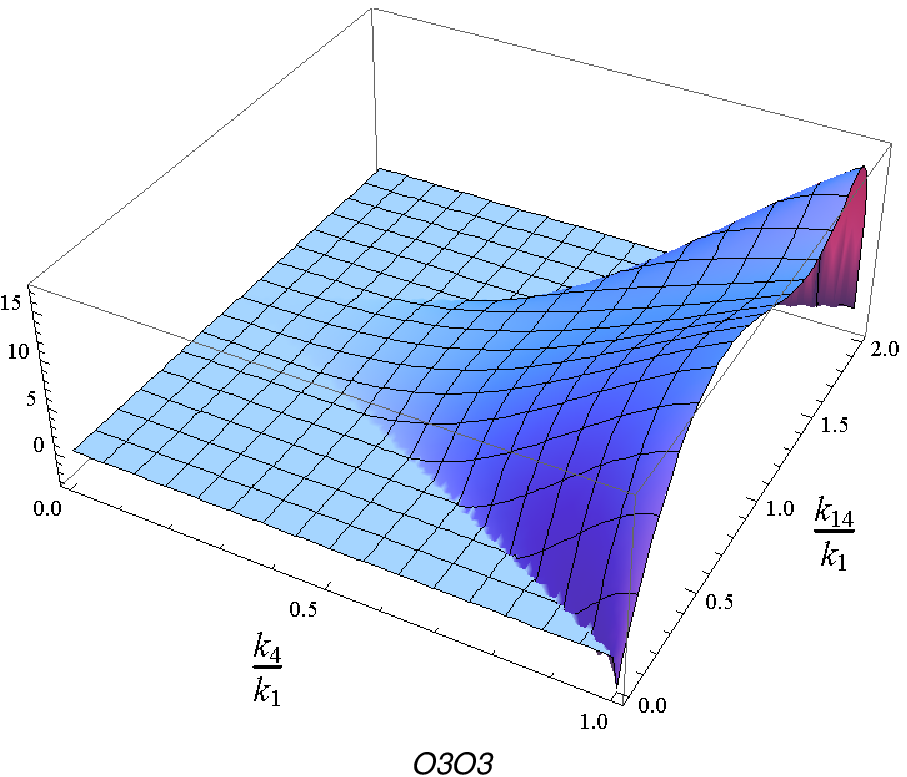}

\includegraphics[width=0.3\textwidth]{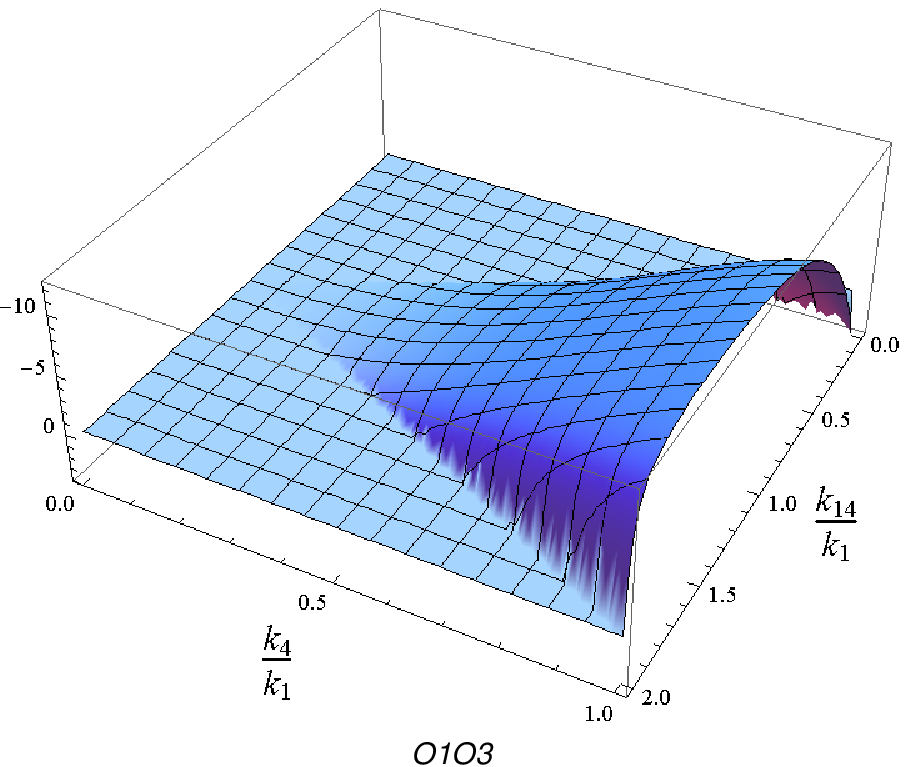}
\hspace{0.05\textwidth}
\includegraphics[width=0.3\textwidth]{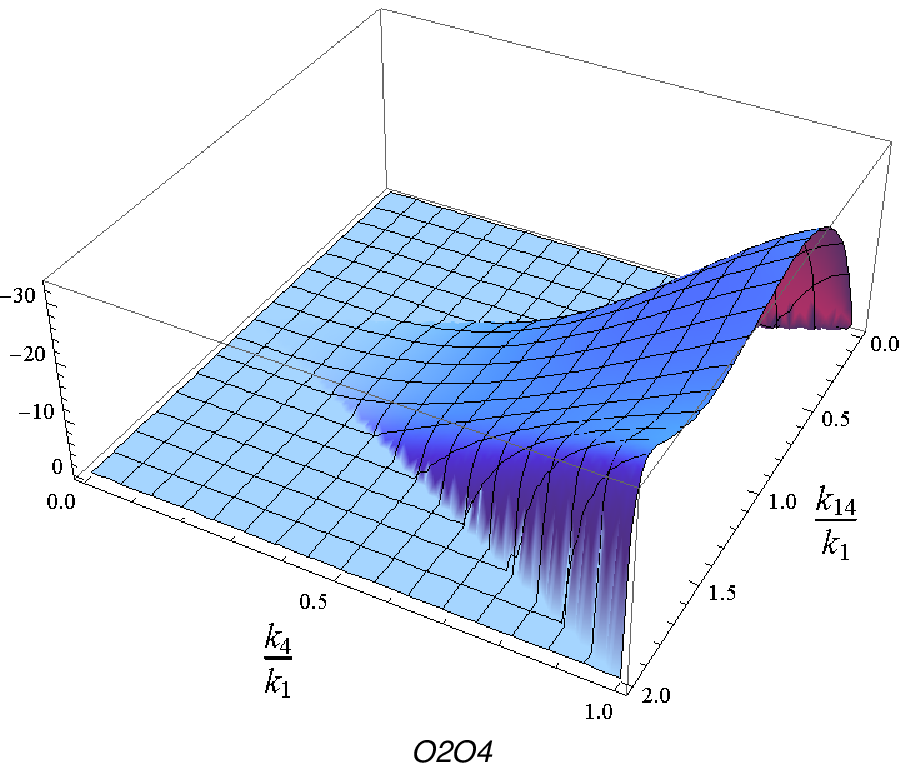}
\caption{\label{SEfoldedlimit} Plotted here are the different SE trispectra shapes in the \textit{folded} limit as functions of $k_{4}/k_1$ and $k_{14}/k_1$. }
\end{figure}
\begin{figure}
\center
\includegraphics[width=0.3\textwidth]{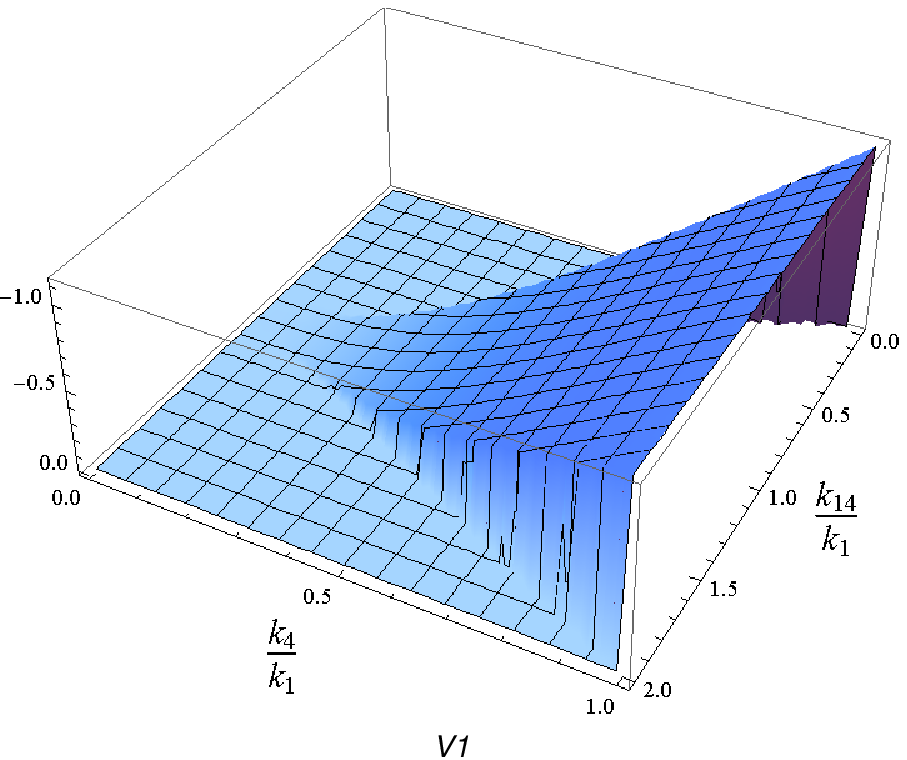}
\hspace{0.05\textwidth}
\includegraphics[width=0.3\textwidth]{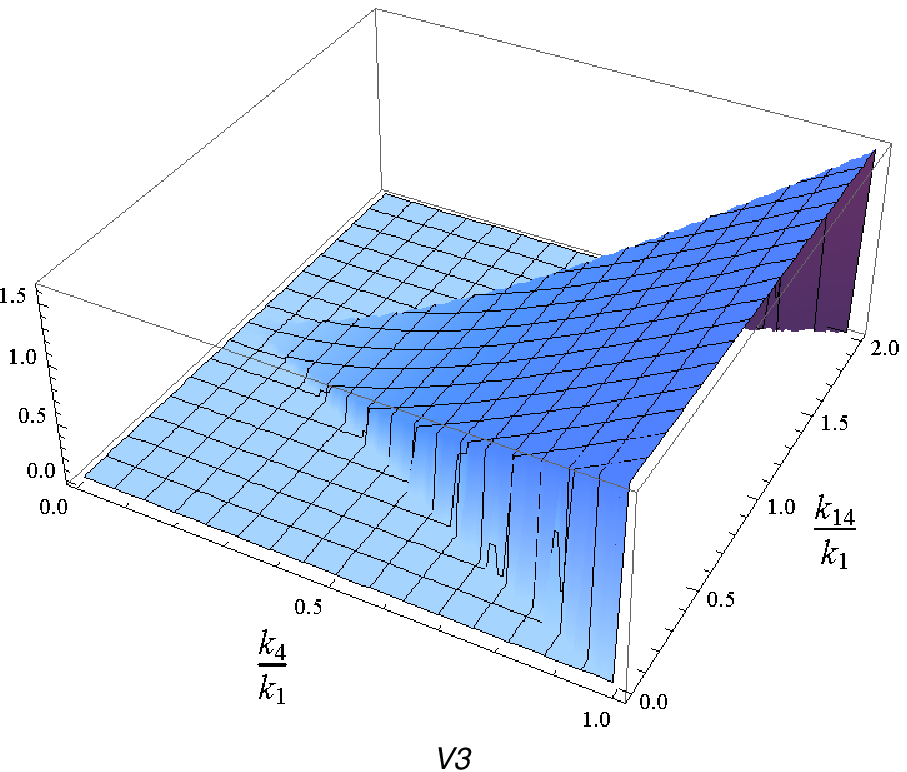}
\caption{\label{CIfoldedlimit1} Some of the different CI trispectra shapes in the \textit{folded} limit as functions of $k_{4}/k_1$ and $k_{14}/k_1$. }
\end{figure}
\begin{figure}
\center
\includegraphics[width=0.3\textwidth]{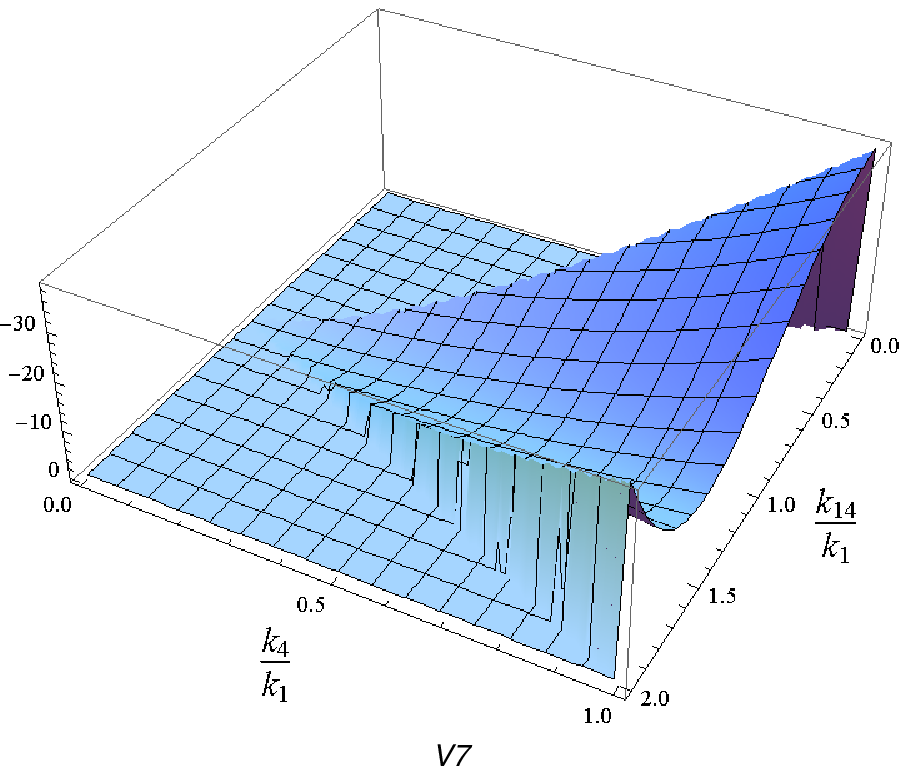}
%\hspace{0.05\textwidth}
\includegraphics[width=0.3\textwidth]{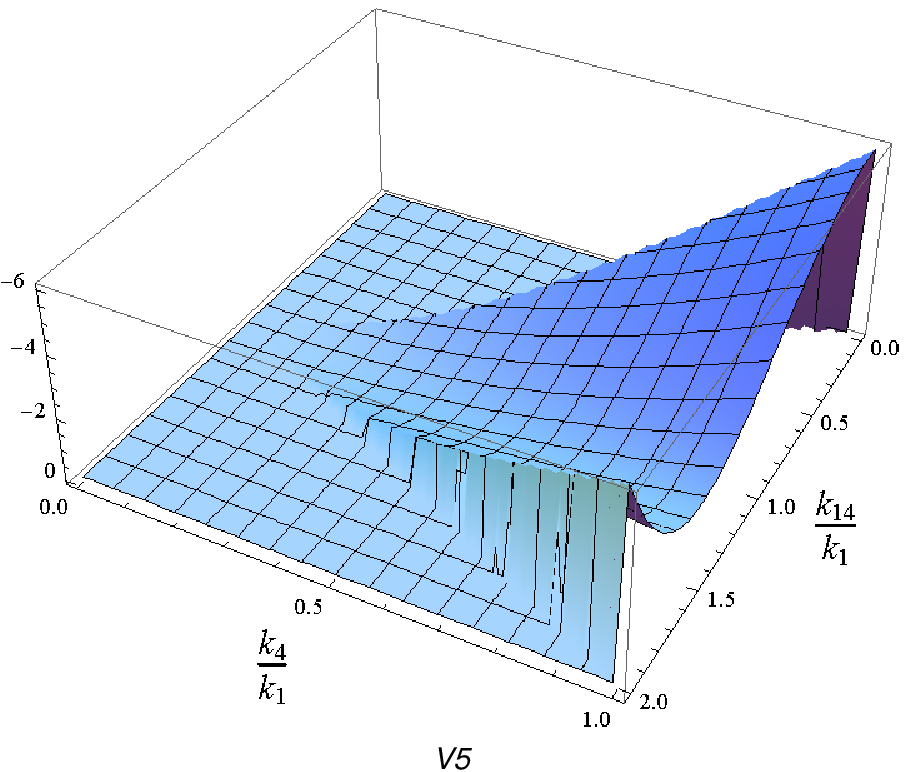}
\includegraphics[width=0.3\textwidth]{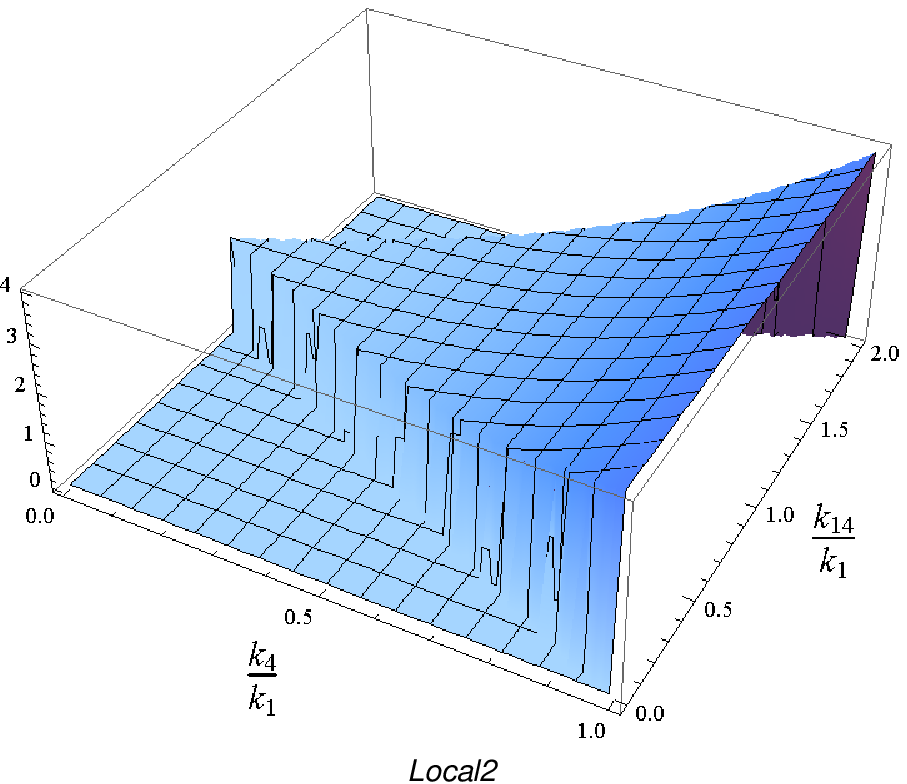}
\caption{\label{CIfoldedlimit2} More trispectra shapes in the \textit{folded} limit as functions of $k_{4}/k_1$ and $k_{14}/k_1$. The last plot is one of the local trispectra shapes. $T_{Local\,1}$ (not plotted)  blows up in this limit and $V_8$ is always zero.}
\end{figure}

\begin{figure}
\center
\includegraphics[width=0.3\textwidth]{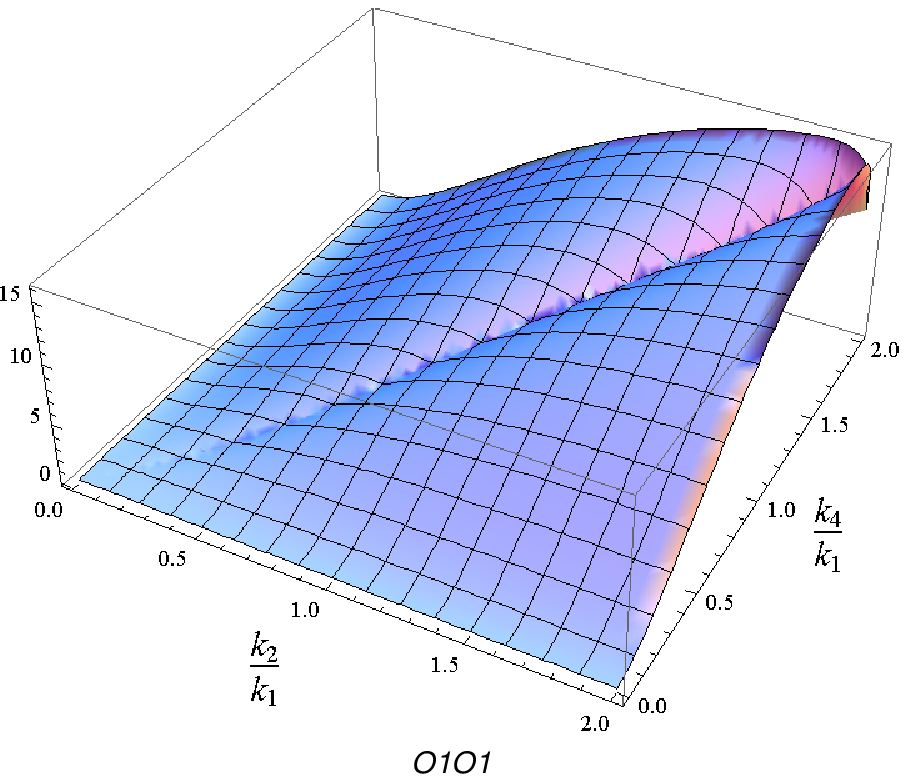}
\hspace{0.05\textwidth}
\includegraphics[width=0.3\textwidth]{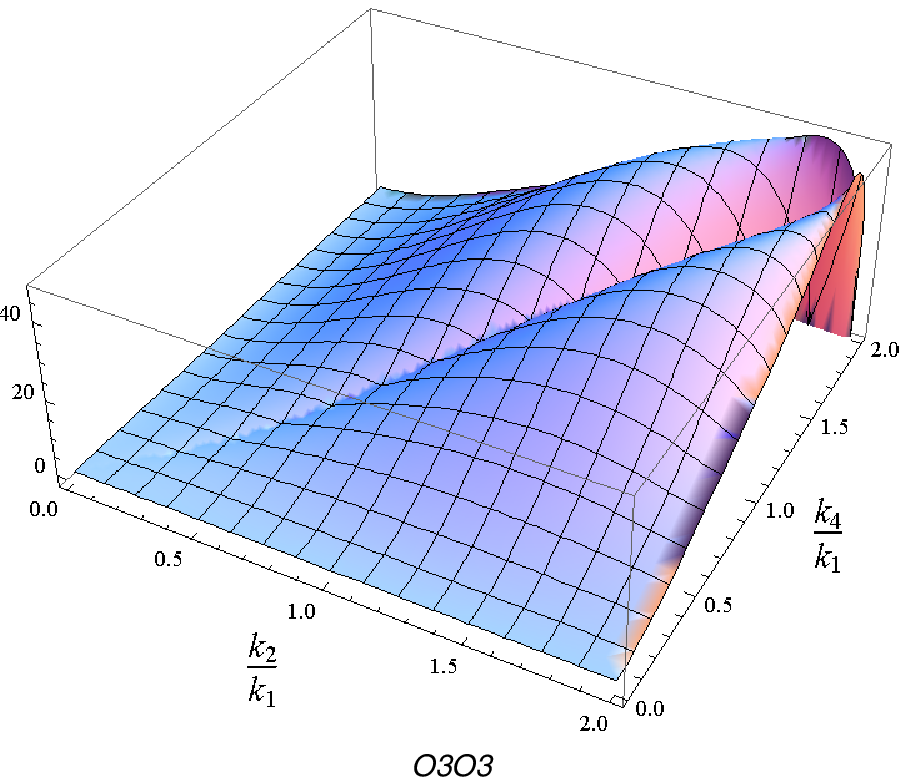}

\includegraphics[width=0.3\textwidth]{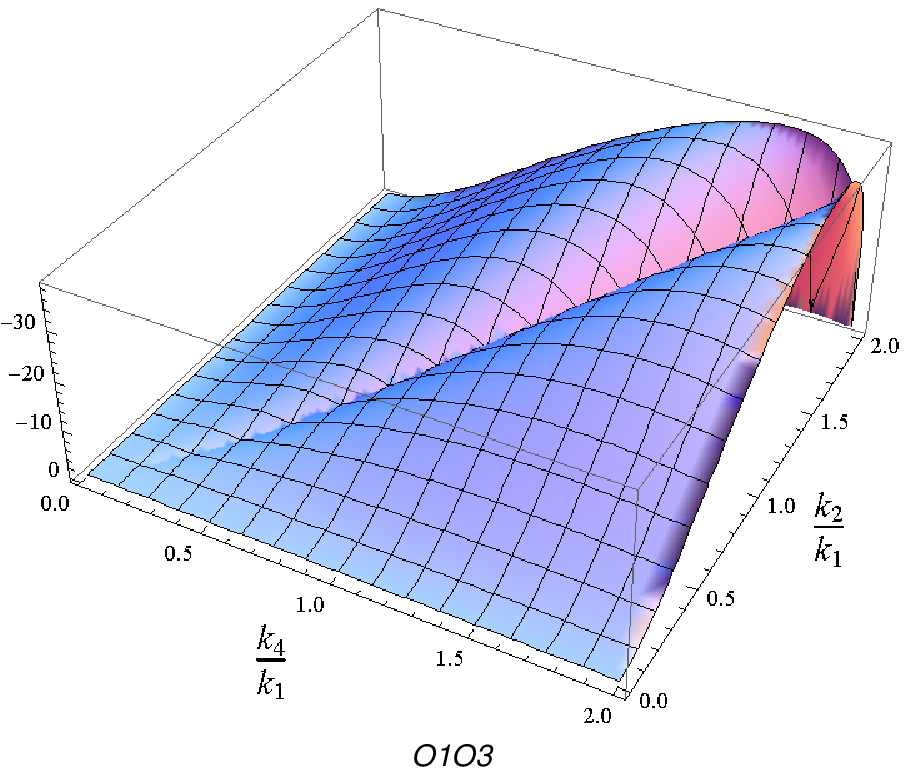}
\hspace{0.05\textwidth}
\includegraphics[width=0.3\textwidth]{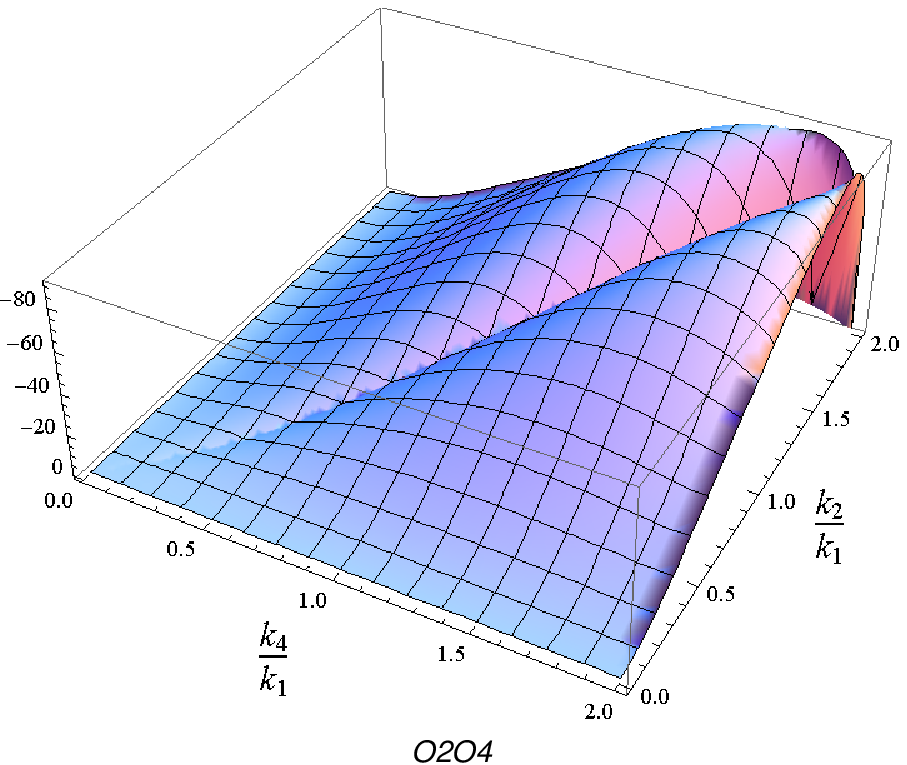}
\caption{\label{SEspecializedplanarlimit} The different SE trispectra shapes in the \textit{specialized planar} limit as functions of $k_2/k_1$ and $k_4/k_1$.}
\end{figure}
\begin{figure}
\center
\includegraphics[width=0.3\textwidth]{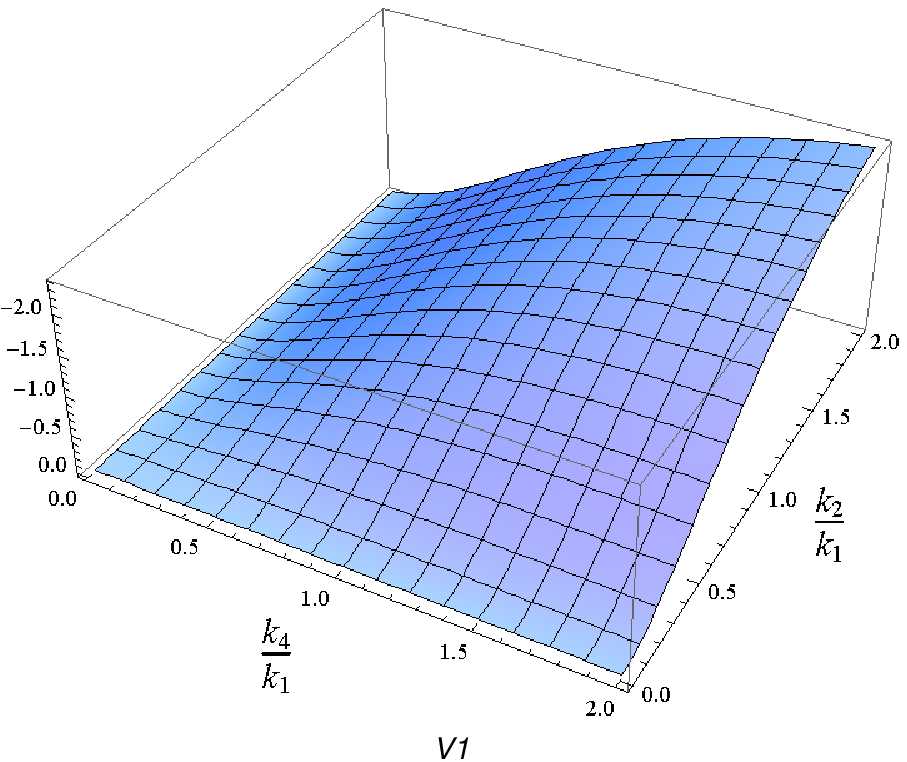}
\hspace{0.05\textwidth}
\includegraphics[width=0.3\textwidth]{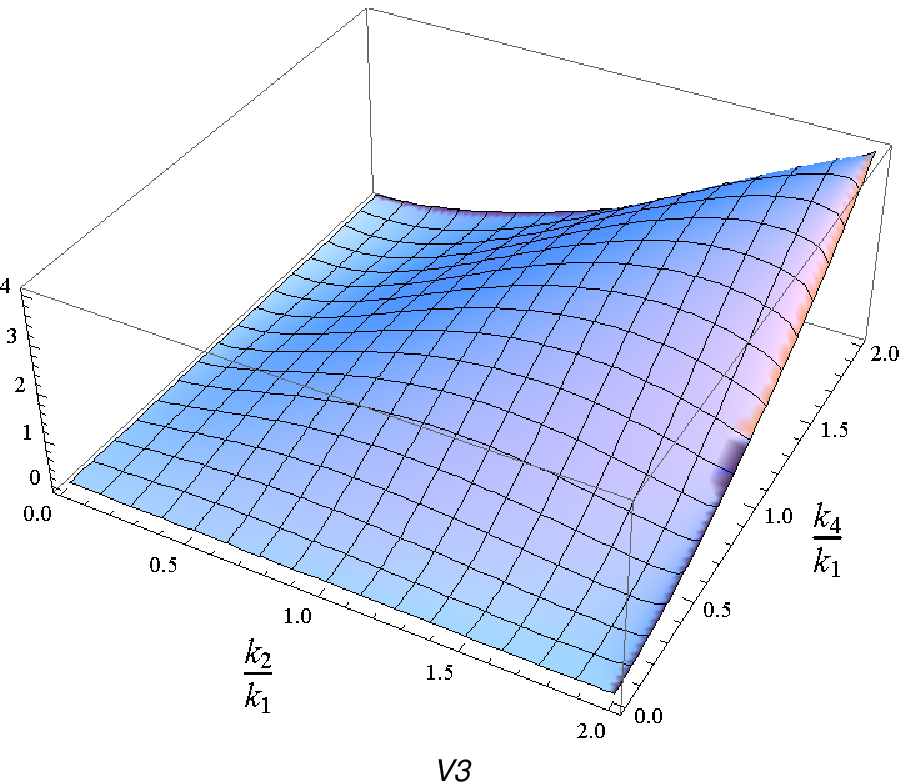}
\caption{\label{CIspecializedplanarlimit1} Some of the different CI trispectra shapes in the \textit{specialized planar} limit as functions of $k_2/k_1$ and $k_4/k_1$.}
\end{figure}
\noindent Much like what we have seen in the \textit{equilateral} configuration plots, we shall show below that the non-Gaussian characterization of Galileon inflation in the \textit{double-squeezed} configuration pays off in that it allows one to distinguish between the Galileon model predictions and those of $P(X,\phi)$ models.

It is perhaps timely at this stage to remind the reader of an interesting fact concerning the \textit{double-squeezed} configurations plots in the $P(X,\phi)$ case: one could say that, for these models, the \textit{double-squeezed} configuration is actually ``aware'' of what sort of contribution, cubic or quartic, is sourcing any given shape-function. This is because in the $(k_{14}=1, k_{12}\rightarrow 0)$ (see e.g.  Fig.~\ref{SEdoublesqueezedlimit}) limit the shape-function is finite but non-zero for each SE-type contribution while it is always finite and precisely zero in the CI case.

One can imagine that, if it were somehow possible to probe observationally  such a configuration, the degeneracy between third and fourth order interactions contribution to the trispectrum could in principle be removed. As we will see, this is not the case for Galileon inflation.

\begin{figure}
\center
\includegraphics[width=0.3\textwidth]{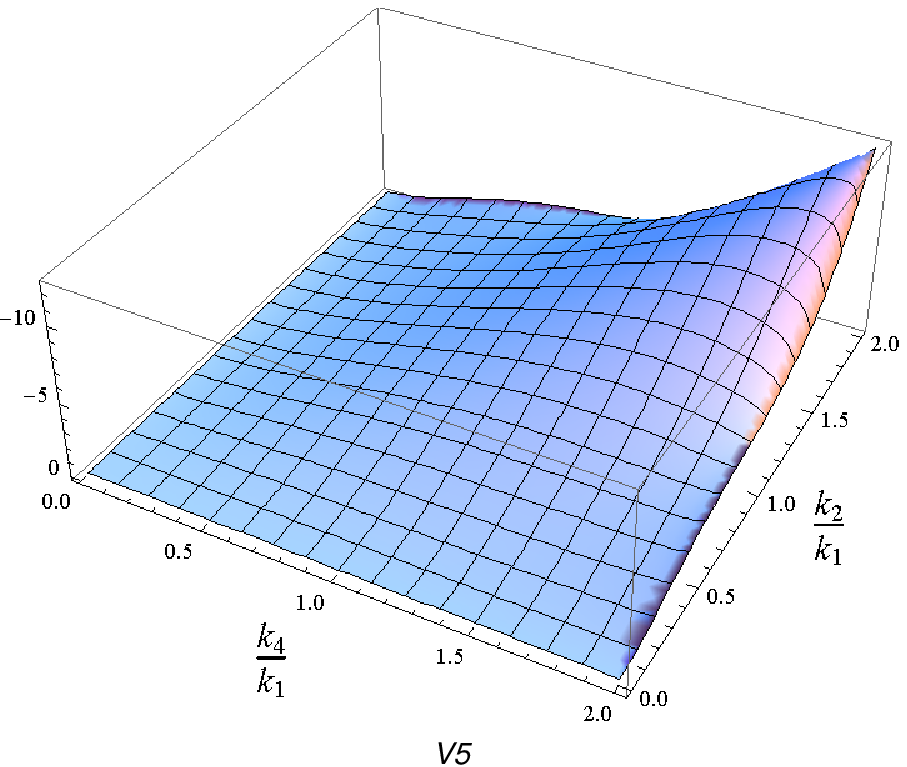}
\includegraphics[width=0.3\textwidth]{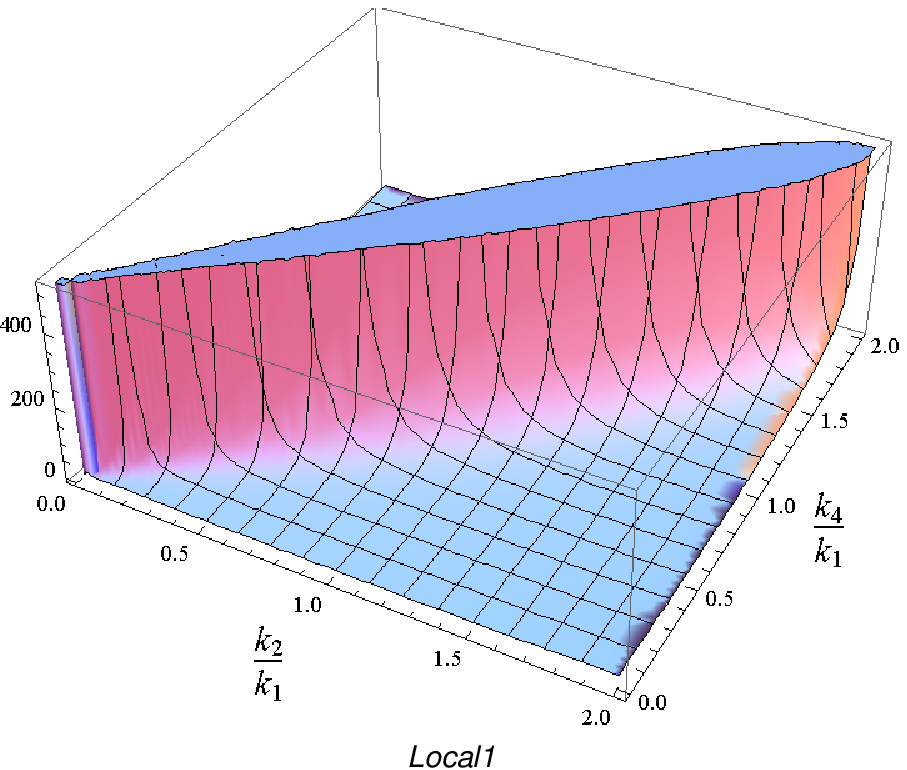}
%\hspace{0.05\textwidth}
\includegraphics[width=0.3\textwidth]{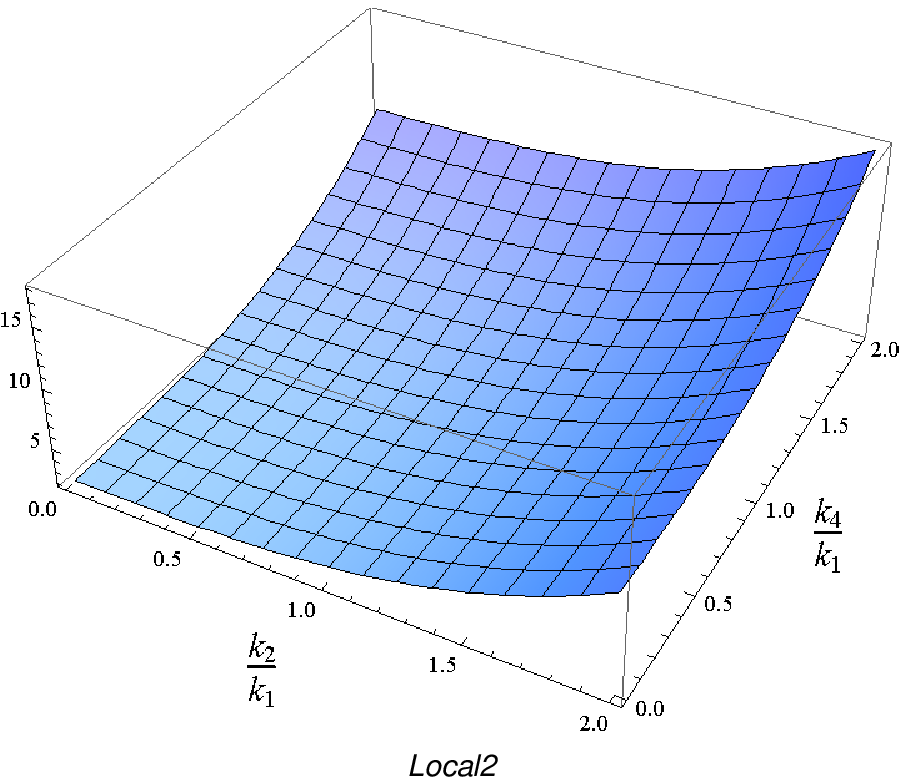}
\caption{\label{CIspecializedplanarlimit2} More of the different CI trispectra shapes in the \textit{specialized planar} limit as functions of $k_2/k_1$ and $k_4/k_1$. The last two plots are the local trispectra shapes. $T_{Local\,1}$ blows up in the limit $k_2\rightarrow k_4$ and the $V_8$-shape is always zero.}
\end{figure}
\begin{figure}
\center
\includegraphics[width=0.3\textwidth]{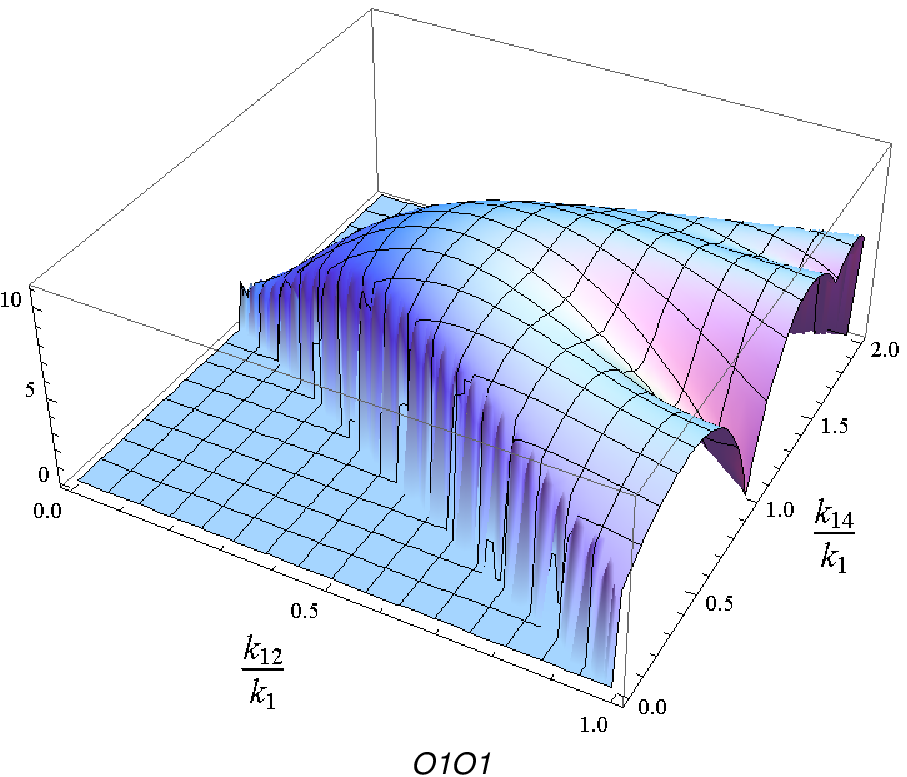}
\hspace{0.05\textwidth}
\includegraphics[width=0.3\textwidth]{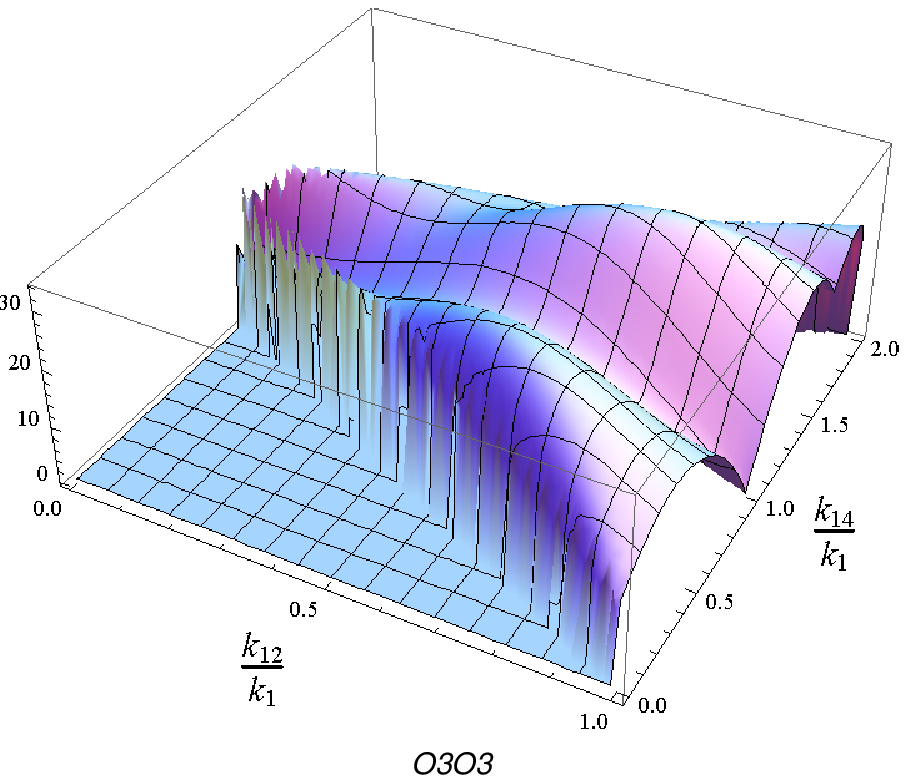}

\includegraphics[width=0.3\textwidth]{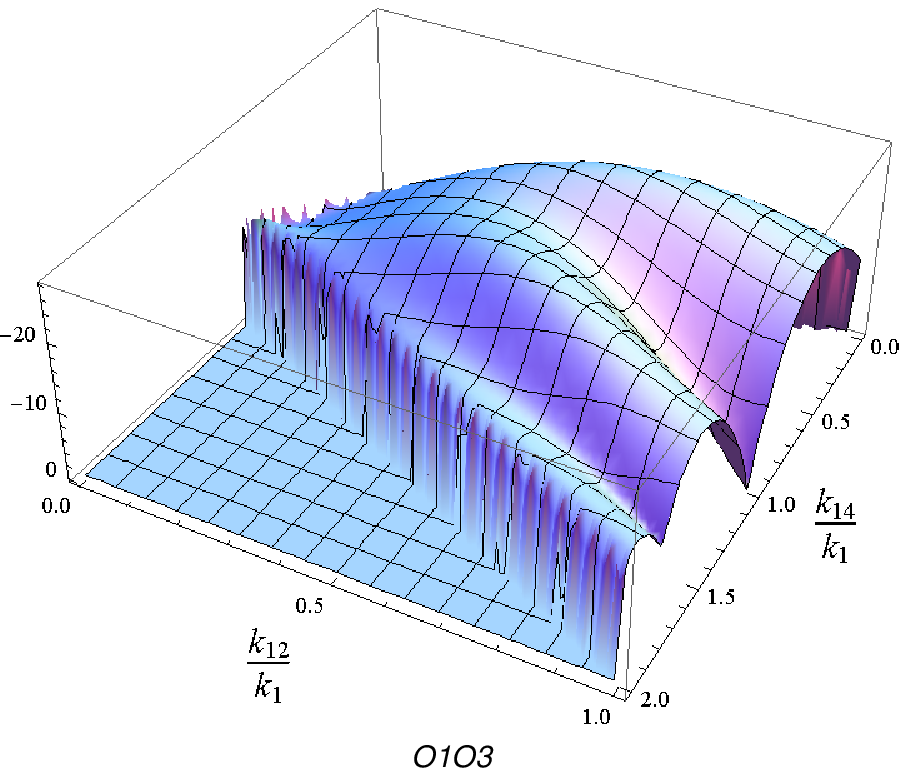}
\hspace{0.05\textwidth}
\includegraphics[width=0.3\textwidth]{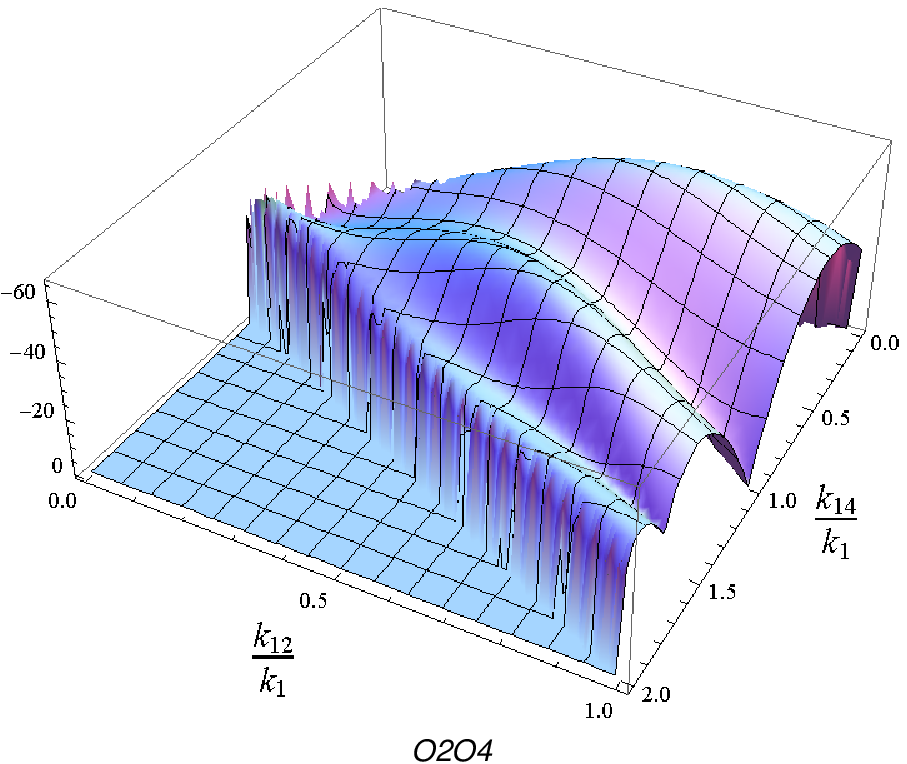}
\caption{\label{SEdoublesqueezedlimit} Plotted are the different SE trispectra shapes divided by $k_1k_2k_3k_4$ in the \textit{double-squeezed} limit as functions of $k_{12}/k_1$ and $k_{14}/k_1$.}
\end{figure}

A quick look at Fig.~(\ref{CIdoublesqueezedlimit1}) reveals how terms such as those CI-type interactions driven by the coefficient $V_3 , V_7$ do indeed have a well-defined and, most importantly, non-zero $k_{12}\rightarrow 0$ limit. This fact then reinstates in Galileon inflation the degeneracy which was lifted in \cite{Chen:2009bc} for $P(X,\phi)$ models.
\noindent This latter characterization adds to the number of distinct features which the setup under scrutiny here  does not share with the results of \cite{Huang:2006eha, Arroja:2008ga,Chen:2009bc,Arroja:2009pd}.\\

\begin{figure}
\center
\includegraphics[width=0.3\textwidth]{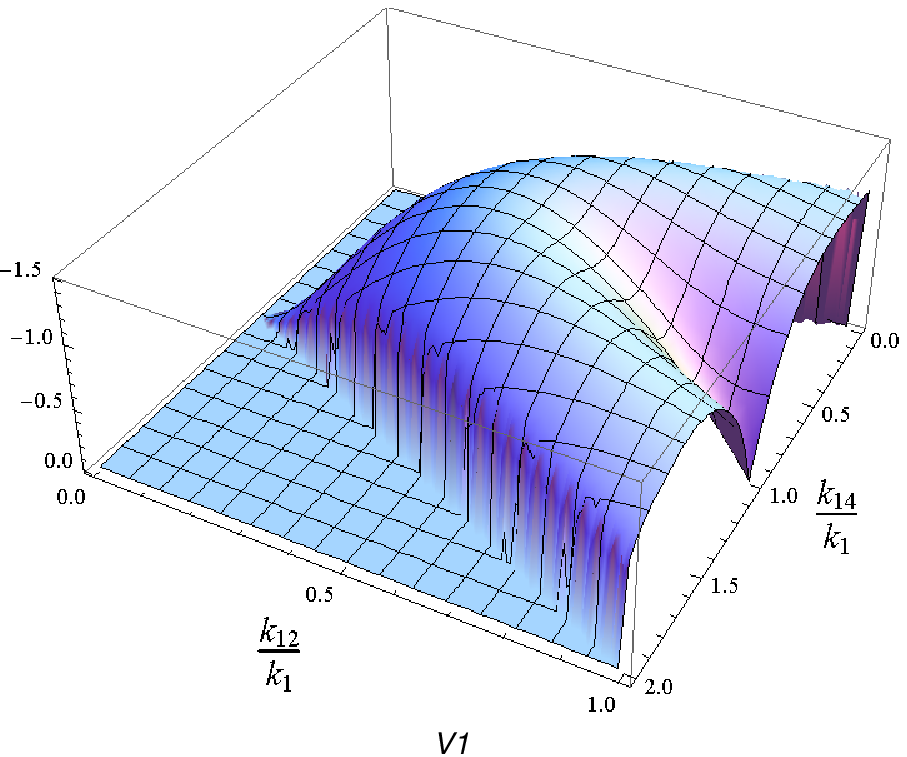}
%\hspace{0.05\textwidth}
\includegraphics[width=0.3\textwidth]{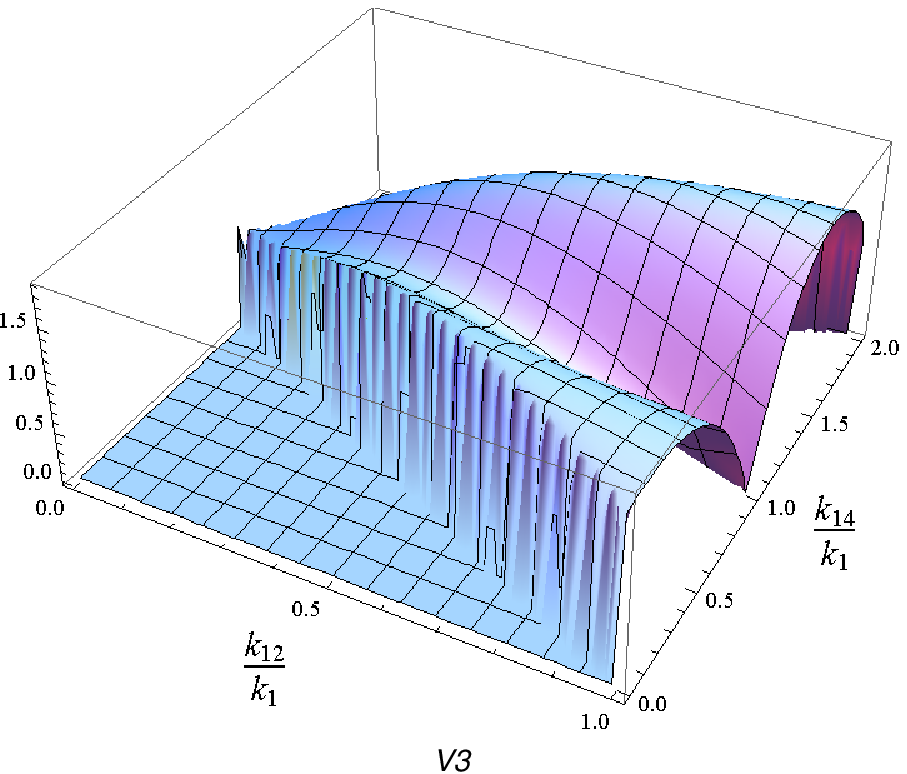}
\includegraphics[width=0.3\textwidth]{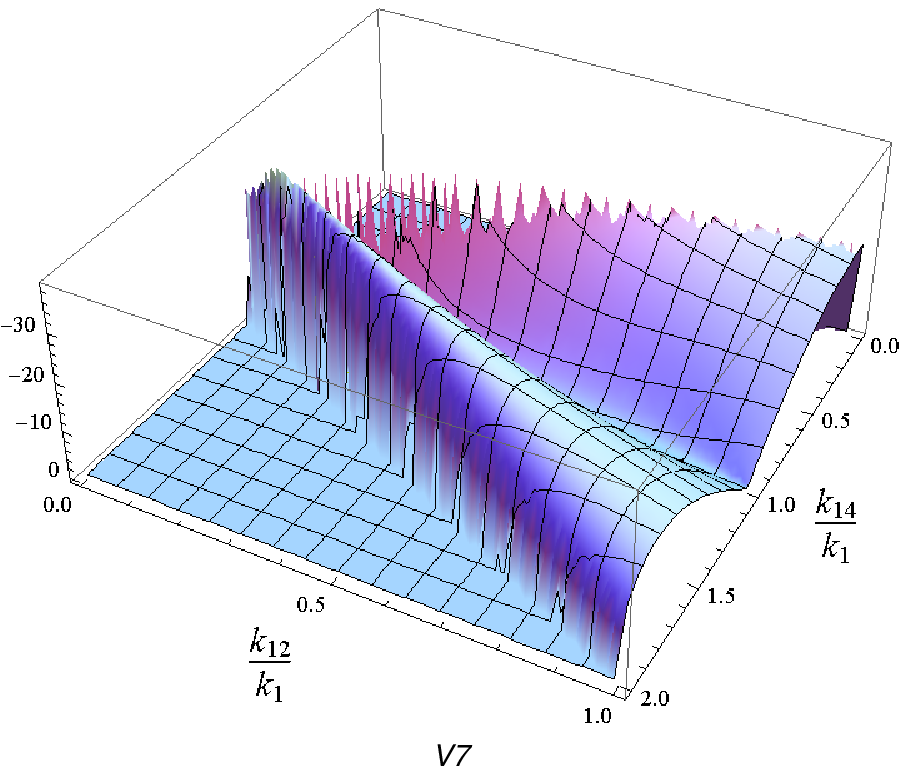}
\caption{\label{CIdoublesqueezedlimit1} The different CI trispectra shapes divided by $k_1k_2k_3k_4$ in the \textit{double-squeezed} limit as functions of $k_{12}/k_1$ and $k_{14}/k_1$.}
\end{figure}
\begin{figure}
\center
\includegraphics[width=0.3\textwidth]{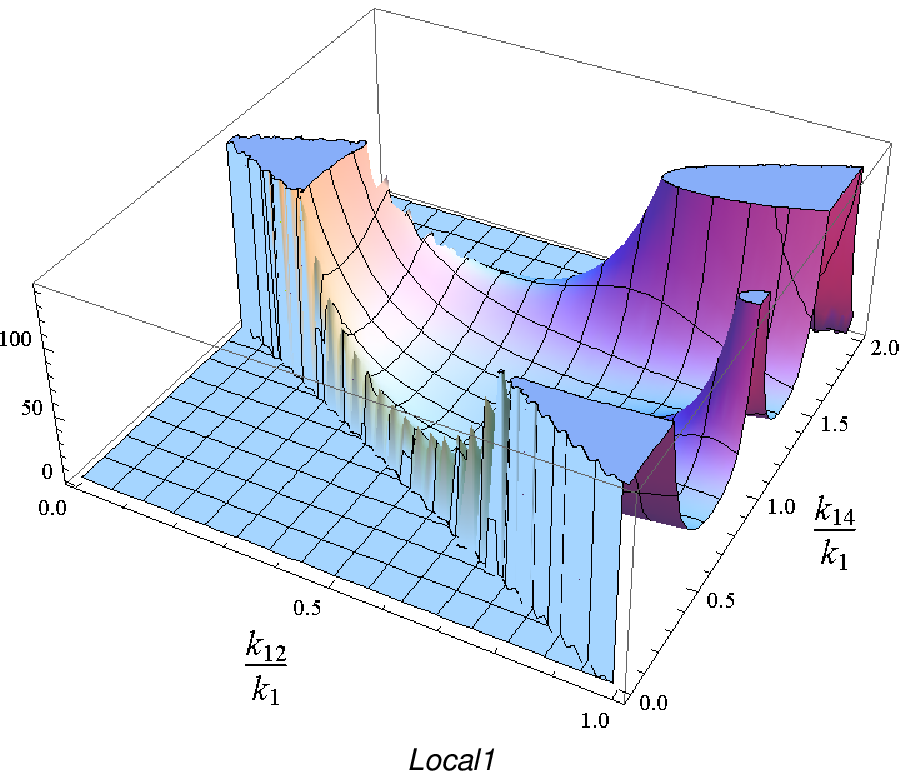}
\hspace{0.05\textwidth}
\includegraphics[width=0.3\textwidth]{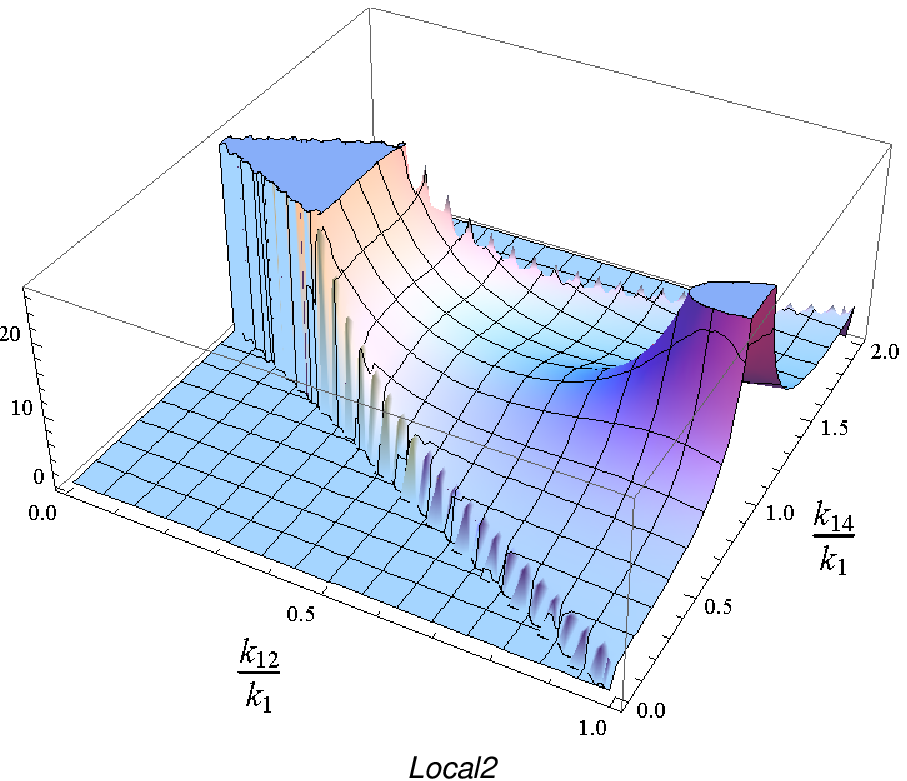}
\caption{\label{CIdoublesqueezedlimit2} For completeness, we plot  here the local trispectra shapes in the \textit{double-squeezed} limit.}
\end{figure}

\newpage
\noindent In interpreting the significance of the various shape-functions one must exert some care in that we must not naively treat all the different interactions as necessarily independent. The bispectrum analysis for example reveals that, of the four initial cubic interactions, one can be field-redefined away \cite{Burrage:2010cu} and one more would give (we verified this explicitly) a bispectrum contribution which is  analytically  identical to a specific linear combination of the contributions from the two remaining interactions (see the comments after Eq.~\ref{5.4}). We have excluded a number of cubic and quartic interactions terms  from the trispectrum plots above because they generate shapes which are, at least qualitatively, indistinguishable from some of those we do plot. It is quite possible that, in doing so, we have restricted the remaining plotted shape-functions to be indeed all independent. It is possible, although very lengthy, to perform a complete ``orthonormalization" of the interactions which contribute to the trispectrum. The difficulty lies in the fact that the $c_n$ coefficients in the interaction terms appear not just in linear combinations, but in quadratic and cubic powers.

There is however a simple way to see that the shapes we plotted are indeed easily (i.e. without fine tuning) generated in Galileon inflation and therefore their signatures  represent a clear cut characterization of this inflationary model. 

The fact that several interaction terms share a specific profile pattern, such as e.g. $V_5,V_7,...$ in the $k_{12}\rightarrow 0$ limit in the double-squeezed configuration, suggests that one may, easily and without fine tuning, arrange for such a shape or, in other words, identify a basis for the various interactions in such a way that one of the basis vector does posses such a behaviour.\\
So far we have been discussing the analytical equivalence between shapes. Although it is always possible to pin down analytically the differences among the various shape-functions, it is not realistic to use such an approach when dealing with actual observables. In this respect, it is quite instructive to look at what happens with the bispectrum. There is indeed a degree to which shape-functions are to be considered hardly distinguishable, when their-cross correlation is sufficiently high for a particular set of data \footnote{In e.g. \cite{Senatore:2009gt} the authors do not mark as ``different" shapes which have an overlap which exceeds $>0.7$.}.
We have been careful to stress here as distinctively different from each other only those shape-functions which, already at the qualitative level, clearly show a distinguishing profile.

\section{Conclusions}
\noindent This paper represents an attempt to navigate part of the landscape of inflationary models guided by observational as well as  more formal criteria. The very recent success of the \textit{Planck} mission has provided us with an improved sensitivity over non-Gaussian observables \cite{Ade:2013ydc}; chief among them, the quantity $f_{NL}$ has been markedly constrained in three of its possible realizations, \textit{local}, \textit{equilateral}, \textit{orthogonal}, and interesting tight constraints have been also put on some other specific non-standard inflationary models.
We have detailed here our results on the study on the model of Galileon inflation, an inflationary mechanism that, for a large region of its parameters space, is easily compatible with \textit{Planck} results, as far as the power spectrum and the bispectrum of curvature perturbations are concerned.

As soon as any model fits within the available observational bounds, it becomes crucial to characterize its properties so as to distinguish it from other realizations. This investigation in our case has been manifold. First we reported on the stability properties which, from a quantum field theory perspective, describe the Galileon inflation model \cite{Burrage:2010cu}. We stressed second order equations of motion and non-renormalization properties that guarantee we are dealing with a predictive model.

We then went on to detail on the study of non-Gaussianities. The bispectrum analysis has already been performed \cite{Burrage:2010cu} and, as it turns out, the shape-function typically peaks in the equilateral limit, just as it does for an array of inflationary models in the literature.

It was therefore essential for the model characterization to go further in perturbation theory and tackle the trispectrum of curvature fluctuations. Doing so pays off in terms of distinguishing this model from other well studied classes of inflationary theories such as $P(X,\phi)$. Since the trispectrum momentum-dependence has too many variables to be plotted simultaneously, we opted for plots in a number of different momenta configurations which proves handy for comparison with other studies. In two of these configurations, \textit{equilateral} and \textit{double-squeezed}, several of the Galileon inflation interactions that contribute to the trispectrum generate a shape function which is strikingly different from its $P(X,\phi)$ counterpart. More specifically, for the most interesting interaction terms, the \textit{equilateral} configuration  shape-function is strongly different w.r.t. $P(X,\phi)$ models: it peaks where the $P(X,\phi)$ counterpart would have either local minima or a plateau. For the \textit{double-squeezed} configuration one finds that often both the scalar-exchange and contact-interactions contributions are characterized by a profile that does not vanish in the $k_{12}\rightarrow 0$ limit. This is in clear contradistinction with what happens in $P(X,\phi)$  models for which there exists a clear cut third-\textit{vs}-fourth order behaviour in that limit.

Having verified that the characterization of Galilean inflation through non-Gaussian observables enables one to identify distinct features which clearly distinguish this model from an entire class of inflationary theories, one should of course point to possible future sources of data that could actually enable such a comparison. Although the results to date of the Planck mission data analysis have been already implemented in the quest for determining the essential features of the most compelling inflationary model, more data is expected and the inclusion of polarization is bound to improve further the constraints on non-Gaussian observables. Further input on (primordial and otherwise) non-Gaussian characterizations of inflationary models is expected to originate from Large Scale Structure sources (see, e.g., \cite{Carbone:2008iz,Euclid}) and possibly from future CMB polarization experiments (see, e.g., \cite{Baumann:2008aq}, \cite{Bouchet:2011ck} and \cite{Andre:2013afa})
Another important probe might well be  21cm cosmology \cite{Cooray:2006km,Pillepich:2006fj,Mao:2013yaa,Lidz:2013tra}.

Within the realm of what one might call the \textit{Galilean approach} to inflation, one should certainly count the work presented here, and, naturally, Ref. \cite{Burrage:2010cu} (see also \cite{Fasiello:2013dla}) where the basis of the model were laid. This model represents a full theory endowed with an inflating background solution and whose stability rests on very firm ground. On the other hand, the appealing properties of Galileon theories have been employed also in a related setup in \cite{Creminelli:2010qf, Bartolo:2013eka} where the fluctuations of an inflating solution around an FLRW background have been equipped with Galilean symmetry. The control over the full theory is  slightly relaxed in this latter case (background behaviour is assumed), more in the spirit of the pioneering work in \cite{Cheung:2007st}. The predictions in terms of non-Gaussian observables are not dissimilar within these two approaches. However, specific interaction terms behaviour presented here (e.g. what we call here the $V_8$-driven interaction) has no corresponding interaction in the related approach \cite{Bartolo:2013eka} and might well provide a sufficiently distinguishing feature.

Having the predictions for  both bispectrum and trispectrum at our disposal, one might well ask if, within the parameters space of the model (and in agreement with observations), there is room for a small-bispectrum \textit{vs} large-trispectrum  region in the parameter space of the model. As it turns out, this region does indeed exist and we refer the interested reader to an upcoming work of ours \cite{bistris} for a detailed analysis. We stress already here that, because of the celebrated non-renormalization properties of Galileon inflation, whatever initial region we select in the parameter space, from there one can proceed safe in the knowledge that the dynamics will not evolve very far upon renormalization and therefore the initial choice will be, in this sense, stable.

\section*{Acknowledgments}
The work of NB has been partially supported by the ASI/INAF Agreement I/072/09/0 for the Planck LFI Activity of Phase E2 and by the PRIN 2009 project ``La Ricerca di non-Gaussianit{\` a} Primordiale''. The work of ED  was partially supported by DOE grant DE-FG02-94ER-40823 at the University of Minnesota. ED is happy to thank the CWRU Physics Department for friendly hospitality during several stages of this work. MF is very grateful to A.J.~Tolley for many enlightening discussions. ED and MF would like to thank the Cosmology Group at the University of Padova, and INFN, Sezione di Padova, for support and for warm hospitality whilst parts of this work were being completed.

\section{Appendix~A. The leading order bispectrum}\vspace{.1cm}

\noindent The result of \cite{Burrage:2010cu} for the leading order (i.e. $\ddot\phi\sim\dot H\sim0$) bispectrum is
\begin{eqnarray}
\langle\zeta(\mathbf{k}_1)\zeta(\mathbf{k}_2)\zeta(\mathbf{k}_3)\rangle&=&-(2\pi)^3\delta^{(3)}(\mathbf{K_b}_t)
\frac{H^9}{4^3\dot\phi_0^3A^3}\frac{1}{c_s^6}\frac{1}{\Pi_ik_i^3}
\Bigg[
24\frac{\mathcal{O}_1+2Hc_s^{-2}\mathcal{O}_2}{H}\frac{\Pi_ik_i^2}{K_b^3}
\nonumber\\
&&+4\frac{\mathcal{O}_3}{Hc_s^2}\frac{1}{K_b}
\Bigg(
k_3^2\mathbf{k}_1\cdot\mathbf{k}_2\left(1+\frac{K_b(k_1+k_2)+2k_1k_2}{K_b^2}\right)
\nonumber\\
&&+k_1^2\mathbf{k}_2\cdot\mathbf{k}_3\left(1+\frac{K_b(k_2+k_3)+2k_2k_3}{K_b^2}\right)+k_2^2\mathbf{k}_1\cdot\mathbf{k}_3\left(1+\frac{K_b(k_1+k_3)+2k_1k_3}{K_b^2}\right)
\Bigg)\nonumber\\
&&+8\frac{\mathcal{O}_4}{c_s^4}\frac{1}{K_b}\left(1+\frac{3k_1k_2k_3+K_b\sum_{j<i}k_ik_j}{K_b^3}\right)
\left(k_1^2\mathbf{k}_2\cdot\mathbf{k}_3+k_2^2\mathbf{k}_1\cdot\mathbf{k}_3+k_3^2\mathbf{k}_1\cdot\mathbf{k}_2\right)\Bigg], \qquad \nonumber \\
&& \label{Bispectrum}
\end{eqnarray}
where $\mathbf{K_b}_t=\mathbf{k}_1+\mathbf{k}_2+\mathbf{k}_3$ and $K_b=k_1+k_2+k_3$.
\noindent One then defines the shapes in the usual way as:
\begin{eqnarray}
Shape_1=(k_1k_2k_3)^2\times\frac{1}{\Pi_ik_i^3}\frac{\Pi_ik_i^2}{K_b^3},
\eea
\bea
Shape_2&=&(k_1k_2k_3)^2\times\frac{1}{\Pi_ik_i^3}\frac{1}{K_b}
\Bigg(
k_3^2\mathbf{k}_1\cdot\mathbf{k}_2\left(1+\frac{K_b(k_1+k_2)+2k_1k_2}{K_b^2}\right)
\\
&&\qquad+k_1^2\mathbf{k}_2\cdot\mathbf{k}_3\left(1+\frac{K_b(k_2+k_3)+2k_2k_3}{K_b^2}\right)+k_2^2\mathbf{k}_1\cdot\mathbf{k}_3\left(1+\frac{K_b(k_1+k_3)+2k_1k_3}{K_b^2}\right)\nonumber
\Bigg),
\eea
\bea
Shape_3&=&(k_1k_2k_3)^2\times\frac{1}{\Pi_ik_i^3}\frac{1}{K_b}\left(1+\frac{3k_1k_2k_3+K_b\sum_{j<i}k_ik_j}{K_b^3}\right)
\left(k_1^2\mathbf{k}_2\cdot\mathbf{k}_3+k_2^2\mathbf{k}_1\cdot\mathbf{k}_3+k_3^2\mathbf{k}_1\cdot\mathbf{k}_2\right). \nonumber \\
\label{5.4}
\eea
It is a well-known fact that these three shapes are highly correlated with the equilateral template. Also one can show that $Shape_3=6\,Shape_1+Shape_2$.

\section{Appendix~B. Contact interaction trispectra}

For the computation of the trispectrum diagrams, we use the leading-order in slow-roll mode function solution for $\delta\phi$
\begin{equation}
u(\tau,k)=\frac{N}{k^{3/2}}\left(1+ikc_s\tau\right)e^{-ikc_s\tau},
\end{equation}
where the normalization factor is $N\equiv H/(2\sqrt{A c_{s}^{3}})$. \\

\noindent We expand $\delta\phi(\tau,\mathbf{k})=u(\tau,k)a(\mathbf{k})+u^*(\tau,k)a^\dag(-\mathbf{k})$ with the standard commutation relations $\left[a(\mathbf{k}_1),a^\dag(\mathbf{k}_2)\right]=(2\pi)^3\delta^{(3)}(\mathbf{k}_1-\mathbf{k}_2)$ and use
\begin{eqnarray}
FPF&\equiv&\langle\Omega|\delta\phi(0,\mathbf{k}_1)\delta\phi(0,\mathbf{k}_2)\delta\phi(0,\mathbf{k}_3)\delta\phi(0,\mathbf{k}_4)|\Omega\rangle\nonumber\\&=&-i\int_{-\infty}^0dt
\langle0|\!\!\left[\delta\phi(0,\mathbf{k}_1)\delta\phi(0,\mathbf{k}_2)\delta\phi(0,\mathbf{k}_3)\delta\phi(0,\mathbf{k}_4),H^{(4)}_{I}(t)\right]\!\!|0\rangle,
\end{eqnarray}
where $H_I=\int d^3x\mathcal{H}_I$ is given at fourth order by (\ref{FourthOrderHamiltonian}).
Then for each of the individual terms in (\ref{FourthOrderHamiltonian}) we obtain
\begin{eqnarray}
FPF_{V_1}&=&-(2\pi)^3\delta^{(3)}(\mathbf{K}_t)\frac{4N^8}{(k_1k_2k_3k_4)^3}12V_1c_s^3\frac{(k_1k_2k_3k_4)^2}{K^5} +23\mathrm{\,perms.},
\label{6.3}
\end{eqnarray}
\begin{eqnarray}
FPF_{V_2}&=&-(2\pi)^3\delta^{(3)}(\mathbf{K}_t)\frac{4N^8}{(k_1k_2k_3k_4)^3}12V_2Hc_s\frac{(k_1k_2k_3k_4)^2}{K^5}\left(1+5\frac{k_4}{K}\right) +23\mathrm{\,perms.},
\end{eqnarray}
\begin{eqnarray}
FPF_{V_3}&=&-(2\pi)^3\delta^{(3)}(\mathbf{K}_t)\frac{4N^8}{(k_1k_2k_3k_4)^3}V_3c_s\frac{k_1^2k_2^2\mathbf{k}_3\cdot\mathbf{k}_4}{K^5}\left(K^2+3K\left(k_3+k_4\right)+12k_3k_4\right)\nonumber\\&+& 23\mathrm{\,perms.},
\end{eqnarray}
\begin{eqnarray}
FPF_{V_4}&=&-(2\pi)^3\delta^{(3)}(\mathbf{K}_t)\frac{4N^8}{(k_1k_2k_3k_4)^3}\frac{12V_4H^2}{c_s}\frac{(k_1k_2k_3k_4)^2}{K^5}\left(1+\frac{5}{K}\left(k_3+k_4\right)+\frac{30 k_3k_4}{K^2}\right)\nonumber\\&+& 23\mathrm{\,perms.},
\end{eqnarray}
\begin{eqnarray}
FPF_{V_5}&=&-(2\pi)^3\delta^{(3)}(\mathbf{K}_t)\frac{4N^8}{(k_1k_2k_3k_4)^3}\frac{12V_5H^2}{c_s}\frac{k_1^2k_2^2(\mathbf{k}_3\cdot\mathbf{k}_4)^2}{K^5}\left(1+\frac{5}{K}\left(k_3+k_4\right)+\frac{30 k_3k_4}{K^2}\right) \nonumber\\&+& 23\mathrm{\,perms.},
\end{eqnarray}
\begin{eqnarray}
FPF_{V_6}&=&-(2\pi)^3\delta^{(3)}(\mathbf{K}_t)\frac{4N^8}{(k_1k_2k_3k_4)^3}\frac{V_6H}{c_s}\frac{k_1^2k_4^2\mathbf{k}_2\cdot\mathbf{k}_3}{K^5}\Big(4K^2-3k_1K+12\left(k_2k_3+k_2k_4+k_3k_4\right)\nonumber\\&&+60\frac{k_2k_3k_4}{K}\Big)+23\mathrm{\,perms.},
\end{eqnarray}
\begin{eqnarray}
FPF_{V_7}&=&-(2\pi)^3\delta^{(3)}(\mathbf{K}_t)\frac{4N^8}{(k_1k_2k_3k_4)^3}\frac{V_7}{c_s}\frac{\mathbf{k}_1\cdot\mathbf{k}_2\mathbf{k}_3\cdot\mathbf{k}_4}{K^5}\Big(K^4+K^2\sum_{i=1}^4\sum_{j>i}^4k_ik_j\nonumber\\&+&3K\sum_{l=1}^4\sum_{m>l}^4\sum_{n>m}^4k_lk_mk_n +12k_1k_2k_3k_4\Big)+23\mathrm{\,perms.},
\end{eqnarray}
\begin{eqnarray}
FPF_{V_8}&=&(2\pi)^3\delta^{(3)}(\mathbf{K}_t)\frac{4N^8}{(k_1k_2k_3k_4)^3}\frac{V_8H^2}{c_s^3}\frac{\mathbf{k}_1\cdot\mathbf{k}_2((\mathbf{k}_3\cdot\mathbf{k}_4)^2-k_3^2k_4^2)}{K^5}\\
&&\times\left(4K^2+12\sum_{i=1}^4\sum_{j>i}^4k_ik_j+\frac{60}{K}\sum_{l=1}^4\sum_{m>l}^4\sum_{n>m}^4k_lk_mk_n+360\frac{k_1k_2k_3k_4}{K^2}\right)+23\mathrm{\,perms.},\nonumber
\label{6.10}
\end{eqnarray}
where $K=k_1+k_2+k_3+k_4$ and $\mathbf{K}_t=\mathbf{k}_1+\mathbf{k}_2+\mathbf{k}_3+\mathbf{k}_4$.\\

\noindent Note that not all the CI contributions to the trispectrum are independent, indeed, after some manipulations one finds:
\bea
FPF_{V_2}|_{V_2=1}=\frac{9H}{4c_s^2}FPF_{V_1}|_{V_1=1},\qquad
FPF_{V_6}|_{V_6=1}=\frac{3H}{4c_s^4}FPF_{V_1}|_{V_1=1}+\frac{5H}{2c_s^2}FPF_{V_3}|_{V_3=1}.\nonumber\\
\eea

%%%%%%%%%%%%%%%%%%%%%%%%%%%%%%%%%%%%%%%%%%%%%%%%%%%%%%%%%%%%%%%%%%%%%

\section{Appendix~C. Scalar exchange trispectra}

\begin{table}
\centering
  \begin{tabular}{ c || c | c | c| c}

                  & $\mathcal{O}_1^R$ & $\mathcal{O}_2^R$ & $\mathcal{O}_3^R$ & $\mathcal{O}_4^R$ \\ \hline\hline
    $\mathcal{O}_1^L$ & $\mathcal{F}_1$ & $\mathcal{F}_5$ & $\mathcal{F}_3$ & $\mathcal{F}_7$ \\ \hline
    $\mathcal{O}_2^L$ & $\mathcal{F}_6$ & $\mathcal{F}_9$ & $\mathcal{F}_{10}$ & $\mathcal{F}_{12}$ \\ \hline
    $\mathcal{O}_3^L$ & $\mathcal{F}_4$ & $\mathcal{F}_{11}$ & $\mathcal{F}_2$ & $\mathcal{F}_{14}$ \\ \hline
    $\mathcal{O}_4^L$ & $\mathcal{F}_8$ & $\mathcal{F}_{13}$ & $\mathcal{F}_{15}$ & $\mathcal{F}_{16}$ \\ \hline
  \end{tabular}
  \caption{Rules to decide which $\mathcal{F}_i$ to use when writing the contribution of a certain diagram according to the diagrammatic rules of \cite{Mizuno:2009mv}. For example, if the lhs vertex of the diagram is the vertex proportional to $\mathcal{O}_3$ and the rhs vertex is the vertex proportional to $\mathcal{O}_4$ then the functions to use is $\mathcal{F}_{14}$.}
  \label{Table1}
\end{table}

For the SE trispectrum contribution, one employs the following:
\bea
FPF^{SE}&\equiv& \langle \Omega| \delta\phi(0,k_1)  \delta\phi(0,k_2)  \delta\phi(0,k_3)  \delta\phi(0,k_4)   |\Omega\rangle  \nonumber \\ &=& -\int_{-\infty}^{0} dt  \int_{-\infty}^{t} d\tilde{t} \langle 0| \Big[ \big[\delta\phi(0,k_1)  \delta\phi(0,k_2)  \delta\phi(0,k_3)  \delta\phi(0,k_4), H_{I}^{(3)}(t)\big], H_{I}^{(3)}(\tilde{t})  \Big]  |0 \rangle . \nonumber \\
\eea

\noindent The different trispecta coming from the different vertices can be written as (using the diagrammatic approach rules described in \cite{Mizuno:2009mv} and Table \ref{Table1})
\begin{eqnarray}
FPF^{SE}_{\mathcal{O}_1\mathcal{O}_2}&=&-2(2\pi)^3\delta^{(3)}(\mathbf{K}_t)\frac{N^4}{(k_1k_2k_3k_4)^\frac{3}{2}}\mathcal{O}_1\mathcal{O}_2
\nonumber\\
&&\times\Big[
-3\mathbf{k}_{12}\cdot\mathbf{k}_{12}\Big(\mathcal{F}_5(k_1,k_2,-k_{12},k_3,k_4,k_{12})-\mathcal{F}_5(-k_1,-k_2,-k_{12},k_3,k_4,k_{12})\Big)
\nonumber\\
&&
\quad\;\,-6k_4^2\Big(\mathcal{F}_5(k_1,k_2,-k_{12},k_3,k_{12},k_4)-\mathcal{F}_5(-k_1,-k_2,-k_{12},k_3,k_{12},k_4)\Big)
\nonumber\\
&&
\quad\;\,-3\mathbf{k}_{12}\cdot\mathbf{k}_{12}\Big(\mathcal{F}_6(k_1,k_2,-k_{12},k_3,k_4,k_{12})-\mathcal{F}_6(-k_1,-k_2,-k_{12},k_3,k_4,k_{12})\Big)
\nonumber\\
&&
\quad\;\,-6k_2^2\Big(\mathcal{F}_6(k_1,-k_{12},k_2,k_3,k_4,k_{12})-\mathcal{F}_6(-k_1,-k_{12},-k_2,k_3,k_4,k_{12})\Big)
\Big]
\nonumber\\&&+23\,\mathrm{perms.\,of\,\{k_1,k_2,k_3,k_4\}},
\label{7.2}
\end{eqnarray}
\begin{eqnarray}
FPF^{SE}_{\mathcal{O}_1\mathcal{O}_4}&=&-2(2\pi)^3\delta^{(3)}(\mathbf{K}_t)\frac{N^4}{(k_1k_2k_3k_4)^\frac{3}{2}}\mathcal{O}_1\mathcal{O}_4
\nonumber\\
&&\times\Big[\left(3k_{12}^2\mathbf{k}_3\cdot\mathbf{k}_4+6k_4^2\mathbf{k}_3\cdot\mathbf{k}_{12}\right)\Big(\mathcal{F}_7(k_1,k_2,-k_{12},k_3,k_4,k_{12})\nonumber\\
&& -\mathcal{F}_7(-k_1,-k_2,-k_{12},k_3,k_4,k_{12})\Big)
\nonumber\\
&&
\quad\;\,+\left(3k_{12}^2\mathbf{k}_1\cdot\mathbf{k}_2-6k_2^2\mathbf{k}_1\cdot\mathbf{k}_{12}\right)\Big(\mathcal{F}_8(k_1,k_2,-k_{12},k_3,k_4,k_{12})\nonumber\\
&&-\mathcal{F}_8(-k_1,-k_2,-k_{12},k_3,k_4,k_{12})\Big)
\Big]+23\,\mathrm{perms.\,of\,\{k_1,k_2,k_3,k_4\}},
\end{eqnarray}
\begin{eqnarray}
FPF^{SE}_{\mathcal{O}_2\mathcal{O}_3}&=&-2(2\pi)^3\delta^{(3)}(\mathbf{K}_t)\frac{N^4}{(k_1k_2k_3k_4)^\frac{3}{2}}\mathcal{O}_2\mathcal{O}_3
\nonumber\\
&&\times\Big[
k_{12}^2\mathbf{k}_{3}\cdot\mathbf{k}_{4}\Big(\mathcal{F}_{10}(k_1,k_2,-k_{12},k_{12},k_3,k_4)-\mathcal{F}_{10}(-k_1,-k_2,-k_{12},k_{12},k_3,k_4)\Big)
\nonumber\\
&&
\quad\;\,+2k_{12}^2\mathbf{k}_{4}\cdot\mathbf{k}_{12}\Big(\mathcal{F}_{10}(k_1,k_2,-k_{12},k_3,k_4,k_{12})-\mathcal{F}_{10}(-k_1,-k_2,-k_{12},k_3,k_4,k_{12})\Big)
\nonumber\\
&&
\quad\;\,+2k_2^2\mathbf{k}_{3}\cdot\mathbf{k}_{4}\Big(\mathcal{F}_{10}(k_1,-k_{12},k_2,k_{12},k_3,k_4)-\mathcal{F}_{10}(-k_1,-k_{12},-k_2,k_{12},k_3,k_4)\Big)
\nonumber\\
&&
\quad\;\,+4k_2^2\mathbf{k}_{4}\cdot\mathbf{k}_{12}\Big(\mathcal{F}_{10}(k_1,-k_{12},k_2,k_3,k_4,k_{12})-\mathcal{F}_{10}(-k_1,-k_{12},-k_2,k_3,k_4,k_{12})\Big)
\nonumber\\
&&
\quad\;\,+k_{12}^2\mathbf{k}_{1}\cdot\mathbf{k}_{2}\Big(\mathcal{F}_{11}(-k_{12},k_1,k_2,k_3,k_4,k_{12})-\mathcal{F}_{11}(-k_{12},-k_1,-k_2,k_3,k_4,k_{12})\Big)
\nonumber\\
&&
\quad\;\,+2k_{4}^2\mathbf{k}_{1}\cdot\mathbf{k}_{2}\Big(\mathcal{F}_{11}(-k_{12},k_1,k_2,k_3,k_{12},k_4)-\mathcal{F}_{11}(-k_{12},-k_1,-k_2,k_3,k_{12},k_4)\Big)
\nonumber\\
&&
\quad\;\,-2k_{12}^2\mathbf{k}_{2}\cdot\mathbf{k}_{12}\Big(\mathcal{F}_{11}(k_1,k_2,-k_{12},k_3,k_4,k_{12})-\mathcal{F}_{11}(-k_1,-k_2,-k_{12},k_3,k_4,k_{12})\Big)
\nonumber\\
&&
\quad\;\,-4k_4^2\mathbf{k}_{2}\cdot\mathbf{k}_{12}\Big(\mathcal{F}_{11}(k_1,k_2,-k_{12},k_3,k_{12},k_4)-\mathcal{F}_{11}(-k_1,-k_2,-k_{12},k_3,k_{12},k_4)\Big)
\Big]
\nonumber\\&&+23\,\mathrm{perms.\,of\,\{k_1,k_2,k_3,k_4\}},
\end{eqnarray}
\begin{eqnarray}
FPF^{SE}_{\mathcal{O}_2\mathcal{O}_4}&=&-2(2\pi)^3\delta^{(3)}(\mathbf{K}_t)\frac{N^4}{(k_1k_2k_3k_4)^\frac{3}{2}}\mathcal{O}_2\mathcal{O}_4
\nonumber\\
&&\times\Big[-k_{12}^2\left(k_{12}^2\mathbf{k}_3\cdot\mathbf{k}_4+2k_4^2\mathbf{k}_3\cdot\mathbf{k}_{12}\right)\Big(\mathcal{F}_{12}(k_1,k_2,-k_{12},k_3,k_4,k_{12})\nonumber\\
&&-\mathcal{F}_{12}(-k_1,-k_2,-k_{12},k_3,k_4,k_{12})\Big)
\nonumber\\
&&
\quad\;\,-2k_2^2\left(k_{12}^2\mathbf{k}_3\cdot\mathbf{k}_4+2k_4^2\mathbf{k}_3\cdot\mathbf{k}_{12}\right)\Big(\mathcal{F}_{12}(k_1,-k_{12},k_2,k_3,k_4,k_{12})\nonumber\\
&&-\mathcal{F}_{12}(-k_1,-k_{12},-k_2,k_3,k_4,k_{12})\Big)
\nonumber\\
&&
\quad\;\,-k_{12}^2\left(k_{12}^2\mathbf{k}_1\cdot\mathbf{k}_2-2k_2^2\mathbf{k}_1\cdot\mathbf{k}_{12}\right)\Big(\mathcal{F}_{13}(k_1,k_2,-k_{12},k_3,k_4,k_{12})\nonumber\\
&&-\mathcal{F}_{13}(-k_1,-k_2,-k_{12},k_3,k_4,k_{12})\Big)
\nonumber\\
&&
\quad\;\,-2k_4^2\left(k_{12}^2\mathbf{k}_1\cdot\mathbf{k}_2-2k_2^2\mathbf{k}_1\cdot\mathbf{k}_{12}\right)\Big(\mathcal{F}_{13}(k_1,k_2,-k_{12},k_3,k_{12},k_4)\nonumber\\
&&-\mathcal{F}_{13}(-k_1,-k_2,-k_{12},k_3,k_{12},k_4)\Big)
\Big]+23\,\mathrm{perms.\,of\,\{k_1,k_2,k_3,k_4\}},
\end{eqnarray}
\begin{eqnarray}
FPF^{SE}_{\mathcal{O}_3\mathcal{O}_4}&=&-2(2\pi)^3\delta^{(3)}(\mathbf{K}_t)\frac{N^4}{(k_1k_2k_3k_4)^\frac{3}{2}}\mathcal{O}_3\mathcal{O}_4
\nonumber\\
&&\times\Big[-\mathbf{k}_1\cdot\mathbf{k}_2\left(k_{12}^2\mathbf{k}_3\cdot\mathbf{k}_4+2k_4^2\mathbf{k}_3\cdot\mathbf{k}_{12}\right)\Big(\mathcal{F}_{14}(-k_{12},k_1,k_2,k_3,k_4,k_{12})\nonumber\\
&&-\mathcal{F}_{14}(-k_{12},-k_1,-k_2,k_3,k_4,k_{12})\Big)
\nonumber\\
&&
\quad\;\,+2\mathbf{k}_2\cdot\mathbf{k}_{12}\left(k_{12}^2\mathbf{k}_3\cdot\mathbf{k}_4+2k_4^2\mathbf{k}_3\cdot\mathbf{k}_{12}\right)\Big(\mathcal{F}_{14}(k_1,k_2,-k_{12},k_3,k_4,k_{12})\nonumber\\
&&-\mathcal{F}_{14}(-k_1,-k_2,-k_{12},k_3,k_4,k_{12})\Big)
\nonumber\\
&&
\quad\;\,-\mathbf{k}_3\cdot\mathbf{k}_4\left(k_{12}^2\mathbf{k}_1\cdot\mathbf{k}_2-2k_2^2\mathbf{k}_1\cdot\mathbf{k}_{12}\right)\Big(\mathcal{F}_{15}(k_1,k_2,-k_{12},k_{12},k_3,k_4)\nonumber\\
&&-\mathcal{F}_{15}(-k_1,-k_2,-k_{12},k_{12},k_3,k_4)\Big)
\nonumber\\
&&
\quad\;\,-2\mathbf{k}_4\cdot\mathbf{k}_{12}\left(k_{12}^2\mathbf{k}_1\cdot\mathbf{k}_2-2k_2^2\mathbf{k}_1\cdot\mathbf{k}_{12}\right)\Big(\mathcal{F}_{15}(k_1,k_2,-k_{12},k_3,k_4,k_{12})\nonumber\\
&&-\mathcal{F}_{15}(-k_1,-k_2,-k_{12},k_3,k_4,k_{12})\Big)
\Big]+23\,\mathrm{perms.\,of\,\{k_1,k_2,k_3,k_4\}},
\end{eqnarray}
\begin{eqnarray}
FPF^{SE}_{\mathcal{O}_2\mathcal{O}_2}&=&-2(2\pi)^3\delta^{(3)}(\mathbf{K}_t)\frac{N^4}{(k_1k_2k_3k_4)^\frac{3}{2}}\mathcal{O}_2^2
\nonumber\\
&&\times\Big[
k_{12}^4\Big(\mathcal{F}_9(k_1,k_2,-k_{12},k_3,k_4,k_{12})-\mathcal{F}_9(-k_1,-k_2,-k_{12},k_3,k_4,k_{12})\Big)
\nonumber\\
&&
\quad\;\,+2k_4^2k_{12}^2\Big(\mathcal{F}_9(k_1,k_2,-k_{12},k_3,k_{12},k_4)-\mathcal{F}_9(-k_1,-k_2,-k_{12},k_3,k_{12},k_4)\Big)
\nonumber\\
&&
\quad\;\,+2k_2^2k_{12}^2\Big(\mathcal{F}_9(k_1,-k_{12},k_2,k_3,k_4,k_{12})-\mathcal{F}_9(-k_1,-k_{12},-k_2,k_3,k_4,k_{12})\Big)
\nonumber\\
&&
\quad\;\,+4k_2^2k_4^2\Big(\mathcal{F}_9(k_1,-k_{12},k_2,k_3,k_{12},k_4)-\mathcal{F}_9(-k_1,-k_{12},-k_2,k_3,k_{12},k_4)\Big)
\Big]
\nonumber\\&&+23\,\mathrm{perms.\,of\,\{k_1,k_2,k_3,k_4\}},
\end{eqnarray}
\begin{eqnarray}
FPF^{SE}_{\mathcal{O}_4\mathcal{O}_4}&=&-2(2\pi)^3\delta^{(3)}(\mathbf{K}_t)\frac{N^4}{(k_1k_2k_3k_4)^\frac{3}{2}}\mathcal{O}_4^2
\nonumber\\
&&\times
\Big(
k_{12}^4(\mathbf{k}_1\cdot\mathbf{k}_2)(\mathbf{k}_3\cdot\mathbf{k}_4)
+2k_4^2k_{12}^2(\mathbf{k}_1\cdot\mathbf{k}_2)(\mathbf{k}_3\cdot\mathbf{k}_{12})
-2k_2^2k_{12}^2(\mathbf{k}_1\cdot\mathbf{k}_{12})(\mathbf{k}_3\cdot\mathbf{k}_4)
\nonumber\\
&&-4k_2^2k_4^2(\mathbf{k}_1\cdot\mathbf{k}_{12})(\mathbf{k}_3\cdot\mathbf{k}_{12})
\Big)
\Big(\mathcal{F}_{16}(k_1,k_2,-k_{12},k_3,k_4,k_{12})\nonumber\\
&&-\mathcal{F}_{16}(-k_1,-k_2,-k_{12},k_3,k_4,k_{12})\Big)+23\,\mathrm{perms.\,of\,\{k_1,k_2,k_3,k_4\}},
\end{eqnarray}
\begin{eqnarray}
FPF^{SE}_{\mathcal{O}_1\mathcal{O}_1}&=&-2(2\pi)^3\delta^{(3)}(\mathbf{K}_t)\frac{N^4}{(k_1k_2k_3k_4)^\frac{3}{2}}9\mathcal{O}_1^2
\Big(\mathcal{F}_1(k_1,k_2,-k_{12},k_3,k_4,k_{12})\nonumber\\
&&-\mathcal{F}_1(-k_1,-k_2,-k_{12},k_3,k_4,k_{12})\Big)+23\,\mathrm{perms.\,of\,\{k_1,k_2,k_3,k_4\}},
\end{eqnarray}
\begin{eqnarray}
FPF^{SE}_{\mathcal{O}_3\mathcal{O}_3}&=&-2(2\pi)^3\delta^{(3)}(\mathbf{K}_t)\frac{N^4}{(k_1k_2k_3k_4)^\frac{3}{2}}\mathcal{O}_3^2
\nonumber\\
&&\times\Big[
(\mathbf{k}_{1}\cdot\mathbf{k}_{2})(\mathbf{k}_{3}\cdot\mathbf{k}_{4})\Big(\mathcal{F}_2(-k_{12},k_1,k_2,k_{12},k_3,k_4)-\mathcal{F}_2(-k_{12},-k_1,-k_2,k_{12},k_3,k_4)\Big)
\nonumber\\
&&
\quad\;\,+2(\mathbf{k}_{1}\cdot\mathbf{k}_{2})(\mathbf{k}_{4}\cdot\mathbf{k}_{12})\Big(\mathcal{F}_2(-k_{12},k_1,k_2,k_3,k_4,k_{12})-\mathcal{F}_2(-k_{12},-k_1,-k_2,k_3,k_4,k_{12})\Big)
\nonumber\\
&&
\quad\;\,-2(\mathbf{k}_{2}\cdot\mathbf{k}_{12})(\mathbf{k}_{3}\cdot\mathbf{k}_{4})\Big(\mathcal{F}_2(k_1,k_2,-k_{12},k_{12},k_3,k_4)-\mathcal{F}_2(-k_1,-k_2,-k_{12},k_{12},k_3,k_4)\Big)
\nonumber\\
&&
\quad\;\,-4(\mathbf{k}_{2}\cdot\mathbf{k}_{12})(\mathbf{k}_{4}\cdot\mathbf{k}_{12})\Big(\mathcal{F}_2(k_1,k_2,-k_{12},k_3,k_4,k_{12})-\mathcal{F}_2(-k_1,-k_2,-k_{12},k_3,k_4,k_{12})\Big)
\Big]
\nonumber\\&&+23\,\mathrm{perms.\,of\,\{k_1,k_2,k_3,k_4\}},
\end{eqnarray}
\begin{eqnarray}
FPF^{SE}_{\mathcal{O}_1\mathcal{O}_3}&=&-2(2\pi)^3\delta^{(3)}(\mathbf{K}_t)\frac{N^4}{(k_1k_2k_3k_4)^\frac{3}{2}}\mathcal{O}_1\mathcal{O}_3
\nonumber\\
&&\times\Big[
-3\mathbf{k}_{3}\cdot\mathbf{k}_{4}\Big(\mathcal{F}_3(k_1,k_2,-k_{12},k_{12},k_3,k_4)-\mathcal{F}_3(-k_1,-k_2,-k_{12},k_{12},k_3,k_4)\Big)
\nonumber\\
&&
\quad\;\,-6\mathbf{k}_{4}\cdot\mathbf{k}_{12}\Big(\mathcal{F}_3(k_1,k_2,-k_{12},k_3,k_4,k_{12})-\mathcal{F}_3(-k_1,-k_2,-k_{12},k_3,k_4,k_{12})\Big)
\nonumber\\
&&
\quad\;\,-3\mathbf{k}_{1}\cdot\mathbf{k}_{2}\Big(\mathcal{F}_4(-k_{12},k_1,k_2,k_3,k_4,k_{12})-\mathcal{F}_4(-k_{12},-k_1,-k_2,k_3,k_4,k_{12})\Big)
\nonumber\\
&&
\quad\;\,+6\mathbf{k}_{2}\cdot\mathbf{k}_{12}\Big(\mathcal{F}_4(k_1,k_2,-k_{12},k_3,k_4,k_{12})-\mathcal{F}_4(-k_1,-k_2,-k_{12},k_3,k_4,k_{12})\Big)
\Big]
\nonumber\\&&+23\,\mathrm{perms.\,of\,\{k_1,k_2,k_3,k_4\}},
\label{7.11}
\end{eqnarray}
where the last three equations are the ones that appear in the $P(X,\phi)$ model and $k_{12}=|\mathbf{k}_{12}|=|\mathbf{k}_1+\mathbf{k}_2|$. Similarly we define $k_{ab}$ as $k_{ab}=|\mathbf{k}_{ab}|=|\mathbf{k}_{a}+\mathbf{k}_{b}|$, where $\mathbf{k}_{a}$ and $\mathbf{k}_{b}$ represent any of the four momentum vectors $\mathbf{k}_{1}$, $\mathbf{k}_{2}$, $\mathbf{k}_{3}$ and $\mathbf{k}_{4}$. Momentum conservation implies $k_{12}=k_{34}$, $k_{13}=k_{24}$ and $k_{14}=k_{23}$.\\

\noindent Using the modified mode function $U(\tau,k)$ defined as \footnote{If the sign of the argument $k$ is positive then $U$ is equal to the mode function, if the sign is negative then $U$ is equal to the complex conjugate of the mode function.}
\begin{equation}
U(\tau,k)\equiv \frac{N}{|k|^{3/2}}(1+ikc_s\tau)e^{-ikc_s\tau},
\end{equation}
we define 16 $\mathcal{F}_i(k_1,k_2,k_3,k_4,k_5,k_6)$ functions as
\begin{eqnarray}
\mathcal{F}_1(k_1,...,k_6)&=&\int^0_{-\infty}d\tau a(\tau)\int^\tau_{-\infty}d\tilde\tau a(\tilde\tau)U^{*'}(\tau,k_1)U^{*'}(\tau,k_2)U^{*'}(\tau,k_3)U^{*'}(\tilde\tau,k_4)U^{*'}(\tilde\tau,k_5)U^{*'}(\tilde\tau,k_6)
\nonumber\\
&=&-4\frac{N^6c_s^6}{H^2}|k_1\cdots k_6|^\frac{1}{2}\frac{1}{\mathcal{A}^3\mathcal{C}^3}\left(1+3\frac{\mathcal{A}}{\mathcal{C}}+6\frac{\mathcal{A}^2}{\mathcal{C}^2}\right),
\label{F1}
\end{eqnarray}
\begin{eqnarray}
\mathcal{F}_2(k_1,...,k_6)&=&\int^0_{-\infty}d\tau a(\tau)\int^\tau_{-\infty}d\tilde\tau a(\tilde\tau)U^{*'}(\tau,k_1)U^{*}(\tau,k_2)U^{*}(\tau,k_3)U^{*'}(\tilde\tau,k_4)U^{*}(\tilde\tau,k_5)U^{*}(\tilde\tau,k_6)
\nonumber\\
&=&-\frac{N^6c_s^2}{H^2}\frac{|k_1k_4|^\frac{1}{2}}{|k_2k_3k_5k_6|^\frac{3}{2}}\frac{1}{\mathcal{A}\mathcal{C}}
\bigg[
      1+\frac{k_5+k_6}{\mathcal{A}}+2\frac{k_5k_6}{\mathcal{A}^2}
      \nonumber\\&&
      +\frac{1}{\mathcal{C}}  \left(k_2+k_3+k_5+k_6+\frac{1}{\mathcal{A}}\left(\left(k_2+k_3\right)\left(k_5+k_6\right)+2k_5k_6\right)+2\frac{k_5k_6\left(k_2+k_3\right)}{\mathcal{A}^2}\right)
      \nonumber\\&&
      +\frac{2}{\mathcal{C}^2}\bigg(k_5k_6+\left(k_2+k_3\right)\left(k_5+k_6\right)+k_2k_3
      \nonumber\\&&
      \qquad\quad+\frac{1}{\mathcal{A}}\left(k_2k_3\left(k_5+k_6\right)+2k_5k_6\left(k_2+k_3\right)\right)+2\frac{k_2k_3k_5k_6}{\mathcal{A}^2}\bigg)
      \nonumber\\&&
      +\frac{6}{\mathcal{C}^3}\left(k_2k_3\left(k_5+k_6\right)+k_5k_6\left(k_2+k_3\right)+2\frac{k_2k_3k_5k_6}{\mathcal{A}}\right)+24\frac{k_2k_3k_5k_6}{\mathcal{C}^4}
\bigg],
\label{F2}
\end{eqnarray}
\begin{eqnarray}
\mathcal{F}_3(k_1,...,k_6)&=&\int^0_{-\infty}d\tau a(\tau)\int^\tau_{-\infty}d\tilde\tau a(\tilde\tau)U^{*'}(\tau,k_1)U^{*'}(\tau,k_2)U^{*'}(\tau,k_3)U^{*'}(\tilde\tau,k_4)U^{*}(\tilde\tau,k_5)U^{*}(\tilde\tau,k_6)
\nonumber\\
&=&2\frac{N^6c_s^4}{H^2}\frac{|k_1k_2k_3k_4|^\frac{1}{2}}{|k_5k_6|^\frac{3}{2}}\frac{1}{\mathcal{A}\mathcal{C}^3}
\left[
      1+\frac{k_5+k_6}{\mathcal{A}}+2\frac{k_5k_6}{\mathcal{A}^2}
      +\frac{3}{\mathcal{C}}\left(k_5+k_6+2\frac{k_5k_6}{\mathcal{A}}\right)
      +12\frac{k_5k_6}{\mathcal{C}^2}
\right],\nonumber\\ &&
\end{eqnarray}
\begin{eqnarray}
\mathcal{F}_4(k_1,...,k_6)&=&\int^0_{-\infty}d\tau a(\tau)\int^\tau_{-\infty}d\tilde\tau a(\tilde\tau)U^{*'}(\tau,k_1)U^{*}(\tau,k_2)U^{*}(\tau,k_3)U^{*'}(\tilde\tau,k_4)U^{*'}(\tilde\tau,k_5)U^{*'}(\tilde\tau,k_6)
\nonumber\\
&=&2\frac{N^6c_s^4}{H^2}\frac{|k_1k_4k_5k_6|^\frac{1}{2}}{|k_2k_3|^\frac{3}{2}}\frac{1}{\mathcal{A}^3\mathcal{C}}
\bigg[1+\frac{\mathcal{A}}{\mathcal{C}}+\frac{\mathcal{A}^2}{\mathcal{C}^2}+\frac{k_2+k_3}{\mathcal{C}}+2\frac{\mathcal{A}\left(k_2+k_3\right)+k_2k_3}{\mathcal{C}^2}
\nonumber\\&&
\qquad\qquad\qquad\qquad\qquad\quad
+3\frac{\mathcal{A}}{\mathcal{C}^3}\left(\mathcal{A}\left(k_2+k_3\right)+2k_2k_3\right)+12k_2k_3\frac{\mathcal{A}^2}{\mathcal{C}^4}\bigg],
\label{F4}
\end{eqnarray}
\begin{eqnarray}
\mathcal{F}_5(k_1,...,k_6)&=&\int^0_{-\infty}d\tau a(\tau)\int^\tau_{-\infty}d\tilde\tau U^{*'}(\tau,k_1)U^{*'}(\tau,k_2)U^{*'}(\tau,k_3)U^{*'}(\tilde\tau,k_4)U^{*'}(\tilde\tau,k_5)U^{*}(\tilde\tau,k_6)
\nonumber\\
&=&
\frac{4c_s^4N^6}{H}\frac{|k_1k_2k_3k_4k_5|^\frac{1}{2}}{\mathcal{A}^4\mathcal{C}^6|k_6|^\frac{3}{2}}
\big(
   30 \mathcal{A}^3 k_6 +
   (6 \mathcal{A}^2 \mathcal{C}+ 3 \mathcal{A} \mathcal{C}^2+
   \mathcal{C}^3) (k_4 + k_5 + 4 k_6)
\big),
\label{F5}
\end{eqnarray}
\begin{eqnarray}
\mathcal{F}_6(k_1,...,k_6)&=&\int^0_{-\infty}d\tau \int^\tau_{-\infty}d\tilde\tau a(\tilde\tau)U^{*'}(\tau,k_1)U^{*'}(\tau,k_2)U^{*}(\tau,k_3)U^{*'}(\tilde\tau,k_4)U^{*'}(\tilde\tau,k_5)U^{*'}(\tilde\tau,k_6)
\nonumber\\
&=&
\frac{4c_s^4N^6}{H}\frac{|k_1k_2k_4k_5k_6|^\frac{1}{2}}{\mathcal{A}^3\mathcal{C}^6|k_3|^\frac{3}{2}}
\big(
   \mathcal{C}^2 (\mathcal{C} + 3 k_3) + 3 \mathcal{A} \mathcal{C} (\mathcal{C} + 4 k_3) +
   6 \mathcal{A}^2 (\mathcal{C} + 5 k_3)
   \big),
\label{F6}
\end{eqnarray}
\begin{eqnarray}
\mathcal{F}_7(k_1,...,k_6)&=&\int^0_{-\infty}d\tau a(\tau)\int^\tau_{-\infty}d\tilde\tau U^{*'}(\tau,k_1)U^{*'}(\tau,k_2)U^{*'}(\tau,k_3)U^{*}(\tilde\tau,k_4)U^{*}(\tilde\tau,k_5)U^{*}(\tilde\tau,k_6)
\nonumber\\
&=&
\frac{-2c_s^2N^6}{H}\frac{|k_1k_2k_3|^\frac{1}{2}}{\mathcal{A}^4\mathcal{C}^6|k_4k_5k_6|^\frac{3}{2}}
\Big[
60 \mathcal{A}^3 k_4 k_5 k_6
\nonumber\\
&&
+2 \mathcal{C}^3 \Big(k_4^3 + 4 k_4 (k_5^2 + 3 k_5 k_6 + k_6^2) + (k_5 + k_6) (4 k_4^2+k_5^2 + 3 k_5 k_6 + k_6^2)\Big)
\nonumber\\
&&
+ 12 \mathcal{A}^2 \mathcal{C} \Big((k_4^2+k_5 k_6) (k_5 + k_6) +
    k_4 (k_5^2 + 6 k_5 k_6 + k_6^2)\Big)
\nonumber\\
&&
+ 3 \mathcal{A} \mathcal{C}^2 \Big(k_4^3 +
    + k_4 (5 k_5^2 + 18 k_5 k_6 + 5 k_6^2)+ (k_5 + k_6) (5k_4^2+k_5^2 + 4 k_5 k_6 + k_6^2)\Big)
\Big],\nonumber\\ &&
\label{F7}
\end{eqnarray}
\begin{eqnarray}
\mathcal{F}_8(k_1,...,k_6)&=&\int^0_{-\infty}d\tau \int^\tau_{-\infty}d\tilde\tau a(\tilde\tau)U^{*}(\tau,k_1)U^{*}(\tau,k_2)U^{*}(\tau,k_3)U^{*'}(\tilde\tau,k_4)U^{*'}(\tilde\tau,k_5)U^{*'}(\tilde\tau,k_6)
\nonumber\\
&=&
\frac{-2c_s^2N^6}{H}\frac{|k_4k_5k_6|^\frac{1}{2}}{\mathcal{A}^3\mathcal{C}^6|k_1k_2k_3|^\frac{3}{2}}
\Big[
\mathcal{C}^3(\mathcal{C}^2+\mathcal{A}\mathcal{C}+\mathcal{A}^2)+\mathcal{C}^2(k_1+k_2+k_3)(\mathcal{C}^2+2\mathcal{A}\mathcal{C}+3\mathcal{A}^2)
\nonumber\\
&&
\qquad\quad+2\mathcal{C}(k_1k_2+k_1k_3+k_2k_3)(\mathcal{C}^2+3\mathcal{A}\mathcal{C}+6\mathcal{A}^2)
+6k_1k_2k_3(\mathcal{C}^2+4\mathcal{A}\mathcal{C}+10\mathcal{A}^2)
\Big],\nonumber\\ &&
\label{F8}
\end{eqnarray}
\begin{eqnarray}
\mathcal{F}_9(k_1,...,k_6)&=&\int^0_{-\infty}d\tau \int^\tau_{-\infty}d\tilde\tau U^{*'}(\tau,k_1)U^{*'}(\tau,k_2)U^{*}(\tau,k_3)U^{*'}(\tilde\tau,k_4)U^{*'}(\tilde\tau,k_5)U^{*}(\tilde\tau,k_6)
\nonumber\\
&=&
-4c_s^2N^6\frac{|k_1k_2k_4k_5|^\frac{1}{2}}{\mathcal{A}^4\mathcal{C}^7|k_3k_6|^\frac{3}{2}}
\Big[
30\mathcal{A}^3(\mathcal{C}+6k_3)k_6
\nonumber\\
&&
\qquad\qquad\qquad+\mathcal{C}(k_4+k_5+4k_6)\Big(\mathcal{C}^2(\mathcal{C}+3k_3)+3\mathcal{A}\mathcal{C}(\mathcal{C}+4k_3)+4\mathcal{A}^2(\mathcal{C}+5k_3)\Big)
\Big],\nonumber\\ &&
\label{F9}
\end{eqnarray}
\begin{eqnarray}
\mathcal{F}_{10}(k_1,...,k_6)&=&\int^0_{-\infty}d\tau \int^\tau_{-\infty}d\tilde\tau a(\tilde\tau)U^{*'}(\tau,k_1)U^{*'}(\tau,k_2)U^{*}(\tau,k_3)U^{*'}(\tilde\tau,k_4)U^{*}(\tilde\tau,k_5)U^{*}(\tilde\tau,k_6)
\nonumber\\
&=&
\frac{-2c_s^2N^6}{H}\frac{|k_1k_2k_4|^\frac{1}{2}}{\mathcal{A}^3\mathcal{C}^6|k_3k_5k_6|^\frac{3}{2}}
\Big[
\mathcal{C} (k_4 + k_5) \Big(\mathcal{C} (\mathcal{C} + 3 k_3) k_4 + \big(2 \mathcal{C} (\mathcal{C} + 3 k_3) +
       3 \mathcal{A} (\mathcal{C} + 4 k_3)\big) k_5\Big)
\nonumber\\
&&
\qquad\qquad+ 3 \Big(4 \mathcal{A}^2 (\mathcal{C} + 5 k_3) k_5 + \mathcal{C}^2 (\mathcal{C} + 3 k_3) (k_4 + 2 k_5) +
    \mathcal{A} \mathcal{C} (\mathcal{C} + 4 k_3) (k_4 + 4 k_5)\Big) k_6
\nonumber\\
&&
\qquad\qquad+ \mathcal{C} \Big(2 \mathcal{C} (\mathcal{C} + 3 k_3) + 3 \mathcal{A} (\mathcal{C} + 4 k_3)\Big) k_6^2
\Big]
,
\label{F10}
\end{eqnarray}
\begin{eqnarray}
\mathcal{F}_{11}(k_1,...,k_6)&=&\int^0_{-\infty}d\tau a(\tau)\int^\tau_{-\infty}d\tilde\tau U^{*'}(\tau,k_1)U^{*}(\tau,k_2)U^{*}(\tau,k_3)U^{*'}(\tilde\tau,k_4)U^{*'}(\tilde\tau,k_5)U^{*}(\tilde\tau,k_6)
\nonumber\\
&=&
\frac{-2c_s^2N^6}{H}\frac{|k_1k_4k_5|^\frac{1}{2}}{\mathcal{A}^4\mathcal{C}^6|k_2k_3k_6|^\frac{3}{2}}
\Big[
3\mathcal{A}^3\left(\mathcal{C}^2+4\mathcal{C}(k_2+k_3)+20k_2k_3\right)k_6
\nonumber\\
&&
\qquad\qquad+(k_4+k_5+4k_6)\Big((\mathcal{C}^2+\mathcal{A}\mathcal{C}+\mathcal{A}^2)\mathcal{C}^3+(k_2+k_3)(\mathcal{C}^2+2\mathcal{A}\mathcal{C}+3\mathcal{A}^2)\mathcal{C}^2
\nonumber\\
&&
\qquad\qquad\qquad\qquad\qquad\qquad+2k_2k_3(\mathcal{C}^2+3\mathcal{A}\mathcal{C}+6\mathcal{A}^2)\mathcal{C}\Big)
\Big]
,
\label{F11}
\end{eqnarray}
\begin{eqnarray}
\mathcal{F}_{12}(k_1,...,k_6)&=&\int^0_{-\infty}d\tau \int^\tau_{-\infty}d\tilde\tau U^{*'}(\tau,k_1)U^{*'}(\tau,k_2)U^{*}(\tau,k_3)U^{*}(\tilde\tau,k_4)U^{*}(\tilde\tau,k_5)U^{*}(\tilde\tau,k_6)
\nonumber\\
&=&
N^6\frac{|k_1k_2|^\frac{1}{2}}{\mathcal{A}^4\mathcal{C}^7|k_3k_4k_5k_6|^\frac{3}{2}}
\Big[
120\mathcal{A}^3(\mathcal{C}+6k_3)k_4k_5k_6
\nonumber\\
&&
+4\mathcal{C}^3(\mathcal{C}+3k_3)
\Big(k_4^3+4k_4(k_5^2+3k_5k_6+k_6^2)+(k_5+k_6)(4k_4^2+k_5^2+3k_5k_6+k_6^2)\Big)
\nonumber\\
&&
+24\mathcal{A}^2\mathcal{C}(\mathcal{C}+5k_3)
\Big(k_4(k_5^2+6k_5k_6+k_6^2)+(k_5+k_6)(k_4^2+k_5k_6)\Big)
\nonumber\\
&&
+6\mathcal{A}\mathcal{C}^2(\mathcal{C}+4k_3)
\Big(k_4(k_4^2+5k_5^2+18k_5k_6+5k_6^2)+(k_5+k_6)(5k_4^2+k_5^2+4k_5k_6+k_6^2)\Big)
\Big]
,\nonumber\\
\label{F12}
\end{eqnarray}
\begin{eqnarray}
\mathcal{F}_{13}(k_1,...,k_6)&=&\int^0_{-\infty}d\tau \int^\tau_{-\infty}d\tilde\tau U^{*}(\tau,k_1)U^{*}(\tau,k_2)U^{*}(\tau,k_3)U^{*'}(\tilde\tau,k_4)U^{*'}(\tilde\tau,k_5)U^{*}(\tilde\tau,k_6)
\nonumber\\
&=&
2N^6\frac{|k_4k_5|^\frac{1}{2}}{\mathcal{A}^4\mathcal{C}^7|k_1k_2k_3k_6|^\frac{3}{2}}
\nonumber\\
&&\times\Big[
3\mathcal{A}^3\Big(\mathcal{C}^3+4\mathcal{C}^2(k_1+k_2+k_3)+20\mathcal{C}(k_1k_2+k_1k_3+k_2k_3)+120k_1k_2k_3\Big)k_6
\nonumber\\
&&+(k_4+k_5+4k_6)\Big(\mathcal{C}^4(\mathcal{C}^2+\mathcal{A}\mathcal{C}+\mathcal{A}^2)
+\mathcal{C}^3(k_1+k_2+k_3)(\mathcal{C}^2+2\mathcal{A}\mathcal{C}+3\mathcal{A}^2)
\nonumber\\
&&+2\mathcal{C}^2(k_1k_2+k_1k_3+k_2k_3)(\mathcal{C}^2+3\mathcal{A}\mathcal{C}+6\mathcal{A}^2)+6k_1k_2k_3\mathcal{C}(\mathcal{C}^2+4\mathcal{A}\mathcal{C}+10\mathcal{A}^2)\Big)
\Big]
,\nonumber\\
\label{F13}
\end{eqnarray}
\begin{eqnarray}
\mathcal{F}_{14}(k_1,...,k_6)&=&\int^0_{-\infty}d\tau a(\tau)\int^\tau_{-\infty}d\tilde\tau U^{*'}(\tau,k_1)U^{*}(\tau,k_2)U^{*}(\tau,k_3)U^{*}(\tilde\tau,k_4)U^{*}(\tilde\tau,k_5)U^{*}(\tilde\tau,k_6)
\nonumber\\
&=&
\frac{N^6}{H}\frac{|k_1|^\frac{1}{2}}{\mathcal{A}^4\mathcal{C}^6|k_2k_3k_4k_5k_6|^\frac{3}{2}}
\Big[
6 \mathcal{A}^3 \Big(\mathcal{C}^2 + 20 k_2 k_3 + 4 \mathcal{C} (k_2 + k_3)\Big) k_4 k_5 k_6
\nonumber\\
&+& 2 \mathcal{C}^3 \Big(\mathcal{C}^2 + 2 k_2 k_3 + \mathcal{C} (k_2 + k_3)\Big) \Big(k_4^3 + 4 k_4^2 (k_5 + k_6) +
    4 k_4 (k_5^2 + 3 k_5 k_6 + k_6^2)
\nonumber\\
&&
\qquad\qquad\qquad\qquad\qquad\qquad\qquad\quad+ (k_5 + k_6) (k_5^2 + 3 k_5 k_6 +
       k_6^2)\Big)
\nonumber\\
&+& 2 \mathcal{A}^2 \mathcal{C} \Big(\mathcal{C}^2 + 12 k_2 k_3 + 3 \mathcal{C} (k_2 + k_3)\Big) \Big(k_4^2 (k_5 + k_6) +
    k_5 k_6 (k_5 + k_6) \nonumber\\
&+& k_4 (k_5^2 + 6 k_5 k_6 + k_6^2)\Big)+ \mathcal{A} \mathcal{C}^2 \Big(\mathcal{C}^2 + 6 k_2 k_3 + 2 \mathcal{C} (k_2 + k_3)\Big)\\
&\times & \Big(k_4^3 +
    5 k_4^2 (k_5 + k_6) + (k_5 + k_6) (k_5^2 + 4 k_5 k_6 + k_6^2)+k_4 (5 k_5^2 + 18 k_5 k_6 + 5 k_6^2)\Big)
\Big],\nonumber
\label{F14}
\end{eqnarray}
\begin{eqnarray}
\mathcal{F}_{15}(k_1,...,k_6)&=&\int^0_{-\infty}d\tau \int^\tau_{-\infty}d\tilde\tau a(\tilde\tau)U^{*}(\tau,k_1)U^{*}(\tau,k_2)U^{*}(\tau,k_3)U^{*'}(\tilde\tau,k_4)U^{*}(\tilde\tau,k_5)U^{*}(\tilde\tau,k_6)
\nonumber\\
&=&
\frac{N^6}{H}\frac{|k_4|^\frac{1}{2}}{\mathcal{A}^3\mathcal{C}^6|k_1k_2k_3k_5k_6|^\frac{3}{2}}
\Bigg[
120 \mathcal{A}^2 k_1 k_2 k_3 k_5 k_6
\nonumber\\
&+& 6 \mathcal{C}^2 \Bigg((k_4 +
       k_5) \bigg(k_1 k_2 k_3 k_4 + (\mathcal{A} k_1 k_2 + 2 k_1 k_2 k_3 +
          \mathcal{A} (k_1 + k_2) k_3) k_5\bigg)
\nonumber\\
&+& \bigg(\mathcal{A}^2 (k_1 + k_2 + k_3) k_5 +
       3 k_1 k_2 k_3 (k_4 + 2 k_5) +
       \mathcal{A} (k_2 k_3 + k_1 (k_2 + k_3)) (k_4 + 4 k_5)\bigg) k_6
\nonumber\\
&+& (\mathcal{A} k_1 k_2 +
       2 k_1 k_2 k_3 + \mathcal{A} (k_1 + k_2) k_3) k_6^2\Bigg) +
 \mathcal{C}^4 \Bigg((k_4 +
       k_5) \bigg(\mathcal{A} k_5 + (k_1 + k_2 + k_3) (k_4 + 2 k_5)\bigg)
\nonumber\\
&+& \bigg(3 (k_1 + k_2 +
          k_3) (k_4 + 2 k_5) + \mathcal{A} (k_4 + 4 k_5)\bigg) k_6 + \bigg(\mathcal{A} +
       2 (k_1 + k_2 + k_3)\bigg) k_6^2\Bigg)
\nonumber\\
&+& 2 \mathcal{C}^3 \Bigg((k_4 +
       k_5) \bigg(\big(k_2 k_3 + k_1 (k_2 + k_3)\big) k_4 + \big(2 k_2 k_3 + 2 k_1 (k_2 + k_3) +
          \mathcal{A} (k_1 + k_2 + k_3)\big) k_5\bigg)
\nonumber\\
&+&  \bigg(\mathcal{A}^2 k_5 +
       3 \big(k_2 k_3 + k_1 (k_2 + k_3)\big) (k_4 + 2 k_5) +
       \mathcal{A} (k_1 + k_2 + k_3) (k_4 + 4 k_5)\bigg) k_6
\nonumber\\
&+&\bigg(2 k_2 k_3 + 2 k_1 (k_2 + k_3) +
       \mathcal{A} (k_1 + k_2 + k_3)\bigg) k_6^2\Bigg)
\nonumber\\
&+& 24 \mathcal{A} \mathcal{C} \Bigg(\mathcal{A} k_2 k_3 k_5 k_6 +
    k_1 \bigg(\mathcal{A} k_3 k_5 k_6 +
       k_2 \big(\mathcal{A} k_5 k_6 + k_3 (k_5 (k_4 + k_5) + (k_4 + 4 k_5) k_6 + k_6^2)\big)\bigg)\Bigg)
\bigg]\nonumber\\ &+&
 \mathcal{C}^5 \bigg(k_4^2 + 3 k_4 (k_5 + k_6) + 2 (k_5^2 + 3 k_5 k_6 + k_6^2)\bigg),
\label{F15}
\end{eqnarray}
\begin{eqnarray}
\mathcal{F}_{16}(k_1,...,k_6)&=&\int^0_{-\infty}d\tau \int^\tau_{-\infty}d\tilde\tau U^{*}(\tau,k_1)U^{*}(\tau,k_2)U^{*}(\tau,k_3)U^{*}(\tilde\tau,k_4)U^{*}(\tilde\tau,k_5)U^{*}(\tilde\tau,k_6)
\nonumber\\
&=&
\frac{N^6}{c_s^2}\frac{1}{\mathcal{A}^4\mathcal{C}^7|k_1k_2k_3k_4k_5k_6|^\frac{3}{2}}
\nonumber\\
&&
\times\Bigg[
\mathcal{A}^3 \bigg(-6 \mathcal{C} \big(\mathcal{C}^2 + 20 k_1 k_2 + 4 \mathcal{C} (k_1 + k_2)\big) -
    24 \big(\mathcal{C}^2 + 30 k_1 k_2 + 5 \mathcal{C} (k_1 + k_2)\big) k_3\bigg) k_4 k_5 k_6
\nonumber\\
&&
\quad\:+
 2 \mathcal{C}^3 \bigg(-\mathcal{C}^3 - 6 k_1 k_2 k_3 - \mathcal{C}^2 (k_1 + k_2 + k_3) -
    2 \mathcal{C} \big(k_1 k_2 + (k_1 + k_2) k_3\big)\bigg)
\nonumber\\
&&
    \qquad\times\bigg(k_4^3 + 4 k_4^2 (k_5 + k_6) +
    4 k_4 (k_5^2 + 3 k_5 k_6 + k_6^2) + (k_5 + k_6) (k_5^2 + 3 k_5 k_6 +
       k_6^2)\bigg)
\nonumber\\
&&
\quad\:+
 \mathcal{A}^2 \mathcal{C} \bigg(-2 \mathcal{C}^3 - 120 k_1 k_2 k_3 - 6 \mathcal{C}^2 (k_1 + k_2 + k_3) -
    24 \mathcal{C} \big(k_2 k_3 + k_1 (k_2 + k_3)\big)\bigg)
\nonumber\\
&&
    \qquad\times\bigg(k_4^2 (k_5 + k_6) + k_5 k_6 (k_5 + k_6) +
    k_4 (k_5^2 + 6 k_5 k_6 + k_6^2)\bigg)
\nonumber\\
&&
\quad\:-
 \mathcal{A} \mathcal{C}^2 \bigg(\mathcal{C}^3 + 24 k_1 k_2 k_3 + 2 \mathcal{C}^2 (k_1 + k_2 + k_3) +
    6 \mathcal{C} \big(k_2 k_3 + k_1 (k_2 + k_3)\big)\bigg)
\nonumber\\
&&
    \qquad\times\bigg(k_4^3 +
    5 k_4^2 (k_5 + k_6) + (k_5 + k_6) (k_5^2 + 4 k_5 k_6 + k_6^2) +
    k_4 (5 k_5^2 + 18 k_5 k_6 + 5 k_6^2)\bigg)
\Bigg]
,\nonumber\\
\label{F16}
\end{eqnarray}
where $\mathcal{A}$ is defined by the sum of the last three arguments of the $\mathcal{F}_i$ functions as  $\mathcal{A}=k_4+k_5+k_6$ and $\mathcal{C}$ is defined by the sum of all the arguments as $\mathcal{C}=k_1+k_2+k_3+k_4+k_5+k_6$.\\

\noindent When plotting the previous expressions for the $FPF^{SE}$ terms in the equilateral configuration one may find divergences. This is because in this configuration $\mathcal{C}=0$ and in the previous expressions $\mathcal{C}$ appears in the denominator. However, these divergences are spurious and they can be shown to disappear if one includes all the permutations as one should. In fact, one can use the following combinations of terms to write the previous results for the $FPF^{SE}$ in a way that is obviously finite in the equilateral limit
\begin{eqnarray}
\mathcal{F}_{1}(-k_a,-k_b,-k_{ab},k_c,k_d,k_{ab})+\mathcal{F}_{1}(-k_c,-k_d,-k_{ab},k_a,k_b,k_{ab})&=&\frac{4N^6c_s^6}{H^2}\frac{k_{ab}\sqrt{k_ak_bk_ck_d}}{\mathcal{A}_1^3\mathcal{A}_2^3},\nonumber\\
\end{eqnarray}
%%%%%%%%%%%%%%%%%%%%%%%%%%%%%%%%%%%%%%%%%%%%%%%%%%%%%%%%%%%%%%%%%%%%%%%%%%%%
\begin{eqnarray}
&&\!\!\!\!\!\!\!\!\!\!\!\!\!\!\!\!\!\!\!\!\!\!\!\!\!\!\!\!\!\!\!\!\mathcal{F}_{2}(-k_{ab},-k_a,-k_b,k_{ab},k_c,k_d)+\mathcal{F}_{2}(-k_{ab},-k_c,-k_d,k_{ab},k_a,k_b)=
\nonumber\\
&&
\frac{N^6c_s^2}{H^2}\frac{k_{ab}}{\mathcal{A}_1^3\mathcal{A}_2^3(k_ak_bk_ck_d)^\frac{3}{2}}
\Big(2 k_a^2  + (3 k_a+ k_{ab} + k_b) (k_{ab} + 2 k_b)\Big) \nonumber\\ &&\times\Big(k_{ab}^2 + 3 k_{ab} (k_c + k_d) + 2 (k_c^2 + 3 k_c k_d + k_d^2)\Big),
\end{eqnarray}
%%%%%%%%%%%%%%%%%%%%%%%%%%%%%%%%%%%%%%%%%%%%%%%%%%%%%%%%%%%%%%%%%%%%%%%%%%%%
\begin{eqnarray}
&&\!\!\!\!\!\!\!\!\!\!\!\!\!\!\!\!\!\!\!\!\!\!\!\!\!\!\!\!\!\!\!\!\mathcal{F}_{2}(-k_{ab},-k_a,-k_b,k_c,k_d,k_{ab})+\mathcal{F}_{2}(-k_c,-k_d,-k_{ab},k_{ab},k_a,k_b)=
\nonumber\\
&&
\frac{N^6c_s^2}{H^2}\frac{k_c^2}{\mathcal{A}_1^3\mathcal{A}_2^3k_{ab}(k_ak_bk_ck_d)^\frac{3}{2}}
\Big(2 k_a^2 + (3 k_a + k_{ab} + k_b) (k_{ab} + 2 k_b)\Big) \nonumber\\ &&\times
\Big(2k_{ab}^2 +(3 k_{ab}+k_c+ k_d)(k_c+2k_d)\Big)
,
\end{eqnarray}
%%%%%%%%%%%%%%%%%%%%%%%%%%%%%%%%%%%%%%%%%%%%%%%%%%%%%%%%%%%%%%%%%%%%%%%%%%%%
\begin{eqnarray}
&&\!\!\!\!\!\!\!\!\!\!\!\!\!\!\!\!\!\!\!\!\!\!\!\!\!\!\!\!\!\!\!\!\mathcal{F}_{2}(-k_b,-k_a,-k_{ab},k_d,k_c,k_{ab})+\mathcal{F}_{2}(-k_d,-k_c,-k_{ab},k_b,k_a,k_{ab})=
\nonumber\\
&&
\frac{N^6c_s^2}{H^2}\frac{k_b^2k_d^2}{\mathcal{A}_1^3\mathcal{A}_2^3k_{ab}^3(k_ak_bk_ck_d)^\frac{3}{2}}
\Big(2 k_a^2 + (3 k_a+k_{ab} + k_b) (2k_{ab} +  k_b)\Big) \nonumber\\ &&\times
\Big(2k_{ab}^2 + (3 k_{ab}+k_c+k_d) (2k_c + k_d)\Big),
\end{eqnarray}
%%%%%%%%%%%%%%%%%%%%%%%%%%%%%%%%%%%%%%%%%%%%%%%%%%%%%%%%%%%%%%%%%%%%%%%%%%%%
\begin{eqnarray}
&&\!\!\!
\mathcal{F}_{3}(-k_a,-k_b,-k_{ab},k_{ab},k_c,k_d)+\mathcal{F}_{4}(-k_{ab},-k_c,-k_d,k_a,k_b,k_{ab})=
\nonumber\\
&&
\qquad\qquad\qquad-\frac{2N^6c_s^4}{H^2}\frac{k_a^2k_{ab}k_b^2}{\mathcal{A}_1^3\mathcal{A}_2^3(k_ak_bk_ck_d)^\frac{3}{2}}
\Big(k_{ab}^2 + 3 k_{ab} (k_c + k_d) + 2 (k_c^2 + 3 k_c k_d + k_d^2)\Big),\nonumber\\&&
\end{eqnarray}
%%%%%%%%%%%%%%%%%%%%%%%%%%%%%%%%%%%%%%%%%%%%%%%%%%%%%%%%%%%%%%%%%%%%%%%%%%%%
\begin{eqnarray}
&&\!\!\!\!\!\!\!\!\!\!\!\!\!\!\!\!\!\!\!\!\!\!\!\!\!\!\!\!\!\!\!\!\!\!\!\!\!\!\!\!
\mathcal{F}_{3}(-k_a,-k_b,-k_{ab},k_c,k_d,k_{ab})+\mathcal{F}_{4}(-k_c,-k_d,-k_{ab},k_a,k_b,k_{ab})=
\nonumber\\
&&
-\frac{2N^6c_s^4}{H^2}\frac{\sqrt{k_ak_bk_ck_d}}{\mathcal{A}_1^3\mathcal{A}_2^3k_{ab}k_d^2}
\Big(2k_{ab}^2 + (3 k_{ab}+k_c+k_d )(k_c + 2k_d)\Big),
\end{eqnarray}
%%%%%%%%%%%%%%%%%%%%%%%%%%%%%%%%%%%%%%%%%%%%%%%%%%%%%%%%%%%%%%%%%%%%%%%%%%%%
\begin{eqnarray}
&&\!\!\!\!\!\!\!\!\!\!\!\!\!\!\!\!\!\!\!\!\!\!\!\!\!\!\!\!\!\!\!\!\!\!\!\!\!\!\!\!
\mathcal{F}_{5}(-k_a,-k_b,-k_{ab},k_c,k_d,k_{ab})+\mathcal{F}_{6}(-k_c,-k_d,-k_{ab},k_a,k_b,k_{ab})=
\nonumber\\
&&
\frac{4N^6c_s^4}{H}\frac{\sqrt{k_ak_bk_ck_d}}{\mathcal{A}_1^4\mathcal{A}_2^3k_{ab}}
\Big(4k_{ab}+k_c+k_d\Big),
\end{eqnarray}
 %%%%%%%%%%%%%%%%%%%%%%%%%%%%%%%%%%%%%%%%%%%%%%%%%%%%%%%%%%%%%%%%%%%%%%%%%%%%
\begin{eqnarray}
&&\!\!\!\!\!\!\!\!\!\!\!\!\!\!\!\!\!\!\!\!\!\!\!\!\!\!\!\!\!\!\!\!\!\!\!\!\!\!\!\!
\mathcal{F}_{5}(-k_a,-k_b,-k_{ab},k_c,k_{ab},k_d)+\mathcal{F}_{6}(-k_c,-k_{ab},-k_d,k_a,k_b,k_{ab})=
\nonumber\\
&&\frac{4N^6c_s^4}{H}\frac{k_{ab}\sqrt{k_ak_bk_ck_d}}{\mathcal{A}_1^4\mathcal{A}_2^3k_d^2}
\Big(k_{ab}+k_c+4k_d\Big),
\end{eqnarray}
%%%%%%%%%%%%%%%%%%%%%%%%%%%%%%%%%%%%%%%%%%%%%%%%%%%%%%%%%%%%%%%%%%%%%%%%%%%%
\begin{eqnarray}
&&\!\!\!\!\!\!\!\!\!\!\!\!\!\!\!\!
\mathcal{F}_{7}(-k_a,-k_b,-k_{ab},k_c,k_d,k_{ab})+\mathcal{F}_{8}(-k_c,-k_d,-k_{ab},k_a,k_b,k_{ab})=
\\
&&
-\frac{4N^6c_s^2}{H}\frac{k_a^2k_b^2}{\mathcal{A}_1^4\mathcal{A}_2^3k_{ab}(k_ak_bk_ck_d)^\frac{3}{2}}
\Big(k_{ab}^3 + 4 k_{ab}^2 (k_c + k_d) + (4k_{ab}+k_c+k_d) (k_c^2 + 3 k_c k_d + k_d^2)\Big),\nonumber
\end{eqnarray}
%%%%%%%%%%%%%%%%%%%%%%%%%%%%%%%%%%%%%%%%%%%%%%%%%%%%%%%%%%%%%%%%%%%%%%%%%%%%
\begin{eqnarray}
&&\!\!\!\!\!\!\!\!\!\!\!\!\!\!\!\!\!\!\!\!\!\!\!\!\!\!\!\!\!\!\!\!\!\!\!\!\!\!\!\!
\mathcal{F}_{9}(-k_a,-k_b,-k_{ab},k_c,k_d,k_{ab})+\mathcal{F}_{9}(-k_c,-k_d,-k_{ab},k_a,k_b,k_{ab})=\nonumber\\
&&
4N^6c_s^2\frac{\sqrt{k_ak_bk_ck_d}}{\mathcal{A}_1^4\mathcal{A}_2^4k_{ab}^3}
\Big(4k_{ab}+k_a+k_b\Big)\Big(4k_{ab}+k_c+k_d\Big),
\end{eqnarray}
%%%%%%%%%%%%%%%%%%%%%%%%%%%%%%%%%%%%%%%%%%%%%%%%%%%%%%%%%%%%%%%%%%%%%%%%%%%%
\begin{eqnarray}
&&\!\!\!\!\!\!\!\!\!\!\!\!\!\!\!\!\!\!\!\!\!\!\!\!\!\!\!\!\!\!\!\!\!\!\!\!\!\!\!\!
\mathcal{F}_{9}(-k_a,-k_b,-k_{ab},k_c,k_{ab},k_d)+\mathcal{F}_{9}(-k_c,-k_{ab},-k_d,k_a,k_b,k_{ab})=\nonumber\\
&&
4N^6c_s^2\frac{\sqrt{k_ak_bk_ck_d}}{\mathcal{A}_1^4\mathcal{A}_2^4k_{ab}k_d^2}
\Big(4k_{ab}+k_a+k_b\Big)\Big(k_{ab}+k_c+4k_d\Big),
\end{eqnarray}
%%%%%%%%%%%%%%%%%%%%%%%%%%%%%%%%%%%%%%%%%%%%%%%%%%%%%%%%%%%%%%%%%%%%%%%%%%%%
\begin{eqnarray}
&&\!\!\!\!\!\!\!\!\!\!\!\!\!\!\!\!\!\!\!\!\!\!\!\!
\mathcal{F}_{9}(-k_a,-k_{ab},-k_b,k_c,k_{ab},k_d)+\mathcal{F}_{9}(-k_c,-k_{ab},-k_d,k_a,k_{ab},k_b)=\\&&\qquad\qquad
4N^6c_s^2\frac{k_{ab}k_a^2k_c^2}{\mathcal{A}_1^4\mathcal{A}_2^4(k_ak_bk_ck_d)^\frac{3}{2}}
\Big(k_{ab}+k_a+4k_b\Big)\Big(k_{ab}+k_c+4k_d\Big),\nonumber\\\nonumber
\end{eqnarray}
%%%%%%%%%%%%%%%%%%%%%%%%%%%%%%%%%%%%%%%%%%%%%%%%%%%%%%%%%%%%%%%%%%%%%%%%%%%%
\begin{eqnarray}
&&\!\!\!\!\!\!\!\!\!\!\!\!\!\!\!\!\!\!
\mathcal{F}_{10}(-k_a,-k_b,-k_{ab},k_{ab},k_c,k_d)+\mathcal{F}_{11}(-k_{ab},-k_c,-k_d,k_a,k_b,k_{ab})=
\\
&&\qquad\qquad
-\frac{2N^6c_s^2}{H}\frac{k_a^2k_b^2(k_a+4k_{ab}+k_b)}{\mathcal{A}_1^3\mathcal{A}_2^4k_{ab}(k_ak_bk_ck_d)^\frac{3}{2}}
\Big(k_{ab}^2 + 3 k_{ab} (k_c + k_d) + 2 (k_c^2 + 3 k_c k_d + k_d^2)\Big),\nonumber\\\nonumber
\end{eqnarray}
%%%%%%%%%%%%%%%%%%%%%%%%%%%%%%%%%%%%%%%%%%%%%%%%%%%%%%%%%%%%%%%%%%%%%%%%%%%%
\begin{eqnarray}
&&\!\!\!\!\!\!\!\!\!\!\!\!\!\!\!\!\!\!\!\!
\mathcal{F}_{10}(-k_a,-k_b,-k_{ab},k_c,k_d,k_{ab})+\mathcal{F}_{11}(-k_c,-k_d,-k_{ab},k_a,k_b,k_{ab})=
\\
&&\qquad\qquad
-\frac{2N^6c_s^2}{H}\frac{\sqrt{k_ak_bk_ck_d}(k_a+4k_{ab}+k_b)}{\mathcal{A}_1^3\mathcal{A}_2^4k_{ab}^3k_d^2}
\Big(2k_{ab}^2 + (3 k_{ab} +k_c + k_d)(k_c + 2k_d)\Big),\nonumber\\\nonumber
\end{eqnarray}
%%%%%%%%%%%%%%%%%%%%%%%%%%%%%%%%%%%%%%%%%%%%%%%%%%%%%%%%%%%%%%%%%%%%%%%%%%%%
\begin{eqnarray}
&&\!\!\!\!\!\!\!\!\!\!\!\!\!\!\!\!\!\!
\mathcal{F}_{10}(-k_a,-k_{ab},-k_b,k_{ab},k_c,k_d)+\mathcal{F}_{11}(-k_{ab},-k_c,-k_d,k_a,k_{ab},k_b)=
\\
&&\qquad\qquad
-\frac{2N^6c_s^2}{H}\frac{k_a^2k_{ab}(k_a+k_{ab}+4k_b)}{\mathcal{A}_1^3\mathcal{A}_2^4(k_ak_bk_ck_d)^\frac{3}{2}}
\Big(k_{ab}^2 + 3 k_{ab} (k_c + k_d) + 2 (k_c^2 + 3 k_c k_d + k_d^2)\Big),\nonumber\\\nonumber
\end{eqnarray}
%%%%%%%%%%%%%%%%%%%%%%%%%%%%%%%%%%%%%%%%%%%%%%%%%%%%%%%%%%%%%%%%%%%%%%%%%%%%
\begin{eqnarray}
&&\!\!\!\!\!\!\!\!\!\!\!\!\!\!\!\!\!\!\!\!\!\!\!\!\!\!
\mathcal{F}_{10}(-k_a,-k_{ab},-k_b,k_c,k_d,k_{ab})+\mathcal{F}_{11}(-k_c,-k_d,-k_{ab},k_a,k_{ab},k_b)=\\
&&\qquad
-\frac{2N^6c_s^2}{H}\frac{k_a^2k_c^2(k_a+k_{ab}+4k_b)}{\mathcal{A}_1^3\mathcal{A}_2^4k_{ab}(k_ak_bk_ck_d)^\frac{3}{2}}
\Big(2k_{ab}^2 + (3 k_{ab}+k_c+k_d) (k_c + 2k_d)\Big),\nonumber\\\nonumber
\end{eqnarray}
%%%%%%%%%%%%%%%%%%%%%%%%%%%%%%%%%%%%%%%%%%%%%%%%%%%%%%%%%%%%%%%%%%%%%%%%%%%%
\begin{eqnarray}
&&\!\!\!\!\!\!\!\!\!\!
\mathcal{F}_{12}(-k_a,-k_b,-k_{ab},k_c,k_d,k_{ab})+\mathcal{F}_{13}(-k_c,-k_d,-k_{ab},k_a,k_b,k_{ab})=
\\
&&
-4N^6\frac{k_a^2k_b^2(k_a+4k_{ab}+k_b)}{\mathcal{A}_1^4\mathcal{A}_2^4k_{ab}^3(k_ak_bk_ck_d)^\frac{3}{2}}
\Big(k_{ab}^3 + 4 k_{ab}^2 (k_c + k_d) + (4k_{ab}+k_c+k_d) (k_c^2 + 3 k_c k_d + k_d^2)\Big),\nonumber\\\nonumber
\end{eqnarray}
%%%%%%%%%%%%%%%%%%%%%%%%%%%%%%%%%%%%%%%%%%%%%%%%%%%%%%%%%%%%%%%%%%%%%%%%%%%%
\begin{eqnarray}
&&\!\!\!\!\!\!\!\!\!\!
\mathcal{F}_{12}(-k_a,-k_{ab},-k_b,k_c,k_d,k_{ab})+\mathcal{F}_{13}(-k_c,-k_d,-k_{ab},k_a,k_{ab},k_b)=
\\
&&
-4N^6\frac{k_a^2(k_a+k_{ab}+4k_b)}{\mathcal{A}_1^4\mathcal{A}_2^4k_{ab}(k_ak_bk_ck_d)^\frac{3}{2}}
\Big(k_{ab}^3 + 4 k_{ab}^2 (k_c + k_d) + (4k_{ab}+k_c+k_d) (k_c^2 + 3 k_c k_d + k_d^2)\Big),\nonumber\\\nonumber
\end{eqnarray}
%%%%%%%%%%%%%%%%%%%%%%%%%%%%%%%%%%%%%%%%%%%%%%%%%%%%%%%%%%%%%%%%%%%%%%%%%%%%
\begin{eqnarray}
&&\!\!\!\!\!\!\!\!\!\!\!\!\!\!\!\!\!\!\!\!
\mathcal{F}_{14}(-k_{ab},-k_a,-k_b,k_c,k_d,k_{ab})+\mathcal{F}_{15}(-k_c,-k_d,-k_{ab},k_{ab},k_a,k_b)=
\\
&&
\frac{2N^6}{H}\frac{1}{\mathcal{A}_1^4\mathcal{A}_2^3k_{ab}(k_ak_bk_ck_d)^\frac{3}{2}}
\Big(2 k_a^2  + (3 k_a+ k_{ab} + k_b) (k_{ab} + 2 k_b)\Big)
\nonumber\\
&&\qquad\qquad\qquad\qquad\qquad\qquad
\times\Big(k_{ab}^3 + 4 k_{ab}^2 (k_c + k_d) + (4k_{ab}+k_c+k_d) (k_c^2 + 3 k_c k_d + k_d^2)\Big),\nonumber\\\nonumber
\end{eqnarray}
%%%%%%%%%%%%%%%%%%%%%%%%%%%%%%%%%%%%%%%%%%%%%%%%%%%%%%%%%%%%%%%%%%%%%%%%%%%%
\begin{eqnarray}
&&\!\!\!\!\!\!\!\!\!\!\!\!\!\!\!\!\!\!\!\!
\mathcal{F}_{14}(-k_a,-k_b,-k_{ab},k_c,k_d,k_{ab})+\mathcal{F}_{15}(-k_c,-k_d,-k_{ab},k_a,k_b,k_{ab})=
\nonumber\\
&&
\frac{2N^6}{H}\frac{k_a^2}{\mathcal{A}_1^4\mathcal{A}_2^3k_{ab}^3(k_ak_bk_ck_d)^\frac{3}{2}}
\Big(k_a^2  + 3 k_a (k_{ab} + k_b)+2(k_{ab}^2+3k_{ab}k_b+k_b^2)\Big)
\\
&&\qquad\qquad\qquad\qquad\qquad\qquad
\times\Big(k_{ab}^3 + 4 k_{ab}^2 (k_c + k_d) + (4k_{ab}+k_c+k_d) (k_c^2 + 3 k_c k_d + k_d^2)\Big),\nonumber\\\nonumber
\end{eqnarray}
%%%%%%%%%%%%%%%%%%%%%%%%%%%%%%%%%%%%%%%%%%%%%%%%%%%%%%%%%%%%%%%%%%%%%%%%%%%%
\begin{eqnarray}
&&\!\!\!\!\!\!\!\!\!\!\!\!\!\!\!\!\!\!\!\!
\mathcal{F}_{16}(-k_a,-k_b,-k_{ab},k_c,k_d,k_{ab})+\mathcal{F}_{16}(-k_c,-k_d,-k_{ab},k_a,k_b,k_{ab})=
\\
&&
\frac{4N^6}{c_s^2}\frac{1}{\mathcal{A}_1^4\mathcal{A}_2^4k_{ab}^3(k_ak_bk_ck_d)^\frac{3}{2}}
\Big(k_a^3  + 4 k_a^2 (k_{ab} + k_b)+(4k_a+k_{ab}+k_b)(k_{ab}^2+3k_{ab}k_b+k_b^2)\Big)\nonumber
\\
&&\qquad\qquad\qquad\qquad\qquad\qquad
\times\Big(k_{ab}^3 + 4 k_{ab}^2 (k_c + k_d) + (4k_{ab}+k_c+k_d) (k_c^2 + 3 k_c k_d + k_d^2)\Big),\nonumber
\end{eqnarray}
where we defined $\mathcal{A}_1$ and $\mathcal{A}_2$ as $\mathcal{A}_1=-(k_{ab}+k_c+k_d)$ and $\mathcal{A}_2=-(k_{ab}+k_a+k_b)$.\nonumber

\section{Appendix~D. $t_{NL}$ amplitudes from the primordial parameters}

The trispectrum amplitudes $t_{NL}$ from the different contributions have the following form
\begin{equation}
t_{NL}^{(i)}\sim \frac{\mathcal{C}_{(i)}}{A^{2}c_{s}^{n_{(i)}}}.\nonumber
\end{equation}
The coefficient $\mathcal{C}_{(i)}$ are linear combinations of products of $c_{2}$, $Z\,c_{3}$, $Z^{2}\,c_{4}$ and $Z^{3}\,c_{5}$. \\More precisely we have
\begin{eqnarray}
&&t_{NL}^{\mathcal{O}_1\mathcal{O}_1}=0.063\frac{\mathcal{C}_{f_{1}f_{1}}}{A^2},\quad
t_{NL}^{\mathcal{O}_3\mathcal{O}_3}=0.31\frac{\mathcal{C}_{f_{3}f_{3}}}{A^2c_s^4},\quad
t_{NL}^{\mathcal{O}_1\mathcal{O}_3}=-0.21\frac{\mathcal{C}_{f_{1}f_{3}}}{A^2c_s^2},\quad
t_{NL}^{\mathcal{O}_1\mathcal{O}_2}=0.25\frac{\mathcal{C}_{f_{1}f_{2}}}{A^2c_s^2},\nonumber
\\
&&
t_{NL}^{\mathcal{O}_1\mathcal{O}_4}=-0.33\frac{\mathcal{C}_{f_{1}f_{4}}}{A^2c_s^4},\quad
t_{NL}^{\mathcal{O}_2\mathcal{O}_3}=-0.37\frac{\mathcal{C}_{f_{2}f_{3}}}{A^2c_s^4},\quad
t_{NL}^{\mathcal{O}_2\mathcal{O}_4}=-0.50\frac{\mathcal{C}_{f_{2}f_{4}}}{A^2c_s^6},\nonumber
\\
&&
t_{NL}^{\mathcal{O}_3\mathcal{O}_4}=0.85\frac{\mathcal{C}_{f_{3}f_{4}}}{A^2c_s^6},\quad
t_{NL}^{\mathcal{O}_2\mathcal{O}_2}=0.23\frac{\mathcal{C}_{f_{2}f_{2}}}{A^2c_s^4},\quad
t_{NL}^{\mathcal{O}_4\mathcal{O}_4}=0.52\frac{\mathcal{C}_{f_{4}f_{4}}}{A^2c_s^8},\nonumber
\end{eqnarray}
for the scalar-exchange contributions  and
\begin{eqnarray}
&&
t_{NL}^{V_1}=-0.035\frac{\mathcal{C}_{V_{1}}}{A^{2}},\quad
t_{NL}^{V_2}=-0.079\frac{\mathcal{C}_{V_{2}}}{A^{2} c_s^2},\quad
t_{NL}^{V_3}=0.051\frac{\mathcal{C}_{V_{3}}}{A^{2} c_s^2},\quad
t_{NL}^{V_4}=-0.19\frac{\mathcal{C}_{V_{4}}}{A^{2} c_s^4},
\nonumber\\
&&
t_{NL}^{V_5}=-0.031\frac{\mathcal{C}_{V_{5}}}{A^{2} c_s^4},\quad
t_{NL}^{V_6}=0.10\frac{\mathcal{C}_{V_{6}}}{A^{2} c_s^4},\quad
t_{NL}^{V_7}=-0.20\frac{\mathcal{C}_{V_{7}}}{A^{2} c_s^4},\quad
t_{NL}^{V_8}=0.16\frac{\mathcal{C}_{V_{8}}}{A^{2} c_s^6},\nonumber
\end{eqnarray}
for the contact interaction trispectra. The coefficients $\mathcal{C}_i$ appearing in the previous equations are listed below
\begin{eqnarray}
&&\mathcal{C}_{f_{1}f_{1}}\equiv 4 \bar{c}_3^2 + 72 \bar{c}_3 \bar{c}_4 + 324 \bar{c}_4^2 + 240 \bar{c}_3 \bar{c}_5 + 2160 \bar{c}_4 \bar{c}_5 + 3600 \bar{c}_5^2 ,\nonumber\\
&&\mathcal{C}_{f_{1}f_{2}}\equiv -4 \bar{c}_3^2 - 60 \bar{c}_3 \bar{c}_4 - 216 \bar{c}_4^2 - 192 \bar{c}_3 \bar{c}_5 - 1368 \bar{c}_4 \bar{c}_5 - 2160 \bar{c}_5^2 ,\nonumber\\
&&\mathcal{C}_{f_{1}f_{3}}\equiv -4 \bar{c}_3^2 - 64 \bar{c}_3 \bar{c}_4 - 252 \bar{c}_4^2 - 192 \bar{c}_3 \bar{c}_5 - 1488 \bar{c}_4 \bar{c}_5 - 2160 \bar{c}_5^2 ,\nonumber\\
&&\mathcal{C}_{f_{1}f_{4}}\equiv 2 \bar{c}_3^2 + 24 \bar{c}_3 \bar{c}_4 + 54 \bar{c}_4^2 + 72 \bar{c}_3 \bar{c}_5 + 288 \bar{c}_4 \bar{c}_5 + 360 \bar{c}_5^2 ,\nonumber\\
&&\mathcal{C}_{f_{2}f_{2}}\equiv 4 \bar{c}_3^2 + 48 \bar{c}_3 \bar{c}_4 + 144 \bar{c}_4^2 + 144 \bar{c}_3 \bar{c}_5 + 864 \bar{c}_4 \bar{c}_5 + 1296 \bar{c}_5^2 ,\nonumber\\
&&\mathcal{C}_{f_{2}f_{3}}\equiv 4 \bar{c}_3^2 + 52 \bar{c}_3 \bar{c}_4 + 168 \bar{c}_4^2 + 144 \bar{c}_3 \bar{c}_5 + 936 \bar{c}_4 \bar{c}_5 + 1296 \bar{c}_5^2 ,\nonumber\\
&&\mathcal{C}_{f_{2}f_{4}}\equiv -2 \bar{c}_3^2 - 18 \bar{c}_3 \bar{c}_4 - 36 \bar{c}_4^2 - 48 \bar{c}_3 \bar{c}_5 - 180 \bar{c}_4 \bar{c}_5 - 216 \bar{c}_5^2 ,\nonumber\\
&&\mathcal{C}_{f_{3}f_{3}}\equiv 4 \bar{c}_3^2 + 56 \bar{c}_3 \bar{c}_4 + 196 \bar{c}_4^2 + 144 \bar{c}_3 \bar{c}_5 + 1008 \bar{c}_4 \bar{c}_5 + 1296 \bar{c}_5^2 ,\nonumber\\
&&\mathcal{C}_{f_{3}f_{4}}\equiv -2 \bar{c}_3^2 - 20 \bar{c}_3 \bar{c}_4 - 42 \bar{c}_4^2 - 48 \bar{c}_3 \bar{c}_5 - 192 \bar{c}_4 \bar{c}_5 - 216 \bar{c}_5^2 ,\nonumber\\
&&\mathcal{C}_{f_{4}f_{4}}\equiv \bar{c}_3^2 + 6 \bar{c}_3 \bar{c}_4 + 9 \bar{c}_4^2 + 12 \bar{c}_3 \bar{c}_5 + 36 \bar{c}_4 \bar{c}_5 + 36 \bar{c}_5^2  ,\nonumber\\
&&\mathcal{C}_{V_{1}}\equiv 9 \bar{c}_3^2 - \frac{9}{4} \bar{c}_2 \bar{c}_4 + 135 \bar{c}_3 \bar{c}_4 + \frac{1215}{2} \bar{c}_4^2 - 15 \bar{c}_2 \bar{c}_5 +
 360 \bar{c}_3 \bar{c}_5 + 3780 \bar{c}_4 \bar{c}_5 + 6300 \bar{c}_5^2 ,\nonumber\\
&&\mathcal{C}_{V_{2}}\equiv -12 \bar{c}_3^2 + 2 \bar{c}_2 \bar{c}_4 - 156 \bar{c}_3 \bar{c}_4 - 540 \bar{c}_4^2 + 12 \bar{c}_2 \bar{c}_5 - 432 \bar{c}_3 \bar{c}_5 -
 3216 \bar{c}_4 \bar{c}_5 - 5040 \bar{c}_5^2 ,\nonumber\\
&&\mathcal{C}_{V_{3}}\equiv -6 \bar{c}_3^2+\frac{7}{2} \bar{c}_2 \bar{c}_4-54 \bar{c}_3 \bar{c}_4-189 \bar{c}_4^2+18 \bar{c}_2 \bar{c}_5-72 \bar{c}_3 \bar{c}_5-840 \bar{c}_4 \bar{c}_5-1080 \bar{c}_5^2 ,\nonumber\\
&&\mathcal{C}_{V_{4}}\equiv 4 \bar{c}_3^2 - \frac{3}{2} \bar{c}_2 \bar{c}_4 + 30 \bar{c}_3 \bar{c}_4 + 63 \bar{c}_4^2 - 6 \bar{c}_2 \bar{c}_5 + 72 \bar{c}_3 \bar{c}_5 +
 360 \bar{c}_4 \bar{c}_5 + 576 \bar{c}_5^2 ,\nonumber\\
&&\mathcal{C}_{V_{5}}\equiv \frac{3}{2} \bar{c}_2 \bar{c}_4 + 18 \bar{c}_3 \bar{c}_4 + 81 \bar{c}_4^2 + 6 \bar{c}_2 \bar{c}_5 + 72 \bar{c}_3 \bar{c}_5 + 504 \bar{c}_4 \bar{c}_5 +
 720 \bar{c}_5^2 ,\nonumber\\
&&\mathcal{C}_{V_{6}}\equiv 4 \bar{c}_3^2 - 3 \bar{c}_2 \bar{c}_4 + 16 \bar{c}_3 \bar{c}_4 + 6 \bar{c}_4^2 - 12 \bar{c}_2 \bar{c}_5 - 72 \bar{c}_4 \bar{c}_5 - 144 \bar{c}_5^2 ,\nonumber\\
&&\mathcal{C}_{V_{7}}\equiv \bar{c}_3^2 - \bar{c}_2 \bar{c}_4 + 2 \bar{c}_3 \bar{c}_4 - 5 \bar{c}_4^2 - 3 \bar{c}_2 \bar{c}_5 - 30 \bar{c}_4 \bar{c}_5 - 36 \bar{c}_5^2 ,\nonumber\\
&&\mathcal{C}_{V_{8}}\equiv \frac{\bar{c}_2 \bar{c}_4}{2} + 6 \bar{c}_3 \bar{c}_4 + 27 \bar{c}_4^2 + 60 \bar{c}_4 \bar{c}_5  ,\nonumber
\end{eqnarray}
where $\bar{c}_{2}\equiv c_{2}$, $\bar{c}_{3}\equiv Z\,c_{3}$, $\bar{c}_{4}\equiv Z^{2}\,c_{4}$ and $\bar{c}_{5}\equiv Z^{3}\,c_{5}$.

\section{Appendix~E. Reproducing the results of the $P(X,\phi)$ inflation model}

%\subsection{The trispectrum of the $P(X,\phi)$ inflation model}

The action for Galilean inflation fluctuations contains all the leading-order vertices of the $P(X,\phi)$ model, which implies that one can use our results to reproduce the results of \cite{Arroja:2009pd,Chen:2009bc}.
This can be simply achieved by using the following expressions for the coupling constants
\begin{eqnarray}
&&A=\frac{P_{,X}}{2\tilde{c}_s^2},\quad
B=-\frac{P_{,X}}{2},\quad
\mathcal{O}_1=\frac{\dot\phi_0^3}{6}P_{,XXX}+\frac{\dot\phi_0^2}{2}P_{,XX},\nonumber \\
&& \mathcal{O}_3=-\frac{\dot\phi_0^2}{2}P_{,XX},\quad \mathcal{O}_2=\mathcal{O}_4=\beta_1=\gamma_1=\gamma_2=\Delta=0 ,\quad \gamma_3=\frac{P_{,XX}}{8},
\nonumber\\
&&
\alpha=\frac{1}{4}\left(\frac{\dot\phi_0^4}{6}P_{,XXXX}+\dot\phi_0^2P_{,XXX}+\frac{1}{2}P_{,XX}\right),\quad
\beta_2=-\frac{1}{4}\left(\dot\phi_0^2P_{,XXX}+P_{,XX}\right) \end{eqnarray}
where the subscript ${}_{,X}$ denotes derivative with respect to $X$ and $\tilde{c}_s$ is defined as usual for the $P(X,\phi)$ lagrangian as $\tilde{c}_s^2\equiv P_{,X}/(P_{,X}+2XP_{,XX})=c_s^2$.

%%%%%%%%%%%%%%%%%%%%%%%%%%%%%%%%%%%%%%%%%%%%%%%%%%%%%%%%%%%%%%%%%%%%
%\bibliography{bibliography}
%\bibliographystyle{JHEP}

\end{document}